\documentclass[12pt,twoside,leftblank]{nuthesis}

\usepackage[utf8]{inputenc}
\usepackage{longtable}
\usepackage{graphicx}
\usepackage[linesnumbered,ruled,vlined]{algorithm2e}
\usepackage{subfig}
\usepackage{cite}
\usepackage{hyperref}
\usepackage{siunitx}
\usepackage[noend]{algpseudocode}
\usepackage{amsmath}
\usepackage{comment}
\usepackage{tikz}
\usepackage{overpic}
\usepackage{booktabs}

\newcounter{customcounter}
\algrenewcommand\algorithmicrequire{\textbf{Input:}}
\algrenewcommand\algorithmicensure{\textbf{Output:}}
\usepackage{lgrind}
\usepackage{cmap}
\usepackage[T1]{fontenc}

\usepackage{multirow}
\usepackage{caption}
\usepackage{soul}
\usepackage{amssymb}
\usepackage{amsfonts}
\usepackage{float}
\usepackage[nolist,nohyperlinks]{acronym}
\usepackage{url} 
\usepackage{setspace}

\begin{document}

%The header pages

\pagestyle{empty}
\begin{titlepage}
    \centering
    
    \LARGE
    NATIONAL UNIVERSITY OF COMPUTER AND EMERGING SCIENCES \\
    \vspace{2cm}
    \Large
   FAST School of Computing \\
    \vspace{2cm}
    
    \includegraphics[width=0.5\textwidth]{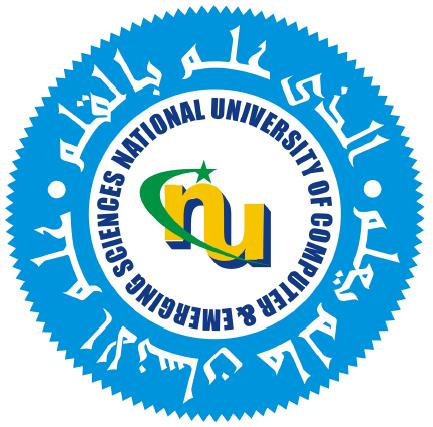}
    
    \vspace{1cm}
    
    \Huge
    \textbf{Abnormalities and Disease Detection in Gastro-Intestinal Tract Images} \\
    \vspace{2cm}

    \Large
    Zeshan Khan \\
    \vspace{1cm}
    PhD (Computer Science) \\
    \vspace{1cm}
    2024 \\
    
\end{titlepage}
\pagestyle{empty}

% -*-latex-*-
% 
% For questions, comments, concerns or complaints:
% thesis@mit.edu
% 
%
% $Log: cover.tex,v $
% Revision 1.8  2008/05/13 15:02:15  jdreed
% Degree month is June, not May.  Added note about prevdegrees.
% Arthur Smith's title updated
%
% Revision 1.7  2001/02/08 18:53:16  boojum
% changed some \newpages to \cleardoublepages
%
% Revision 1.6  1999/10/21 14:49:31  boojum
% changed comment referring to documentstyle
%
% Revision 1.5  1999/10/21 14:39:04  boojum
% *** empty log message ***
%
% Revision 1.4  1997/04/18  17:54:10  othomas
% added page numbers on abstract and cover, and made 1 abstract
% page the default rather than 2.  (anne hunter tells me this
% is the new institute standard.)
%
% Revision 1.4  1997/04/18  17:54:10  othomas
% added page numbers on abstract and cover, and made 1 abstract
% page the default rather than 2.  (anne hunter tells me this
% is the new institute standard.)
%
% Revision 1.3  93/05/17  17:06:29  starflt
% Added acknowledgements section (suggested by tompalka)
% 
% Revision 1.2  92/04/22  13:13:13  epeisach
% Fixes for 1991 course 6 requirements
% Phrase "and to grant others the right to do so" has been added to 
% permission clause
% Second copy of abstract is not counted as separate pages so numbering works
% out
% 
% Revision 1.1  92/04/22  13:08:20  epeisach

% NOTE:
% These templates make an effort to conform to the MIT Thesis specifications,
% however the specifications can change.  We recommend that you verify the
% layout of your title page with your thesis advisor and/or the MIT 
% Libraries before printing your final copy.

\title{Abnormalities and Disease Detection in Gastro-Intestinal Tract Images}

\author{Zeshan Khan}
% If you wish to list your previous degrees on the cover page, use the 
% previous degrees command:
%       \prevdegrees{A.A., Harvard University (1985)}
% You can use the \\ command to list multiple previous degrees
%       \prevdegrees{B.S., University of California (1978) \\
%                    S.M., Massachusetts Institute of Technology (1981)}
\department{FAST-School of Computing}

% If the thesis is for two degrees simultaneously, list them both
% separated by \and like this:
% \degree{Doctor of Philosophy \and Master of Science}
\degree{Doctor of Philosophy in Computer Science}

% As of the 2007-08 academic year, valid degree months are September, 
% February, or June.  The default is June.

\degreemonth{December}
\degreeyear{2023}
\thesisdate{December, 2023}

%% By default, the thesis will be copyrighted to MIT.  If you need to copyright
%% the thesis to yourself, just specify the `vi' documentclass option.  If for
%% some reason you want to exactly specify the copyright notice text, you can
%% use the \copyrightnoticetext command.  
%\copyrightnoticetext{\copyright IBM, 1990.  Do not open till Xmas.}

% If there is more than one supervisor, use the \supervisor command
% once for each.
\supervisor{Dr. Muhammad Atif Tahir}{Professor}
%\cosupervisor{Dr. Muhammad Rafi}{Assistant Professor}

% This is the department committee chairman, not the thesis committee
% chairman.  You should replace this with your Department's Committee
% Chairman.

\chairman{Dr. Ghufran Ahmed}{Chair, Campus Committee on Graduate Theses}

% Make the titlepage based on the above information.  If you need
% something special and can't use the standard form, you can specify
% the exact text of the titlepage yourself.  Put it in a titlepage
% environment and leave blank lines where you want vertical space.
% The spaces will be adjusted to fill the entire page.  The dotted
% lines for the signatures are made with the \signature command.

%\maketitle

\begin{titlepage}
\begin{center}
\Huge{Abnormalities and Disease Detection in Gastro-Intestinal Tract Images}\\[1cm]
\large{Submitted by \\}
\Large{Zeshan Khan \\ (17K-3501)}\\[2cm]
\large{Submitted to the FAST-School of Computing\\in partial fulfillment of the requirements for the degree of\\}
{\setstretch{1.5}
\large{Doctor of Philosophy in Computer Science\\}
\large{at the\\}
\large{National University of Computer and Emerging Sciences (FAST-NUCES),\\Karachi\\2024\\~\\~\\} %{\today}}
}
\large{Thesis Supervisor\\Prof. Dr. Muhammad Atif Tahir\\~\\}
%\large{Thesis Co-Supervisor\\Dr. Muhammad Rafi}
\end{center}
\end{titlepage}

\pagestyle{empty}
% First copy: start a new page, and save the page number.
\cleardoublepage
% Uncomment the next line if you do NOT want a page number on your
% abstract and acknowledgments pages.

\cleardoublepage
\pagestyle{empty}
\setcounter{savepage}{\thepage}
\pagenumbering{roman}

\chapter*{Declaration}\label{CH:DECLARATION}
\thispagestyle{empty}
I, 'Zeshan Khan', hereby state that this Ph.D. thesis titled 'Abnormalities and Disease Detection in Gastro-Intestinal Tract Images' is my own work and it has not been previously submitted by me for taking partial or full credit for the award of any degree at this University or anywhere else in the world. 

 \begin{flushright}
        \line(1,0){100} \\
        Zeshan Khan \\
        17K-3501
    \end{flushright}
\chapter*{Plagiarism Undertaking}\label{CH:PLAG}
\thispagestyle{empty}
I take full responsibility of the research work conducted during this PhD thesis titled 'Abnormalities and Disease Detection in Gastro-Intestinal Tract Images'. I solemnly declare that the research work presented in the thesis is done solely by me with no significant help from any other person; however, small help wherever taken is duly acknowledged. I have also written the complete thesis by myself. Moreover, I have not presented this thesis (or substantially similar research work) or any part of the thesis previously to any other degree awarding institution within Pakistan or abroad.

I understand that the management of National University of Computer and Emerging Sciences has a zero-tolerance policy towards plagiarism. Therefore, I as an author of the above-mentioned thesis, solemnly declare that no portion of my thesis has been plagiarized and any material used in the thesis from other sources is properly referenced. Moreover, the thesis does not contain any literal citing of more than 70 words (total) even by giving a reference unless I have the written permission of the publisher to do so. Furthermore, the work presented in the thesis is my own original work and I have positively cited the related work of the other researchers by clearly differentiating my work from their relevant work.

I further understand that if I am found guilty of any form of plagiarism in my thesis work even after my graduation, the University reserves the right to revoke my PhD degree. Moreover, the University will also have the right to publish my name on its website that keeps a record of the students who plagiarized in their thesis work.

 \begin{flushright}
        \line(1,0){100} \\
        Zeshan Khan \\
        17K-3501
    \end{flushright}

%\begin{abstractpage}
\section*{Abstract}
Gastrointestinal (GI) tract image analysis plays a crucial role in medical diagnosis. This research addresses the challenge of accurately classifying and segmenting GI tract images for real-time applications. This research delves into the complexities of GI image interpretation, where traditional methods often fall short due to the diverse and intricate nature of abnormalities encountered. The high computational demands and complex nature of GI abnormalities necessitate efficient and adaptable methods.

This PhD thesis explores a multifaceted approach to GI image analysis. Initially, texture-based feature extraction and classification methods were investigated. This approach achieved remarkable processing speed (over 4000 FPS) and promising metrics (F1-score: 0.76, Accuracy: 0.98) on the Kvasir V2 dataset.

Transitioning to deep learning, the research meticulously selected and optimized a deep learning algorithm using data bagging techniques. This resulted in significant improvements, achieving an accuracy of 0.92 and F1-score of 0.60 on the HyperKvasir dataset, alongside a strong F1-score of 0.88 on Kvasir V2.

Furthermore, to address real-time detection needs, a streamlined neural network utilizing texture and local binary patterns was developed. This system tackled challenges like inter-class similarity and intra-class variations by incorporating a learned threshold from training data. It achieved exceptional processing speed (41 FPS) with high accuracy (0.99) and robust F1-score (0.91) on the HyperKvasir dataset.

To enhance the usability of the system two image segmentation tools are proposed in the research with the objectives of good detection and a faster detection using Depth-Wise Separable Convolution and ensemble of Neural Networks for the better detection in low FPS.

In conclusion, this research not only introduces novel methodologies for GI image analysis but also highlights the adaptability of various approaches in tackling the domain's unique complexities. The progression from texture-based methods to deep learning and ensemble strategies offers a valuable framework for advancing medical image analysis in the context of GI abnormalities.

\cleardoublepage
%\end{abstractpage}

\chapter*{Acknowledgements}\label{CH:ACK}
\thispagestyle{empty}
In the commencement of this expression of gratitude, I extend my heartfelt thanks to the Almighty, whose benevolence has afforded me the opportunity, requisite resources, and fortitude to undertake and complete this endeavor.

Acknowledgment is also due to my esteemed supervisor, Prof. Dr. Muhammad Atif Tahir, whose profound expertise, unwavering guidance, mentorship, support, and motivational influence have been consistently invaluable throughout the entirety of this scholarly odyssey.

Furthermore, I wish to express my gratitude to all my educators, whose unwavering dedication and diligent efforts have significantly contributed to my academic and intellectual development.

I am also indebted to several individuals within the FAST-NUCES community, both past and present, whose mentorship, guidance, and support have played a pivotal role in shaping the course of my academic pursuit. Special appreciation is extended to Dr. Farrukh Hassan Syed, Dr. Waqas, Mr. Musawir, Mr. Shoaib, Dr. Shahbaz Siddiqui, Dr. Hur Jalal Bhayo, and Dr. Hanif Soomro for their indispensable assistance and motivational encouragement.

In conclusion, profound appreciation is extended to my family, to whom no measure of gratitude can truly suffice. Their unwavering support, encouragement, understanding, patience, and continuous prayers have been instrumental at every juncture of this journey. This scholarly undertaking would have been inconceivable without their steadfast involvement.
\cleardoublepage
\pagestyle{plain}
  % -*- Mode:TeX -*-
%% This file simply contains the commands that actually generate the table of
%% contents and lists of figures and tables.  You can omit any or all of
%% these files by simply taking out the appropriate command.  For more
%% information on these files, see appendix C.3.3 of the LaTeX manual. 

\tableofcontents

%\newpage
\listoffigures
%\newpage
\listoftables

%\newpage

\chapter*{List of Abbreviations}

\begin{longtable}{lllll}
GI & Gastrointestinal\\
GI-Tract & Gastrointestinal Tract\\
UC & Ulcerative Colitis\\
BBPS & Boston Bowel Preparation Scale\\
EAD & Endoscopic Artefact Detection \\
ICPR & International Conference on Pattern Recognition \\
ICPR EndoTech & ICPR Task of Endoscopic Abnormality Detection \\
SSL & Semi-supervised Learning \\
ARMSE& Average Root Mean Squared Error \\
ARRMSE & Average Relative Root Mean Squared Error \\
GA & Genetic Algorithm \\
GA-FS & Genetic Algorithm based Feature Selection \\
ML & Machine Learning \\
MLC & Multi-label Classification \\
RRMSE & Relative Root Mean Squared Error \\
RMSE & Root Mean Squared Error \\
SAFER & SAFE semi-supervised Regression \\
LBP & Local Binary Patterns \\
CLBPS & Complete Local Binary Patterns Sign \\
CLBPM & Complete Local Binary Patterns Magnitude \\
CLBP & Complete Local Binary Patterns \\
DLBP & Dominant Local Binary Patterns \\
RILBP & Rotation invariant LBP \\
MLBP & Multiscale LBP \\
NULBP & The non-uniform LBP \\
LQP & Local Quinary Pattern \\
NRLBP & Noise Resistant LBP \\
LDOBP & Local Derivative Ordinal Binary Pattern \\
CSLBP & Center-Symmetric Local Binary Patterns \\
EBP & Elliptical Binary Pattern \\
SOILBP & Scale and Orientation Invarient LBP \\
ALBP & Adaptive Local Binary Pattern \\
LTP & Local Ternary Patterns \\
GF & Gabor Filters \\
LCF & Local Color Features \\
CL & Color Layout \\
EH & Edge Histogram \\
JCD & Joint Composite Descriptor \\
PHOG & Pyramid Histogram of Oriented Gradients\\
TCCD & Tamura, Coarseness, Contrast and Directionality\\
CEDD & Color and Edge Directivity Descriptor\\
FCTH & Fuzzy Color and Texture Histogram \\
ACC & Auto Color Correlation \\
GLCM & Gray Level Co-Occurrence Features \\
HTF & Haralick Texture Features \\
GTSD & gray-tone spacial dependence \\
BCC & Binary Class Classification \\
TCC & Ternary Class Classification \\
CNN & Convolutional Neural Network \\
FPS & Frame Per Second\\
Acc. & Accuracy\\
P & Precision\\
R & Recall \\
F1 & F1-Score\\
MCC & Matheos Corelation Coefienct \\
\end{longtable}

\chapter*{List of Publications}\label{CH:PUBLICATIONS}

\section*{Journal Publications}
\begin{enumerate}
\item Khan Z, Tahir MA. 2023. Real time Anatomical Landmarks and Abnormalities Detection in Gastrointestinal Tract. PeerJ Computer Science 9:e1685 https://doi.org/10.7717/peerj-cs.1685.
\item Jha Debesh, Khan Zeshan and others. "A comprehensive analysis of classification methods in gastrointestinal endoscopy imaging." Medical image analysis 70 (2021): 102007.
\item Jha Debesh, Khan Zeshan and others. "An objective validation of polyp and instrument segmentation methods in colonoscopy through Medico 2020 polyp segmentation and MedAI 2021 transparency challenges" Medical image analysis, 2024
\end{enumerate}

\section*{Conference Publications}
\begin{enumerate}
\item Khan, Zeshan, and Muhammad Atif Tahir. "Majority Voting of Heterogeneous Classifiers for Finding Abnormalities in the Gastro-Intestinal Tract." MediaEval. 2018.
\item Khan, Zeshan, et al. "Medical diagnostic by data bagging for various instances of neural network." Pattern Recognition. ICPR International Workshops and Challenges: Virtual Event, January 10-15, 2021, Proceedings, Part VIII. Springer International Publishing, 2021.
\item Ali, Syed Muhammad Faraz, et al. "Depth-Wise Separable Atrous Convolution for Polyps Segmentation in Gastro-Intestinal Tract." MediaEval. 2020.
\item Khan, Zeshan, et al. "Medico 2021: Medical Image Augmentation and Segmentation using Combination of Segmentation Neural Networks." (2021).
\item Khan, Zeshan, et al. "Voting Neural Network (VNN) for Endoscopic Image Segmentation." 2022 International Conference on Emerging Trends in Smart Technologies (ICETST). IEEE, 2022.
\end{enumerate}

\cleardoublepage
\pagenumbering{arabic}

%The content of the thesis starts here

\chapter{Introduction}

The health care systems are been improved by the use of multimedia applications. There are some applications to assist the doctors for better and quicker understanding the situations of the patients by using health care systems \cite{pogorelov2018medico,riegler2017multimedia,tahir2015enhancing,ul2016telemedicine,ul2015evaluating}. 2.8 million luminal Gastrointestinal Tract (GI-Tract) cancers are detected globally every year, many of these cancer patients can be prevented by the use of improved endoscopic performance using high quality and systematic screening and detection of the areas of the disease \cite{brenner2014pox}. These type of cancers have a mortality rate of $65\%$ \cite{who2019}. The detection of the such type of cancers have a limitations of the operator or the observer variant. The different GI detectors or experts may detect different diseases with $80\%$ similarity and a miss rate of $20\%$.

Advanced image analysis offers a powerful tool for healthcare advancement. By leveraging multimedia research techniques, it can revolutionize computer-aided diagnosis. This technology holds immense potential to detect and interpret various abnormalities across the body, leading to earlier disease identification and improved treatment outcomes. Early detection is particularly crucial in areas like the gastrointestinal tract, where it can significantly enhance successful treatment and patient survival. Detection of some of the diseases needs a lot of experience for a good accuracy for the doctor. The polyps detection by an experienced colonoscopist is usually at $90\%$ but the the colonoscopist with lesser experience can have accuracy below $80\%$. The detection is also observer dependant \cite{hewett2012validation,ignjatovic2009optical}. A. Rastogi et al. shown that the different observers/colonoscopists can detect same polyp $70\%$ times while the same observer/colonoscopist can detect different polyp $20\%$ times \cite{rastogi2009recognition}.

The abnormalities detection tasks includes the image classification for the abnormality detection for disease prediction and categorization of the abnormality. There is a strong work available for the image classification, using handcrafted features and classification and the neural network approaches \cite{naqvi2017ensemble,tan2010enhanced,krizhevsky2012imagenet,he2016deep,zeiler2014visualizing,simonyan2014very,szegedy2015going,huang2017densely}.

The classification of the images tasks becomes more difficult when we have to classify with a smaller amount of labeled dataset. In the problem of abnormalities detection there are just some thousands of labeled images available in different datasets. \cite{liu2017hkbu,pogorelov2018medico}.

There are several diseases relevant to the GI-tract. Many disease can be detected by analysing GI-tract images or videos \cite{naqvi2017ensemble,liu2017hkbu,pogorelov2018medico}. There are several techniques available for the classification of the images but the recent history shows that the performance accuracy is higher in case of neural network architectures. There are a set of neural network architectures that can be used for abnormality detection or image classification. The models are as 
K. Fukushima 1983 \cite{fukushima1983neocognitron}, LeCun nwtwork \cite{lecun1998gradient}, AlexNet \cite{krizhevsky2012imagenet}, ResNet \cite{he2016deep},Clarifia \cite{zeiler2014visualizing}, VGG \cite{simonyan2014very}, Google LeNet \cite{szegedy2015going} and DenseNet \cite{huang2017densely} etc.

There are some approaches that uses the textural features with deep features to get better performance in terms of classification accuracy \cite{khan2018majority,ko2018weighted}. The results shows a good accuracy for even with a smaller set of labeled images datasets. The medical image analysis is also done on the dataset of a few thousands and the results have more than $90\%$ accuracy \cite{naqvi2017ensemble}. The majour challenge is that there are some of the GI tract disease where there is a minor inter-class difference and major intra-class differences \cite{hewett2012validation}.

\section{Problem Statement}
\label{ch1:prob_statement}

This thesis deals with exploring the potential of multimedia applications to enhance the healthcare system, particularly in the context of improving the early detection and diagnosis of gastrointestinal (GI) tract diseases. The integration of multimedia technology into healthcare has shown promise in assisting doctors to gain a better and quicker understanding of patients' conditions. However, despite the advancements in this field, a significant challenge remains in the accurate and consistent detection of abnormalities in GI tract images and videos, which has crucial implications for the early diagnosis of diseases like luminal GI cancers. With approximately 2.8 million cases of luminal GI cancers diagnosed globally each year and a high mortality rate of 65\%, there is a pressing need to address the limitations associated with operator or observer-dependent detection. Different GI detectors or experts exhibit variations in disease detection, with a similarity rate of 80\% and a miss rate of 20\%. This research aims to explore innovative solutions to improve the accuracy, consistency, and efficiency of GI tract disease detection through multimedia-based image analysis and classification methods.

Incorporating multimedia research knowledge and methods into healthcare holds significant promise, especially for diseases like GI tract abnormalities, where early detection is critical for successful treatment and patient survival. Gastrointestinal diseases often require experienced medical professionals for accurate detection, and the accuracy can vary greatly between experts. For instance, experienced colonoscopists can achieve a detection rate of 90\% for polyps, while less-experienced ones may fall below 80\%. Moreover, the detection process is observer-dependent, further complicating the consistency of results. This study aims to harness advanced image classification techniques, including neural network architectures, to address the challenges in abnormalities detection in the GI tract. Additionally, it tackles the problem of limited labeled datasets, which is common in the field of medical image analysis. With just a few thousand labeled images available in various datasets, there is a pressing need to develop methods that can achieve high classification accuracy, even with a smaller amount of labeled data. Furthermore, the research confronts the specific challenges posed by GI tract diseases, characterized by minor inter-class differences and major intra-class differences, which necessitates innovative approaches to improve classification accuracy. 

\subsection{Aims and Objectives} The aims and objectives of this thesis are as follows:

\begin{enumerate}
    \item  Investigate image quality enhancement and augmentation techniques specific to medical image analysis, with a focus on gastro-intestinal images, to improve the accuracy and reliability of disease detection.
    \item Investigate and identify relevant features and approaches for the medical analysis of gastro-intestinal images, with a particular emphasis on detecting gastrointestinal diseases.
    \item Investigate the features that either facilitate or hinder the detection of abnormalities and the identification of landmarks within gastro-intestinal images, aiming to enhance diagnostic precision.
    \item Conduct an investigation into the utilization of deep learning techniques for medical image analysis, with a focus on their potential applications in gastro-intestinal disease detection.
    \item Investigate faster detection approaches to expedite the development of a real-time detection system, optimizing its speed and efficiency in identifying abnormalities.
    \item Investigate and establish optimal thresholds for abnormalities in images characterized by distinct inter-class differences and intra-class similarities, ensuring more precise and consistent detection within the medical context.
\end{enumerate}

In the quest for more robust and accurate gastro-intestinal abnormalities detection, researchers have turned to ensemble learning and deep learning techniques. Ensemble learning leverages the collective decision-making of multiple models to improve overall performance. In the context of GI abnormalities detection, ensemble techniques like Random Forest, AdaBoost, and Gradient Boosting, combine the outputs of multiple classifiers, each with their own strengths and weaknesses, to enhance the accuracy and reliability of the diagnosis. This approach is particularly valuable when dealing with the scarcity of labeled data in medical image analysis, as it can mitigate overfitting and enhance generalization.

Deep learning, on the other hand, offers a more data-centric approach. Convolutional Neural Networks (CNNs) have revolutionized the field by automatically learning hierarchical features from medical images, enabling the models to adapt and improve their performance as more data becomes available. The advent of deep learning architectures like U-Net for semantic segmentation, and various CNN variants for image classification, have significantly advanced the state-of-the-art in gastro-intestinal abnormalities detection. Deep learning models have the potential to uncover intricate patterns in the GI tract images and can be fine-tuned to tackle inter-class differences and intra-class similarities, contributing to a higher accuracy and consistency in detecting abnormalities.

\section{Background}
\label{ch1:background}

Machine learning is concerned with the methods and techniques which allows machines to learn and make inferences on given data by using algorithms that can find patterns and relationships between the data elements. This generally involves providing some data to the algorithms so they can model these relationships. In the future, if the machine receives similar data, it can make inferences based on the model it already has of the world that is described by the original data. The power of a model resides in how much of the seen world it can remember and how accurately and how much of the unseen world it can infer.

\subsection{Data-Driven Learning Techniques}
In the context of machine learning, the activity of learning is generally categorized into the following sub-categories:
\begin{enumerate}
  \item Supervised learning.
  \item Unsupervised learning.
  \item Semi-supervised learning.
  \item Reinforcement learning.
\end{enumerate}
This research deals with only two types of learning; supervised and un-supervised.

\subsubsection{Supervised Learning}
Supervised learning is the type of learning which takes place with the help of labeled data. That is, the learning algorithm is given a set of values for each of the feature and a corresponding output or label, that is, the value of the dependent variable that occurs for that particular combination of the feature values. It is the job of the algorithm to use its underlying logic and develop a rule which defines this connection between the features and the label. In this way, the algorithm is given a number of these feature values and corresponding label values. The algorithm is then given a set of feature values that is has not seen before, and tasked with inferring its label. The performance of the algorithm is then evaluated based on some performance criteria. Eventually, the algorithm should learn to infer the labels correctly in increasing numbers. Therefore, in order words, in supervised learning, the algorithm learns to approximate the world described by the data by using a set of features which contain certain characteristics of the world, and the provided labels, which describe what the combination of feature values given in the feature set stands for.

\subsubsection{Unsupervised Learning}
Unsupervised learning is on the opposite spectrum of learning mechanisms.  In unsupervised learning, the algorithm is not provided with a label. Instead, it is given some information about how to determine or calculate the scale of similarity or dissimilarity between the feature values. It is then given some data and tasked to organize or group the data based on the described similarity measure. This organization or group is usually referred to as a cluster.
Historically, more research has been focused on supervised learning. As more and more data has become available, and the cost of assigning a label to this data has remained high, unsupervised learning has become more popular as well.

Any of the learning types can be applied to machine learning problems relating to both classification and regression problems. But as far as GI-Track medical image classification is concerned, most of the work has been done on supervised learning, and un-supervised learning for the image preprocessing and augmentation tasks. Objective $1$ of this thesis deals with this research.

\subsubsection{Dimensionality Reduction}
All machine learning tasks consist of finding the underlying relationships in the given data by attempting to relate the different combinations of the feature values to the values given in the labels. Consequently, it is quite intuitive that as the number of features and feature values increase, the number of possible combinations of feature values also increases. This, in turn, on the one hand, will make finding the underlying relationships harder, and on the other hand, it would also require more and more examples of the data to effectively approximate and model the state space presented by the combinations. Along with these problems, the larger number of features (also called dimensions) introduce a myriad of issues. Collectively, a number of such issues which arise with the increase in the number of features are referred to as the `Curse of Dimensionality' \cite{tahir2013dimensionality, bellman1966dynamic}.
Various techniques exist which can be employed to reduce the dimensionality of the data, thus providing processing benefits as well as mitigating the issues that occur due to the high dimensionality of the feature space. These techniques are called dimensionality reduction techniques \cite{van2009dimensionality}. Two commonly used techniques are feature selection and feature extraction \cite{khalid2014survey}. 
In feature extraction, the dimensions of the data are reduced to smaller dimensions by applying different methods, which do so by transforming and combining the features into a smaller number of features without losing the information available in them. However, due to the transformations, the original features are lost. In addition, the new set of features are uninterpretable. Thus, analysts are unable to explain how features contribute to the dependent variable.
The second group of techniques, called feature selection, is the set of techniques in which features from the original set are identified which contribute more towards explaining the dependent variable. Inversely, they also attempt to remove those features that are redundant and do not provide any additional information. Thus, feature selection methods attempt to reduce the dimensionality of the data just by trying to select a subset of the original features without modifying them. This, in turn, makes them more interpretable and allows business and domain decisions to be made, such as which attributes to change to increase sales of a particular product.
This research is concerned with feature selection only, and therefore, this concept will be discussed in detail.

\subsubsection{Feature Selection for Dimensional Reduction}
\begin{figure}
\includegraphics[scale = 1]{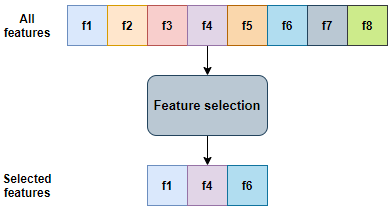}
\centering
\caption{Example of Feature Selection.}\label{feature_selection}
\end{figure}

Figure~\ref{feature_selection} shows the process of feature selection. From the set of all available features, a subset of features is selected through some feature selection mechanism.

Formally, feature selection can be defined as follows: If for any given data having $d$ dimensions and containing $n$ instances, its dimensions need to be reduced using feature selection, then there needs to be defined a selection criterion, given by a function $J$. This function either needs to be minimized, or maximized depending on whether the function defines an error driven metric or an accuracy driven metric. The objective is to find feature subset of dimensions $\bar{d}$ such that $\bar{d} \subseteq d$, and $J(\bar{d}) \leq J(d)$ (minimization problem) or $J(\bar{d}) \geq J(d)$ (maximization problem) \cite{tahir2008feature}. 
Much research has been carried out related to feature selection techniques.

Feature selection techniques are generally categorized into the following three groups \cite{jovic2015review}:

\begin{enumerate}
  \item Filter methods
  \item Wrapper methods
  \item Embedded methods
\end{enumerate}

In filter methods, the feature selection is carried out independently of any machine learning algorithm. As the name implies, these techniques are added as a filtering stage that is usually carried out during preprocessing. A number of filter methods exist. Some use correlations between features and targets to identify and keep the features that are highly correlated with the targets and discard the features that are redundant or do not correlate with the target. Others use the concept of information gain and provide a ranking of the features based on the amount of information they contact related to the target.
Wrapper methods, on the other hand, use machine learning algorithms themselves as the evaluation criteria for selecting the features. A feature subset is selected by the wrapper method and sent to a machine learning algorithm. The algorithm obtains the results based on the subset of features. In this way, different subsets of features are evaluated. The feature subset providing the best performance is selected for future use.
Embedded feature selection methods are embedded or integrated into the machine learning algorithm themselves. Decision trees are the best example of algorithms that use embedded feature selection by calculating how well any feature separates the data points from each other and using the features which score well as nodes higher up in the tree.
In this thesis, a wrapper approach has been used based on genetic algorithms. The following section discusses the application of genetic algorithm for feature selection in detail.

\subsubsection{Genetic Algorithm for Feature Selection}
\begin{figure}
\centering
\fbox{\includegraphics[scale = 0.94]{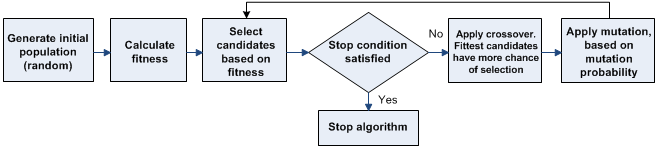}}
\caption{\small{Flowchart of Genetic Algorithm.}}
\label{ga_flowchart_2.png}
\end{figure}

In practice, feature selection problems are computationally very expensive problems as they are in essence combinatorial problems and require a search through a huge state space, i.e. the different combinations of features and feature values, presented by the features set, especially if there is also a requirement to model the feature relationships. Solving such problems would require a very long time using any exhaustive search method. Therefore, feature selection methods use different techniques to explore the state-space and look for promising solutions. One method is the use of evolutionary algorithms \cite{xue2015survey}. Evolutionary algorithms are metaheuristic based approaches to solving optimization problems \cite{back1996evolutionary}. They are inspired by the process of evolution in living organisms. The intuition behind using evolutionary algorithms is that using these mechanisms, living organisms learn to adapt to a great many different conditions, which present an even larger state space. The organisms face an environment where their survival depends on changing some of their characteristics to suit the environment. The organisms that have suitable characteristics to traverse the environment successfully survive and carry on. This concept is referred to as the survival of the fittest and is inspired by the works of Charles Darwin. Different evolutionary algorithms exist and have been used for feature selection. One family of evolutionary algorithm that has been used by many researchers is called genetic algorithm (GA) \cite{kim2006genetic}.
In genetic algorithm, a combination of different operations inspired by nature are applied to a set of solutions. These operators transform the solutions and form new solutions. The fitness of each solution is evaluated, and only those that pass a fitness threshold are kept while the rest are discarded. These new solutions then pass through the same cycle. The operators which genetic algorithms use are selection, crossover, and mutation. Figure \ref{ga_flowchart_2.png}  shows a flowchart of steps for the genetic algorithm. Each step and the operators will be discussed one by one shortly. 

At the start of the algorithm, a set of solutions are generated. These constitute the candidate solutions and can be represented or encoded as strings of $0's$ and $1's$ or length $n$ where $n$ is the number of features. Such strings are called chromosomes in the terminology of GA, and each $0$ or $1$ is referred to as a gene. Therefore, each gene represents one feature. A chromosome represents one candidate solution with a $0$ in place of a feature indicates that the feature is not present in the candidate solution, and a $1$ indicates that it is. 

  \begin{equation}
    Feature_{i}(Feature Subset_{j})=
    \begin{cases}
      1, & \text{if feature exists in feature subset}\  \\
      0, & \text{otherwise}
    \end{cases}
  \end{equation}

Where $i=\{1,2,...n\}$, such that $n$ is the number of features, and $j=\{1,2,...k\}$ such that $k$ is the number of population of candidate solutions.

\begin{figure*}[ht]
\centering
%\captionsetup{justification=centering}
\includegraphics[scale=1]{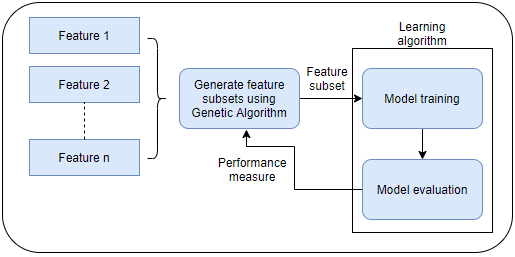}
%\decoRule
\caption{Flow of GA based feature selection method using wrapper approach.}
\label{GA_for_FS}
\end{figure*}

There are a many ways to generate the initial set of solutions. They may be generated using a simple feature selection technique, or they could be generated using domain knowledge. The most commonly used method is a random generation. The intuition is that the genetic algorithm will itself converge to promising solutions using its solution reproduction mechanism. This process is sometimes referred to as initialization. The selection operator is then applied to all candidate solutions. The selection operator checks the viability of each candidate solution by evaluating the solution based on a fitness function. Those candidate solutions which do not pass the threshold are discarded. The candidate functions that remain are then passed through the crossover operation. In crossover, two (in some variations more) candidate solutions are selected and combined to produce more candidate solutions. The original candidate solutions are referred to as `parents', while the solutions resulting from their crossover are referred to as their `children' or `offspring'. This cycle of selection and crossover giving birth to new candidate solutions is called a `generation'. There are variations in how the crossover operator is applied across different variations of the genetic algorithm and how the individuals to crossover are selected, but in most cases, the probability of selection for crossover is directly proportional to the fitness function of any candidate solution. This is also intuitive as we want to focus our efforts on the candidate solutions, which give us better results. This part of the algorithm is also called the exploitation part as promising candidate solutions are paid more focus to. In this way, each generation produces candidate solutions resulting from the combination of promising candidate solutions from the previous generation. Over the course of the process, it is quite possible that due to the randomness inherently present in the processes such as initialization and crossover, some good candidate solutions have either not been explored at all or have been lost. The mutation operator tries to overcome this issue. Mutation is a process in which, over the course of the genetic algorithm, one or more solutions are modified at random. The rate of how frequently mutation occurs is generally configured by defining a mutation rate for the process. This part of the algorithm is referred to as exploration. Generally, the mutation rate is kept small so that more time is spent in exploitation and less in exploration. Some research has suggested that the mutation rate could also be kept variable \cite{doerr2017fast, pham1997genetic}, similar to simulated annealing \cite{van1987simulated}, so that at the start, the mutation rate is relatively high, thus allowing exploration to take place. Gradually as the generations progress, this rate is brought down. Figure \ref{GA_for_FS} illustrates the wrapper based feature selection cycle using GA.

\subsection{Medical Image Classification}
Medical image classification is the process of assigning predefined labels or categories to medical images based on their visual content. It involves the extraction of relevant features from the images, followed by the application of machine learning algorithms, particularly deep learning, for accurate classification. In the context of GI disease detection and GI landmark identification, medical image classification plays a crucial role in automating the diagnosis process, reducing human error, and increasing the efficiency of healthcare.

\subsection{Gastro-Intestinal (GI) Disease Detection}

GI disease detection aims to automatically identify abnormalities, such as polyps, tumors, or other pathologies, in GI tract images, typically obtained through procedures like endoscopy. The key steps involved in GI disease detection through medical image classification are as follows:
\begin{itemize}
    \item \textbf{Image Acquisition:} GI images are acquired using endoscopic procedures or other imaging modalities. These images serve as the input data for the classification system.
    \item \textbf{Preprocessing:} Images are preprocessed to enhance their quality, remove noise, and standardize the format. This may involve techniques such as noise reduction, contrast enhancement, and color correction.
    \item \textbf{Feature Extraction:} Features are extracted from the preprocessed images to represent the relevant information. In the case of GI disease detection, features may include texture, color, and shape characteristics. The process can be represented by Equation~\ref{eq:ch1-extraction}.
    \begin{equation}
        \label{eq:ch1-extraction}
        P(i, j) = \frac{N(i, j)}{N}
    \end{equation}
    Where:
    \begin{itemize}
        \item $P(i, j)$ is the normalized GLCM.
        \item $N(i, j)$ is the number of occurrences of pixel pairs with values $i$ and $j$ at a certain distance and orientation.
        \item $N$ is the total number of pixel pairs at that distance and orientation.
    \end{itemize}
    
    \item \textbf{Classification:} Machine learning models, particularly deep neural networks, are trained using labeled data to classify images into categories. The categories may represent different GI diseases or the absence of disease. The detection of Neural Networks can be expressed with Equation~\ref{eq:ch1-clf}.
    \begin{equation}
        \label{eq:ch1-clf}
        Z_l = f(W_l * A_{l-1} + b_l)
    \end{equation}
    Where:
    \begin{itemize}
        \item $Z_l$ is the output of layer $l$.
        \item $W_l$ is the weight matrix for layer $l$.
        \item $A_{l-1}$ is the activation from the previous layer.
        \item $b_l$ is the bias for layer $l$.
        \item $f$ is the activation function.
    \end{itemize}
    \item \textbf{Post-processing:} Post-processing techniques, such as thresholding or filtering, are applied to the classification results to refine the diagnosis and reduce false positives.
\end{itemize}

\subsubsection{Applications}
Medical image classification has a wide range of applications in the field of healthcare and medical research. Here are some key applications of medical image classification:
\begin{description}
    \item [Disease Detection and Diagnosis:] The detection of various diseases as follows:
    \begin{itemize}
        \item Cancer Detection: Medical image classification is widely used for the early detection of various cancers, such as breast cancer, lung cancer, and skin cancer, by identifying tumors and abnormalities in medical images.
        \item Cardiovascular Disease: It can help in identifying heart diseases by analyzing images of the heart, blood vessels, and the circulatory system.
        \item Neurological Disorders: Identifying brain abnormalities, lesions, and tumors in neuroimaging data is crucial for the diagnosis of neurological disorders like Alzheimer's disease or brain tumors.
    \end{itemize}
    \item [Anomaly Detection:] The detection of various anomalies as follows:
    \begin{itemize}
        \item Radiology: Detecting abnormalities in X-rays, CT scans, and MRIs is a common application. This includes identifying fractures, bone diseases, and soft tissue abnormalities.    
        \item Endoscopy: Identifying anomalies such as polyps, ulcers, or other lesions in gastrointestinal endoscopy images.
    \end{itemize}
    \item [Organ and Tissue Segmentation:] Automatic segmentation of organs and tissues is vital for surgical planning, radiation therapy, and tracking disease progression.
    \item [Medical Data Analysis:] The analysis of various medical issues from medical data such as:
    \begin{itemize}
        \item Analyzing and categorizing pathology slides in pathology images, helping pathologists identify and classify various diseases.
        \item Identifying and classifying retinal diseases, such as diabetic retinopathy, to assist ophthalmologists in early intervention.
    \end{itemize}
    \item [Radiomics and Precision Medicine:] Extracting quantitative features from medical images to provide insights for personalized treatment decisions.
    \item [Drug Discovery and Development:] Identifying and categorizing cell structures and disease models in drug discovery research.
    \item [Prosthesis and Implant Design:] Classifying bone structures and tissue types to aid in the design and placement of prosthetics and implants.
    \item [Monitoring Disease Progression:] Tracking disease progression over time through repeated medical imaging to assess the effectiveness of treatment.
    \item [Telesurgery and Robotic Surgery:] Assisting surgeons during minimally invasive procedures, such as laparoscopic or robotic surgery, by providing real-time analysis of images.
    \item [Quality Control and Image Enhancement:] Ensuring the quality and accuracy of medical images, and enhancing image quality for better diagnosis.
    \item [Image Retrieval and Archive Management:] Efficiently managing vast collections of medical images, including retrieval and categorization based on content and metadata.
    \item [Telediagnosis and Remote Healthcare:] Enabling remote diagnosis and healthcare by sending medical images for expert analysis and diagnosis.
    \item [Education and Training:] Training medical professionals through simulated cases and providing educational materials.
    \item [Research and Clinical Trials:] Supporting medical research by providing quantitative data for clinical trials and studies.
    \item [Public Health Surveillance:] Monitoring the prevalence and spread of diseases, such as infectious diseases, through image analysis.
    \item [Pulmonary and Respiratory Health:] Identifying and classifying lung conditions, such as pneumonia, tuberculosis, and chronic obstructive pulmonary disease (COPD), from chest X-rays and CT scans.
    \item [Orthopedics:] Detecting and categorizing orthopedic conditions and injuries, including fractures, joint disorders, and degenerative bone diseases.
    \item [Dental and Oral Health:] Identifying dental and oral health issues like caries, periodontal diseases, and oral tumors through dental radiographs.
    \item [Mammography:] Aiding in the early detection of breast cancer by analyzing mammograms for suspicious lesions and calcifications.
    \item [Fetal Health and Obstetrics:] Monitoring fetal development and health during pregnancy through ultrasound and MRI scans, detecting anomalies or developmental issues.
    \item [Oncology:] Assessing tumor characteristics, tumor grading, and staging to guide treatment decisions in cancer patients. 
    \item [Radiation Therapy Planning:] Assisting in the planning and targeting of radiation therapy for cancer treatment by outlining tumor regions and healthy tissue.
    \item [Preoperative Planning:] Providing insights into patient anatomy and pathology to aid surgeons in planning complex surgeries.
    \item [Infection Detection:] Detecting infections and abscesses in various parts of the body using medical imaging.
    \item [Gait Analysis and Movement Disorders:] Analyzing movement and gait patterns for the diagnosis and treatment of orthopedic and neurological disorders.
    \item [Psychiatry and Neurology:] Supporting the diagnosis and monitoring of neurological conditions like multiple sclerosis, epilepsy, and Alzheimer's disease through brain imaging.
    \item [Emergency Medicine:] Rapid assessment of traumatic injuries and acute conditions in emergency departments through imaging, guiding immediate medical interventions.
    \item [Telemedicine and Mobile Health:] Enabling remote medical consultations and diagnostics through mobile apps and telemedicine platforms.
    \item [Veterinary Medicine:] Extending medical image classification to the diagnosis and treatment of diseases in animals through veterinary imaging.
    \item [Dermatology:] Classifying skin conditions and dermatological disorders based on images for early diagnosis and treatment.
\end{description}

\section{Goals of the Thesis}

The objective of this thesis is to conduct an examination of existing methodologies employed in the field of medical image classification, with a specific emphasis on addressing challenges related to the identification of gastro-intestinal abnormalities and landmarks.

Furthermore, this research endeavors to address the issue of constructing efficient learning models in scenarios where there is a scarcity of annotated data. The constraints posed by real-time analysis requirements, particularly in terms of detection time, are also considered in the context of this investigation.

To delineate the specific aims and objectives outlined in Section \ref{ch1:prob_statement}, the following goals are pursued within the framework of this thesis:

\begin{enumerate}
  \item Investigate current methodologies utilized in medical image classification and segmentation, with a focus on their application to gastro-intestinal abnormalities and landmarks detection. \label{aim1}
  \item Address the challenge of developing effective learning models in situations where the availability of labeled data is limited. \label{aim2}
  \item Explore strategies to optimize the detection time of models, particularly emphasizing real-time analysis requirements. \label{aim3}
  \item Evaluate the applicability and performance of existing approaches in the specific domain of gastro-intestinal abnormalities and landmarks detection. \label{aim4}
  \item Propose and implement novel techniques or enhancements to existing models to overcome limitations related to data scarcity and real-time analysis. \label{aim5}
  \item Validate the effectiveness of the developed models through rigorous experimentation and comparison with established benchmarks. \label{aim6}
  \item Provide insights and recommendations for the improvement of medical image classification in the context of gastro-intestinal abnormalities, with a consideration for practical deployment in real-world scenarios.
\end{enumerate}

\section{Contributions}
As more and more real-world fields move towards data analytics to analyze and solve real-world problems, they are faced with novel situations and circumstances. As the ability to generate and store data is increasing, the expectations of data-driven solutions also increase. One such area of study is that of medical image classification. This thesis attempts to make a contribution towards the fields of Data Science and Machine Learning by proposing various techniques and solutions that can be used to tackle medical image classification problems as a whole, and also when the availability of labeled data is an issue, which is true for most real-world problems. To summarize the contributions:

\begin{enumerate}
  \item Improve image classification prediction reliability, accuracy, and efficiency by proposing feature subset selection methods for supervised learning.
  \item Identify the best image quality enhancement technique.
  \item Determine the most useful and feasible features for GI-Tract image analysis in combination.
  \item Leverage the benefits of neural network architectures to achieve higher accuracy by combining texture features. Additionally, consider lightweight networks suitable for real-time execution.
  \item Establish optimal thresholds for each class based on the varying intra-class similarities and inter-class differences. 
  \item Using the power of Decision trees and iterative learning solutions for the domain of medical image analysis. 
\end{enumerate}

\section{Dissertation Organization}
The subsequent sections of this thesis are structured as follows:

The initial chapter, Chapter~\ref{literature}, delves into a comprehensive review of existing literature and relevant works directly connected to the chosen research topic. This foundational chapter establishes a strong theoretical grounding for the subsequent analysis.

Chapter~\ref{sec:data_collection} focuses on the practical implementation of the research. Here, the details of the experimental analysis are outlined, including the specific benchmark datasets employed for the study.

In Chapters~\ref{sec:approaches} and~\ref{sec:resultsall}, the focus shifts to the methodological aspects of the research. These chapters introduce and discuss various classification methods specifically designed for the rapid classification of medical images related to the GI-tract. The potential trade-offs associated with these methods, particularly the balance between speed and accuracy, are also critically examined.

The concluding chapter, Chapter~\ref{sec:conclusion}, serves as a culmination of the entire research endeavor. It encapsulates a concise summary of the key findings and insights gleaned from the study. Additionally, this chapter offers valuable recommendations for potential avenues of future research, effectively concluding the exploration of the dissertation's objectives and outcomes.
\chapter{Literature Review}
\label{literature}
\section{Image Classification Techniques}
Image classification, a fundamental task in computer vision, holds profound implications for a wide array of applications. At its core, image classification seeks to imbue machines with the capability to discern and categorize visual content with a level of accuracy akin to human perception. By employing sophisticated algorithms and neural networks, image classification systems traverse the intricate landscape of pixels, extracting distinctive features and patterns.

There is a significant work available on the disease prediction by classification of the medical images using visual and texture based analysis. The researches are available for the diseases of tumor detection \cite{karkanis2003computer}, endoscopic disease detection using informative and distinct frames identification and feature extraction for classification \cite{oh2007informative,arnold2009indistinct,vasantha2010medical,guo2010completed}. The classification of the images is based on various types of image features including Local Binary Pattern with variations \cite{guo2010completed,shrivastava2014effective,wolf2008descriptor,liu2016medical,george2018breast} and the neural network based architectures \cite{wang2015unified,twinanda2016endonet,zhang2016deep,ahmad2017endoscopic,pogorelov2017comparison,pogorelov2018deep,zhuang2019structured}.

The statement discusses the presence of substantial research in the field of gastrointestinal (GI) tract disease detection through the utilization of image processing techniques. Specifically, this research is focused on the categorization and classification of images related to the GI tract for the purpose of detecting abnormalities \cite{naqvi2017ensemble,tan2010enhanced,khan2018majority,krizhevsky2012imagenet,he2016deep,zeiler2014visualizing,simonyan2014very,szegedy2015going,huang2017densely}. 

\section{Medical Image Features}

This section delves into the discussion of specific features that play a crucial role in image classification. Features are the essential building blocks that empower image classification algorithms to discriminate between various objects and patterns within an image. These features can range from basic attributes like color and texture to more complex elements such as edges, corners, and shapes. The selection and extraction of relevant features are pivotal in defining the performance and accuracy of image classification systems. Engineers and researchers often employ techniques like convolutional neural networks (CNNs) to automatically learn and identify discriminative features from raw image data, rendering the process more efficient and capable of handling a wide range of image classification tasks. Understanding the nuances of these features is fundamental to advancing the field of computer vision and enhancing the capacity of machines to comprehend the visual world. Some of the major feature categories are listed below:

\begin{enumerate}
    \item Local Binary and Ternary Patterns (LBTP) Features
    \item Spectral Features
    \item Textural Features (Distribution of Darkness)
    \item Contextual Features
    \item Gray Level Co-Occurrence Features
\end{enumerate}

\subsection{Local Binary and Ternary Patterns (LBTP)}
Local Binary Patterns (LBP) is a texture representation method for an image or a region within the image. It captures the gray level relationships of a pixel with its neighboring pixels. The LBP algorithm encodes the local patterns formed by the intensity variations in a specified neighborhood around each pixel, providing a compact and informative description of the texture characteristics within the analyzed region \cite{wang1990texture,ojala2002multiresolution, ojala1996comparative,ahonen2004face,tan2010enhanced,tan2010enhanced}.

\subsubsection{Local Binary Patterns (LBP)}
The LPBs are calculated by the computation of the following operations on any multi-scale (3 or more) image by applying the following steps  \cite{ojala2002multiresolution, ojala1996comparative}:

\begin{itemize}
    \item \textbf{Step 1: Image Grayscale Conversion:} Convert the input image to grayscale if it's in color. LBP is typically applied to grayscale images.
    \item \textbf{Step 2: Select a Neighborhood:} Define a neighborhood around each pixel in the image. The neighborhood is typically a square or circular region of pixels with a specific radius (e.g., a 3x3 or 5x5 square or a circle with an appropriate radius). The size of the neighborhood influences the level of detail captured by LBP. The process can be understood by the Equation~\eqref{eq:distance_2d} and Equation~\eqref{eq:lbp_binary_code}.
    
    \begin{equation}
    \label{eq:distance_2d}
    distance((x,y),(z,w)) \leftarrow max(\vert {x-z} \vert,\vert{y-w}\vert)
    \end{equation}
    
    \begin{equation}
    \label{eq:lbp_binary_code}
    X(x,y) \leftarrow {I(i,j) \forall_{i,j} \: I(i,j) \in Image \wedge distance((x,y),(i,j)) = radius}
    \end{equation}
    \item \textbf{Step 3: Center Pixel Value:} For each pixel in the image, you choose a center pixel and compare it with the surrounding pixel values in the defined neighborhood.
    \item \textbf{Step 4: Thresholding:} Compare the intensity value of each pixel in the neighborhood to the intensity value of the center pixel. If the surrounding pixel's value is greater or equal to the center pixel's value, assign it a value of 1; otherwise, assign it a value of 0. The process can be understood by the Equation~\eqref{eq:lbp_binary_set}.
    \begin{equation}
    \label{eq:lbp_binary_set}
    P(i) \leftarrow 
    \begin{cases}
    1,& \text{if } X(x,y) \geq I(i,j)\\
    0,              & \text{otherwise}
    \end{cases}
    \end{equation}
    \item \textbf{Step 5: Binary to Decimal Conversion:} Now, we have a binary string (a series of 0s and 1s) based on the pixel comparisons in the neighborhood. Convert this binary string to a decimal value. This decimal value is the LBP code for the center pixel. The process in expressed in the Equation~\eqref{eq:lbp}.
    
    \begin{equation}
    \label{eq:lbp}
    LBP(i,j) \leftarrow \sum_{\forall p \in P}2^p
    \end{equation}
    \item \textbf{Step 6: Repeat for All Pixels:} Repeat the process described in steps 4-6 for all pixels in the image. This results in a new image where each pixel represents an LBP code.
    \item \textbf{Step 7: Histogram Calculation:} Once you have LBP codes for all pixels in the image, you can calculate a histogram. The histogram shows how many times each LBP code occurs in the image. The histogram provides a compact representation of the texture information in the image.
    \item \textbf{Step 8: Feature Vector:} The final LBP feature vector is the normalized histogram. This feature vector can be used for various tasks, such as texture classification, object recognition, or face recognition.
\end{itemize}

Additional steps can be considered to enhance the quality of results. These optional measures encompass image preprocessing and histogram normalization, which can significantly contribute to the refinement of the LBP feature extraction process. Image preprocessing allows for the improvement of image quality, aiding in noise reduction and contrast enhancement. Histogram normalization, on the other hand, plays a pivotal role in rendering LBP features robust to variations in illumination and contrast, ensuring a more reliable and stable representation of texture information. These supplementary steps are valuable tools for optimizing LBP-based applications, enhancing the overall effectiveness of texture analysis and object recognition tasks \cite{zhang2005local}.

To grasp the concept of Local Binary Pattern (LBP) computation, consider the following illustrative example:

\begin{table}[htbp]
    \caption{Image and Local Binary Patterns}
    \begin{minipage}{.5\linewidth}
      \caption{Image in Gray-scale}
      \centering
      \setlength{\arrayrulewidth}{0.5pt}
        \begin{tabular}{|c|c|c|}
         \hline
         15 & 50 & 30 \\
         \hline
         18 & 20 & 40 \\
         \hline
         35 & 12 & 10 \\
         \hline
        \end{tabular}
    \end{minipage}%
    \begin{minipage}{.5\linewidth}
      \centering
        \caption{Binary Codes of the Image}
      \setlength{\arrayrulewidth}{0.5pt}
        \begin{tabular}{|c|c|c|}
         \hline
         0 & 1 & 1 \\
         \hline
         0 &  & 1 \\
         \hline
         1 & 0 & 0 \\
         \hline
        \end{tabular}
    \end{minipage} 
\end{table}

LBP value is $01110010 \leftarrow 114$. The value for the middle point of the image in LBP is 114 for above example.

There are some variants of the LBP using the neighbouring elements in ellipse, parabola, hyperbola, archimedean spiral instead of a circle of neighboring elements \cite{nanni2010local}.
There are several alternate versions of the LBPs as follows:

\subsubsection{Complete Local Binary Patterns Sign (CLBPS)}
\label{sec:clbps}

The Complete Local Binary Patterns Sign (CLBPS) is an extension and enhancement of the Local Binary Patterns (LBP) texture analysis operator, designed for improved texture classification. Developed by Zhenhua Guo, Lei Zhang, and David Zhang, this technique addresses some of the limitations of traditional LBP. In their paper, "A Completed Modeling of Local Binary Pattern Operator for Texture Classification," the authors introduce the concept of CLBPS, which aims to provide a more comprehensive representation of texture patterns in an image \cite{guo2010completed}.

CLBPS enhances the LBP operator by incorporating both the sign and magnitude information of the texture patterns. In traditional LBP, only the sign information is considered, which can lead to a loss of valuable information. By including the magnitude of the difference between pixel values in the neighborhood, CLBPS captures a richer description of local texture patterns. This means that not only the transitions between pixel values are encoded, but also the intensity variations are considered, providing a more discriminating representation of textures.

The integration of both sign and magnitude information in CLBPS results in a more robust and descriptive texture operator, making it particularly well-suited for tasks such as texture classification. The research by Guo, Zhang, and Zhang presents an important advancement in the field of texture analysis, offering a more complete modeling of local texture patterns, which can be crucial for various computer vision applications, including object recognition, image segmentation, and quality assessment \cite{guo2010completed}. The CLBPS computation can be written as Equation~\eqref{eq:clbps}.
\begin{equation*}
diff_s(x,y) \leftarrow 
\begin{cases}
+1,& \text{if } y \geq x\\
-1,              & \text{otherwise}
\end{cases}
\end{equation*}
\begin{equation}
\label{eq:clbps}
CLBPS(i,j) \leftarrow Vector_{\forall p \in P} [diff_s(val_{center},val_{p})]
\end{equation} \cite{liao2009dominant,guo2010completed}

\subsubsection{Complete Local Binary Patterns Magnitude (CLBPM)}
\label{sec:clbpm}

Complete Local Binary Patterns Magnitude (CLBPM) is an advanced texture analysis technique introduced in the research paper titled "A Completed Modeling of Local Binary Pattern Operator for Texture Classification" by Zhenhua Guo, Lei Zhang, and David Zhang \cite{guo2010completed}. This method is an extension of the original Local Binary Patterns (LBP) and is designed to provide a more robust and discriminative representation of textures in images.

In CLBPM, the authors enhance the LBP operator by considering the magnitude of the differences between the center pixel and its neighbors. This addition allows CLBPM to capture not only the local binary patterns but also the variations in intensity within the neighborhood. It effectively encodes both structural and textural information. By incorporating these magnitude differences into the LBP framework, CLBPM can provide a more comprehensive description of texture patterns, making it highly suitable for texture classification tasks.

CLBPM's ability to capture both the binary patterns and their associated intensity variations empowers it to handle a wider range of textures, including those with subtle variations and irregularities. This technique has found applications in various fields, such as image analysis, object recognition, and quality control, where accurate texture discrimination is critical. CLBPM represents a valuable advancement in the field of texture analysis, contributing to improved performance in numerous computer vision applications. The CLBPM can be computed as in equation \ref{eq:clbpm}.
\begin{equation}
\label{eq:clbpm}
CLBPS(i,j) \leftarrow Vector_{\forall p \in P} [val_{center}-val_{p}]
\end{equation} \cite{liao2009dominant,guo2010completed}

\subsubsection{Complete Local Binary Patterns (CLBP)}
The LBPs are the representation of an image pixel in terms of its gray level difference from its equidistant neighboring pixels. A lot of research is available to show that these structure representations can be more helpful, for image classification, than the original image. Z. Guo et al. presented another similar texture feature named Complete Local Binary Patterns (CLBP) \cite{guo2010completed}.

The CLBP features are computed by using the neighboring pixels of the image similar to the LBPs. The CLBP of a pixel is a vector of the difference of the gray values from neighboring pixels. Z. Guo et al. again divided that vector into two components i.e. the direction (+,-) and magnitude of the vector value. The computation of the Direction and magnitude can be represented as in equation \ref{eq:clbp_base} \cite{guo2010completed}.
 
\begin{equation}
\label{eq:clbp_base11}
Direction(n) \leftarrow 
\begin{cases}
    +1,& \text{if } LBP_{val} \geq 0\\
    0,              & \text{otherwise}
\end{cases}
\end{equation} \cite{guo2010completed}
\begin{equation}
\label{eq:clbp_base}
Magnitude(n) \leftarrow \vert Point_{center} - Point_{neighbour}\vert
\end{equation} \cite{guo2010completed}

CLBP distributed into magnitude and direction with a combination of the LBP provided good classification on different image datasets \cite{guo2010completed}.

\subsubsection{Dominant Local Binary Patterns (DLBP)}
The standard LBP is good for capturing the texture information of the image having sharp and curvature edges. The curvature edges' information can not be captured in the neighboring 0 and 1 values or the LBP. The histogram in the uniform LBP will be nearly a uniform line for high curvature edges. Uniform LBPs may not have dominating proportions. 
S. Liao et al. proposed a technique of using the most frequent local binary patterns as dominant local binary patterns (DLBPs) \cite{liao2009dominant}. According to the Pareto principle, the image texture can be effectively represented by using only the top $80\%$ of the most occurring patterns. The LDBPs can be computed by taking the union of the part, of the LBP histogram that has the $80\%$ combined information out of the whole LBP histogram information, for all images \cite{liao2009dominant}. The DLBPs can be computed as in equation \ref{eq:dlbp_base}.

\begin{equation}
\label{eq:dlbp_base}
Part_{80\%} \leftarrow LBP(i,j) \forall(i,j) : i \in Images \wedge j \in LBP_{i} \wedge \dfrac {\sum_{k=0}^{j} LBP}{\sum_{k=0}^{n} LBP} \geq 80\%
\end{equation} \cite{liao2009dominant}

%\subsection{D B Local Binary Patterns (DBLBP)}
%\label{sec:dblbp}

%\subsection{(AGLD)}
%\label{sec:agld}

%\subsection{(SGLD)}
%\label{sec:sgld}

\subsubsection{Rotation invariant LBP}
Rotation invariant LBP extends the basic LBP by making it invariant to rotation. It involves circularly shifting the binary pattern and selecting the minimum value among all possible rotations as the representative LBP code \cite{liu2016median,li2011scale}.

The basic idea is to compute LBP for multiple rotation angles and select the minimum value as the representative LBP code. Let's denote the following variables and parameters:

    P: The number of sampling points in a circular neighborhood.
    R: The radius of the circular neighborhood.
    I(x, y): The intensity value of the pixel at location (x, y) in the image.

The algorithm for computing Rotation Invariant LBP can be shown in Algorithm~\ref{algo:lilbp}: 

\begin{algorithm}
\SetKwData{P}{P}
\SetKwData{R}{R}
\SetKwData{LBPcodes}{LBP\_codes}
\SetKwData{RotationIndices}{rotation\_indices}
\SetKwData{minRotation}{min\_rotation}
\SetKwData{DeltaI}{delta\_I}
\SetKwData{RotationInvariantLBP}{rotation\_invariant\_LBP}
\SetKwFunction{cos}{cos}
\SetKwFunction{sin}{sin}
\SetKwFunction{argmin}{argmin}

\KwData{Image, \P (number of sampling points), \R (radius of the circular neighborhood)}

\For{each pixel $(x, y)$ in the image}{
    \LBPcodes[\P] $\leftarrow$ \{0\}\;
    \RotationIndices[\P] $\leftarrow$ \{0\}\;
    
    \For{$i \leftarrow 0$ \KwTo \P - 1}{
        $x_i \leftarrow x + \R \cdot \cos(2\pi i / \P)$\;
        $y_i \leftarrow y - \R \cdot \sin(2\pi i / \P)$\;
        
        \DeltaI $\leftarrow$ Image($x_i$, $y_i$) - Image($x$, $y$)\;
        
        \If{\DeltaI $\geq 0$}{
            \LBPcodes[$i$] $\leftarrow 1$\;
        }
        \Else{
            \LBPcodes[$i$] $\leftarrow 0$\;
        }
        
        \RotationIndices[$i$] $\leftarrow i$\;
    }
    
    \minRotation $\leftarrow$ \argmin{\LBPcodes}\;
    \RotationInvariantLBP $\leftarrow$ \LBPcodes[\minRotation]\;
}
\label{algo:lilbp}
\caption{Rotation Invariant Local Binary Patterns Algorithm \cite{li2011scale}}
\end{algorithm}

\subsubsection{Multi scale LBP}

In contrast, the Local Binary Pattern (LBP) operator, though effective in capturing texture features, is limited in its ability to consider color information. The LBP's computation relies on grayscale images, neglecting the valuable cues provided by color, particularly in distinguishing between objects within natural scenes. Real-world scenes often exhibit diverse lighting and viewing conditions, introducing significant variations in surface illumination, scale, and other factors that pose challenges to the recognition task.

The inherent limitation of the LBP operator is its invariance solely to monotonic changes in gray-level lighting, rendering it less adept at handling the complex variations prevalent in natural scenes. Recognizing these shortcomings, an innovative approach is introduced to address these limitations and enhance the LBP operator's performance \cite{zhu2010multi}. This approach involves the development of six novel multi-scale color LBP operators, aiming to incorporate color information, improve photometric invariance, and enhance the discriminative power of the original LBP operator.

The proposed multi-scale color LBP operators are specifically tailored for Visual Object Classes (VOC) recognition tasks, demonstrating their suitability for such applications. Experimental analyses conducted on the PASCAL VOC 2007 image benchmark assess the performances of these novel operators, providing empirical evidence of their enhanced capabilities in handling color information and addressing the challenges posed by variations in real-world scenes.

\subsubsection{The non-uniform LBP}
Non-uniform Local Binary Patterns (LBP) represent an extension and enhancement of the basic LBP methodology by incorporating considerations for spatial relationships among neighboring pixels. This modification is particularly valuable in capturing complex texture patterns, as it introduces a more nuanced understanding of the inter-pixel dynamics within an image \cite{nanni2012simple}. The spatial relationships considered in non-uniform LBP contribute to a more sophisticated representation of textures, allowing for improved discrimination and characterization of intricate patterns within images.

\subsubsection{Local Quinary Pattern (LQP)}
Local Quinary Pattern extends the concept of Local Binary Patterns (LBP) by considering five possible values (instead of two in LBP) for the comparison of pixel values within a neighborhood. The LQP operator aims to capture more complex texture patterns by encoding patterns based on five intensity levels \cite{nanni2012survey}. The computation of the LQP can be understood by the Equation ~\ref{eq:lqp}.

Equation (for a 3x3 neighborhood with the center pixel as P):
\begin{equation}
    \label{eq:lqp}
    LQP(P) = \sum_{i} (I_i > P - I_c)
\end{equation}
Where $I_i$ represents the intensity values of neighboring pixels, and $I_c$ is the intensity value of the center pixel.

\subsubsection{Noise Resistant LBP (NRLBP)}
Noise-resistant LBP is an extension of Local Binary Patterns designed to be robust against noise in images. It achieves this by introducing a threshold that adapts to local image characteristics \cite{ren2013noise}. The computation of the NRLBP can be understood by the Equation ~\ref{eq:nrlbp}.

\begin{equation}
    \label{eq:nrlbp}
    NRLBP(P) = \sum_{i} (I_i - I_c \geq T)
\end{equation}
Where $I_i$ represents neighboring pixel intensities, $I_c$ is the center pixel intensity, and T is the adaptive threshold.

\subsubsection{Local Derivative Ordinal Binary Pattern (LDOBP)}
LDOBP is a texture analysis method that captures local derivative information. It encodes information about the magnitude and sign of local gradients, which is useful for texture classification tasks \cite{shang2015robust}. The computation of the LDOBP can be understood by the Equation ~\ref{eq:LDOBP}.

\begin{equation}
    \label{eq:LDOBP}
    LDOBP(P) = sign(I_i - I_c) * (|I_i - I_c| > T)
\end{equation}

Where $I_i$ represents neighboring pixel intensities, $I_c$ is the center pixel intensity, and T is a threshold.

\subsubsection{Center-Symmetric Local Binary Patterns (CSLBP)}
Center-Symmetric Local Binary Patterns (CSLBP) is a variant of Local Binary Patterns (LBP) designed to capture texture information in a rotation-invariant manner. It was introduced to address the issue of standard LBP's sensitivity to rotation. CSLBP computes LBP codes for circular neighborhoods around each pixel and takes into account the symmetrical patterns \cite{zhu2010multi}. Here's a simplified equation to compute CSLBP:
\begin{enumerate}
    \item Select a circular neighborhood of pixels around the center pixel.
    \item For each neighbor pixel, compare its intensity to the center pixel's intensity.
    \item If the neighbor's intensity is greater than or equal to the center pixel's intensity, assign a value of 1; otherwise, assign a value of 0.
    \item Create a binary string from these 1s and 0s.
    \item Convert the binary string to a decimal value.
    \item The CSLBP code is obtained as the decimal value.
\end{enumerate}

\subsubsection{Elliptical Binary Pattern (EBP)}
Elliptical Binary Pattern (EBP) is a texture analysis method that builds upon the principles of Local Binary Patterns (LBP) while introducing a more adaptable approach to capturing texture features. EBP is especially useful in scenarios where textures exhibit elliptical or elongated patterns. EBP extends the basic LBP concept by allowing for varying radii in different directions, making it suitable for capturing features in textures that are not strictly circular \cite{liao2007face}.

EBP defines a circular neighborhood around a central pixel, similar to LBP. However, instead of using a fixed radius for the entire neighborhood, EBP employs two radii, one for the vertical direction (major axis) and another for the horizontal direction (minor axis). This is particularly useful for analyzing textures with anisotropic properties.

The EBP calculation process involves several steps:
\begin{enumerate}
    \item Selecting a Neighborhood: Define an elliptical neighborhood with major and minor axes' radii.
    \item Center Pixel Value: Choose a center pixel in the neighborhood.
    \item Thresholding: Compare the intensity values of the surrounding pixels to the center pixel's intensity in both the major and minor axes. If a surrounding pixel's intensity is greater or equal to the center pixel's intensity in the major axis, assign it a value of 1; otherwise, assign it a value of 0. Do the same for the minor axis.
    \item Binary to Decimal Conversion: Convert the binary strings from the major and minor axes to decimal values. These values are used to form an EBP code for the central pixel.
    \item Repeat for All Pixels: Repeat these calculations for all pixels in the image, resulting in an image of EBP codes.
    \item Histogram Calculation: Compute a histogram of EBP codes, capturing the distribution of texture patterns.
\end{enumerate}
   
EBP is versatile, as you can adapt the radius of the major and minor axes based on the texture's characteristics. This adaptability allows EBP to excel in scenarios where traditional LBP might not adequately capture the texture's features.

\subsubsection{Scale and Orientation Invariant LBP}
Texture patterns may appear at different scales within an image. To make LBP scale-invariant, the neighborhood size or the radius of the circular neighborhood is allowed to vary. In other words, the LBP operator should be able to capture texture patterns regardless of their size \cite{hegenbart2015scale}.

Texture patterns can also appear in various orientations within an image. To make LBP orientation-invariant, a circular neighborhood should be rotated to align with the predominant orientation of the texture pattern, ensuring that the LBP descriptor can capture the pattern regardless of its orientation \cite{hegenbart2015scale}.

\subsubsection{Adaptive Local Binary Pattern (ALBP)}
Adaptive Local Binary Pattern (ALBP) is a texture descriptor used in image processing and computer vision, particularly in the context of texture analysis and image classification \cite{liu2016medical}. It is a variation of the Local Binary Pattern (LBP) method, which characterizes the texture of an image by comparing each pixel in the image to its neighboring pixels.

ALBP extends the idea of LBP by introducing adaptiveness in selecting the neighborhood configuration for each pixel in an image. This adaptiveness allows ALBP to better capture the local variations in texture, making it particularly useful for applications such as medical image classification where subtle variations can be crucial for accurate diagnosis.

\subsubsection{Moment invariant and Fourier descriptor shape features}
Moment invariants and Fourier descriptors are methods used for shape analysis and recognition. Moment invariants capture the geometric properties of shapes, and Fourier descriptors represent a shape as a combination of sinusoidal waves. These techniques are often used for object recognition and image analysis \cite{pourghassem2008content}.

\subsubsection{Local Ternary Patterns}
LBPs are resistant to the darkness effects. They are invariant to the transformation. LBPs are sensitive to the noise in the nearly or uniform regions of the image. The medical images are usually nearly uniform in most of the regions so, There is a need to tolerate the noise.

The noise can be handled by Local Ternary Patterns (LTP) that use a three-level representation of the values at any pixel i.e. the $+1$, $0$, and $-1$ instead of $0$ and $+1$ only. The values are $+1$ if the pixel value is greater than that of the sum of the center value and the predefined constant d. The value is 0 if the pixel value is greater than $center value - constant$ but lesser than $(center value + constant)$. The value is -1 if the pixel value is less than $(center value - constant)$.
The LTBs are calculated by the computation of following operations on grayscale image:
\begin{enumerate}
	\item Binary code of pixel difference from the centroid for each pixel as centroid as in equation \ref{eq:lbp_binary_code}.
	\item Representing all the pixel values as a set of 0,+1,-1 numbers as in equation \ref{eq:ltp_pos}
	\item Building two pattern square matrices of dimensions $((r*2)+1) \times (r*2)+1)$ for each pixel by using -1 as zero in one and 1 in other by using equations \ref{eq:ltp_pos} and \ref{eq:ltp_neg}.
	\item Computation of two decimal values by using pixel values as a series of bits of binary number as in equation \ref{eq:lbp}.
\end{enumerate}

\begin{equation}
\label{eq:ltp_pos}
X_1(i,j) \leftarrow 
\begin{cases}
+1,& \text{if } X_1(i,j) == +1\\
0,& \text{if } X_1(i,j) == -1\\
0,& \text{if } X_1(i,j) == 0
\end{cases}
\end{equation}
\begin{equation}
\label{eq:ltp_neg}
X_0(i,j) \leftarrow 
\begin{cases}
0,& \text{if } X_1(i,j) == +1\\
1,& \text{if } X_1(i,j) == -1\\
0,& \text{if } X_1(i,j) == 0
\end{cases}
\end{equation}

The LTPs are split into non-negative and non-positive parts into two LBPs. The two LBPs will give two histograms that can be used as a feature as these are tolerated towards the noise more than that of the one LBP histogram \cite{tan2010enhanced}.

\subsection{Spectral Features}
Tonal variations in a grayscale image are a fascinating facet of digital visual representation. These features, often subtle and nuanced, encapsulate the gradations of light and shadow within the image, serving as a rich tapestry of information. Tonal variations are governed by the pixel intensities, ranging from the brightest whites to the deepest blacks and all the shades of gray in between. They convey depth, texture, and dimensionality in a monochromatic canvas, and play a pivotal role in the visual storytelling of a photograph or artwork. These variations can highlight intricate details, create mood and atmosphere, and bring a sense of realism to the image. Understanding and manipulating these tonal variations is a crucial skill for photographers and digital artists, as it allows them to evoke specific emotions, direct the viewer's gaze, and craft compelling visual narratives in grayscale imagery. The range of the values shows the contrast of the image. These features are a good representation of the grayscale level of the image. and works better when the part of the image under observation has a good gray level or combined RGB values.
There is an assumption in spectral features that the texture information in the RGB image is directly proportional to the textual information in the gray-scale image. The relationship can be analyzed better in Equation~\ref{eq:rgbpropgray}.
\begin{equation}
\label{eq:rgbpropgray}
Texture(RGB) \propto Texture(Gray)
\end{equation}

\subsection{Textural Features}
Textural features are the representation of the distribution of the tonal variation in the different sections or pixels of the image. These features are the statistical distribution of the darkness. The features are very powerful for locating an object inside the image. The textural is very helpful for the analysis of grayscale images. The features are powerful when the image contains a higher variance of color or high contrast.

There is an assumption in spectral features that the texture information in the RGB image is directly proportional to the textual information in the gray-scale image. The relationship can be analyzed better in Equation~\ref{eq:rgbpropgray}.

\subsubsection{Gabor Filters (GF)}
\label{sec:gf}
Gabor Filter (GF) is a linear filter for the texture classification. The GF is a representation of any existing pattern or sequence of the content. In the case of an image, the specific content is the sequence of a specific color in a specific direction. The GF is based on the idea of human visibility of different color patterns. A Gabor plan or the Gabor feature that can be computed by the image from any direction and the part of the image controlled by the tuneable parameters \cite{fogel1989gabor,bovik1990multichannel,mehrotra1992gabor}.
 
The Gabor filter is defined in the spatial domain as a complex sinusoidal wave modulated by a Gaussian function. The formula for a 2D Gabor filter in the spatial domain is shown in Equation \ref{eq:gabour}.

\begin{equation}
    \label{eq:gabour}
    g(x, y) = \exp\left(-\frac{x^2 + \gamma^2 y^2}{2\sigma^2}\right) \cos(2\pi f x)
\end{equation}

Where:

    $x$ and $yy$ represent spatial coordinates.
    $\sigma$ is the standard deviation of the Gaussian envelope, controlling the filter's size.
    $f$ is the frequency of the sinusoidal carrier wave.
    $\gamma$ is the aspect ratio, which determines the ellipticity of the filter.

Gabor filters can be tuned to extract specific frequency and orientation information from an image. By varying the values of $\alpha$, $f$, and $\gamma$, you can create a bank of Gabor filters that are sensitive to different spatial frequencies and orientations. These filters are then applied to an image using convolution.

\subsubsection{Local Color Features (LCF)}
\label{sec:lcf}

Local Color Features, often used in image classification tasks, are an essential component of the visual representation of an image. These features focus on capturing information about the distribution and variation of colors in local regions of an image. One notable technique for extracting local color features is the Local Color Contrastive Descriptor. This method extends the principles of traditional Local Binary Patterns (LBP) from grayscale to color images, offering an effective means to encode local color information \cite{guo2015local}.

The Local Color Contrastive Descriptor operates by examining color information within a defined neighborhood of each pixel. It assesses the contrast between the central pixel and its neighboring pixels in terms of color differences, enabling it to identify distinctive color patterns and texture variations. This technique is particularly valuable in applications where color plays a significant role, such as object recognition in natural scenes, where the unique combination of colors can be an essential discriminative factor. By incorporating local color features like those obtained through the Local Color Contrastive Descriptor, image classification models gain the ability to distinguish objects or scenes based on their specific color characteristics, contributing to the accuracy and robustness of the classification process.

\subsubsection{Color Layout}
\label{sec:cl}
Color Layout is a visual descriptor that focuses on the spatial distribution of colors in an image. It provides information about the arrangement of colors in an image and is useful for tasks like image retrieval and content-based image analysis \cite{sikora2001mpeg,lux2008lire}.

The Color Layout descriptor is typically computed using the following steps \cite{sikora2001mpeg}:
\begin{enumerate}
    \item Color Quantization: The image is first divided into a grid, and each grid cell is quantized into a fixed number of color bins (typically 256 or 512). The quantization can be performed in different color spaces, such as RGB or HSV.
    \item Color Histogram: For each grid cell, a color histogram is computed, which represents the distribution of colors within that cell. The color histogram is a vector of color bin counts.
    \item Concatenation: The color histograms from all grid cells are concatenated to form a single feature vector, which is the Color Layout descriptor.
\end{enumerate}

Mathematically, the Color Layout descriptor can be represented as follows:

Let:

    $N$ is the total number of grid cells.
    $C$ is the number of color bins.
    $H_i$ is the color histogram for grid cell $i$.

The Color Layout descriptor CL can be represented as a feature vector:
\begin{equation}
    \label{eq:cl}
    CL = [H_1, H_2, \ldots, H_N]
\end{equation}

\subsubsection{Edge Histogram (EH)}
\label{sec:EH}
Edge Histogram is a visual descriptor that focuses on the distribution of edges in an image. It characterizes the presence and strength of edges within an image, which can be essential for tasks like object recognition and shape analysis \cite{sikora2001mpeg,lux2008lire}.

The Edge Histogram descriptor is typically computed using the following steps \cite{sikora2001mpeg}:
\begin{enumerate}
    \item Edge Detection: Edge detection techniques are applied to the image to identify edges. Common methods include Sobel, Canny, or Prewitt edge detectors.
    \item Edge Orientation Quantization: The orientations of detected edges are quantized into a fixed number of bins (e.g., 8 bins). This quantization step groups similar edge orientations together.
    \item Edge Histogram: For each image, an edge histogram is computed, where each bin in the histogram represents the count of edges within a specific orientation range.
\end{enumerate}
Mathematically, the Edge Histogram descriptor can be represented as follows:

Let:
    $E$ is the number of edge orientation bins.
    $HE$ is the edge histogram.

The Edge Histogram descriptor $EH$ can be represented as a feature vector:
\begin{equation}
    EH = [HE_1, HE_2, \ldots, HE_E]
\end{equation}

The edge histogram $HE_i$ for each orientation bin $i$ counts the number of edges in that orientation range.

\subsubsection{Tamura, Coarseness, Contrast and Directionality (TCCD)}
\label{sec:tccd}

Tamura texture features are a set of three measures that capture different aspects of texture perception \cite{tamura1978textural,lux2008lire}. These are:

\textbf{Coarseness:}
Coarseness describes the scale or size of the texture elements in an image. It is computed as the inverse of the average size of the elements in the image. Mathematically, it can be defined as:

\begin{equation}
    T1 = 1 / \Sigma(1 + d(i, j))
\end{equation}

where $d(i, j)$ is the distance between adjacent pixel pairs $(i, j)$ with similar intensity values.

\textbf{Contrast:}
Contrast quantifies the difference in brightness or intensity between neighboring regions of an image. It is calculated as the variance of the local intensity values. Mathematically, it can be defined as:

\begin{equation}
    T2 = \sigma^2 / \mu
\end{equation}

where $\sigma^2$ is the local variance and $\mu$ is the local mean of intensity values.

\textbf{Directionality:}
Directionality measures the predominant direction or orientation of the texture elements in an image. It is computed using the gray-level co-occurrence matrix (GLCM) and typically involves calculating the statistical properties of the GLCM. The exact formula for T3 varies, but it usually represents the dominant direction.

\subsubsection{Color and Edge Directivity Descriptor (CEDD)}
\label{sec:cedd}
The Color and Edge Directivity Descriptor (CEDD) is a compact feature descriptor used in image indexing and retrieval. It's especially useful in the context of medical image analysis, where efficient and effective feature extraction is crucial. CEDD combines color and edge information to create a concise representation of an image \cite{chatzichristofis2008cedd,lux2008lire}.

The CEDD descriptor has two main components: color and edge information. It divides the color space into several bins and computes color histogram statistics. Additionally, it analyzes the edge orientation in the image.
\textbf{Color Component:}
The color component of CEDD starts by dividing the RGB color space into $n$ bins. Most of the authors used $n=8$ \cite{chatzichristofis2008cedd}. Each bin represents a range of colors. To calculate the CEDD color descriptor, follow these steps:
\begin{enumerate}
    \item Calculate the color histogram $H_color$ for each channel (Red, Green, and Blue). Each channel's histogram can be represented as a one-dimensional array with $n$ bins.
    
    $H-color-r[i]$ is the number of pixels in the Red channel falling into the i-th color bin.    
    $H-color-g[i]$ is the number of pixels in the Green channel falling into the i-th color bin.
    $H-color-b[i]$ is the number of pixels in the Blue channel falling into the i-th color bin.

    \item Normalize the histograms to sum to 1:
    \begin{equation}
        \begin{split}
            HColorR_{normalized}[i] = HColorR[i] / pixels_{total} \\
            HColorG_{normalized}[i] = HColorG[i] / pixels_{total} \\
            HColorB_{normalized}[i] = HColorB[i] / pixels_{total}
        \end{split}
    \end{equation}
\end{enumerate}

\textbf{Edge Component:}
The edge component captures edge information using the Canny edge detector or any edge detection algorithm of your choice. You calculate the edge histogram, $H_edge$ by dividing the edge orientations into $m$ bins. Most of the authors used $n=8$ \cite{chatzichristofis2008cedd}.

$HEdge[i]$ is the number of edge pixels with orientations falling into the i-th edge bin.

%Normalize the edge histogram as follows:
The histogram can be normalized by using Equation ~\ref{eq:norm-histogram}.

\begin{equation}
    \label{eq:norm-histogram}
    HEdge_{normalized}[i] = HEdge[i] / TotalEdgePixels
\end{equation}
CEDD descriptor is a combination of the normalized color and edge histogram that can be computed by using Equation ~\ref{eq:cedd}

\begin{equation}
    \label{eq:cedd}
    CEDD = [HColorR_{normalized}, HColorG_{normalized}, HColorB_{normalized}, HEdge_{normalized}]
\end{equation}

\subsubsection{Fuzzy Color and Texture Histogram (FCTH)}
\label{sec:fcth}
The Fuzzy Color and Texture Histogram (FCTH) is a feature descriptor used for image indexing and retrieval, primarily in the context of medical image analysis. It combines color and texture information to represent the content of an image effectively \cite{chatzichristofis2008fcth,lux2008lire}.
FCTH is calculated by first dividing an image into a grid and then quantizing the colors and textures within each grid cell. It uses fuzzy sets to represent the distribution of color and texture features. The computation of FCTH can be represented as Equation ~\ref{eq:fcth}.

\begin{equation}
    \label{eq:fcth}
    FCTH = \sum (\mu_c(i, j) \cdot \mu_t(i, j) \cdot \delta(i, j))
\end{equation}

Where, $\mu_c(i, j)$ and $\mu_t(i, j)$ represent the membership values of color and texture for the ith grid cell and jth feature, while $\delta(i, j)$ is the quantized value of the feature. The summation is performed over all grid cells and features, combining color and texture information in a fuzzy manner. The resulting FCTH provides a compact representation of the image's content that can be used for indexing and retrieval.

In the context of CEDD, which is another image descriptor, FCTH can be used in conjunction with CEDD to create a more comprehensive image representation. CEDD focuses on edge information, and FCTH enhances it with color and texture features. By combining these two descriptors, you can improve the performance of image retrieval systems, especially in medical image analysis, where both color and edge information can be crucial for identifying and matching medical images effectively \cite{chatzichristofis2008fcth}.

\subsubsection{Auto Color Correlation (ACC)}
\label{sec:acc}
Huang et al. introduced a novel image feature capturing the color distribution and spatial relationships between colors. This feature, called a correlogram, encodes the probability of encountering specific color differences between neighboring pixels. The calculation of the correlogram is detailed in Equation~\ref{eq:acc_base} \cite{huang1997image,lux2008lire}.

\begin{equation}
\label{eq:acc_base1}
ACC(P_{x,y},C,k)= count(\forall_{i,j} q \in I_{i,j} : (d(q,P)=k \wedge Col(q)=C))/ count(\forall_{i,j} q \in I_{i,j} : (d(q,P)=k)
\end{equation}
\begin{equation}
\label{eq:acc_base}
d(X_{i,j},Y_{k,l})=max(|i-k|,|j-l|)
\end{equation}
\cite{huang1997image}
The $P_{x,y}$ is the pixel from where the probability is been computed at distance $k$ for the colour $C$. The distance is been used as the city block distance in the feature computation \cite{huang1997image}.

\subsubsection{Colour Vector Field}
A color vector serves as a concise representation of color information in an image, encapsulating the combination of various color channels. Commonly used color spaces for this purpose include RGB (Red, Green, Blue), HSV (Hue, Saturation, Value), or Lab (Lightness, a, b). This compact representation enables the encoding of the color attributes of a pixel or a region within an image \cite{hafner2012color}. The color vector provides a convenient and efficient means to describe and analyze the chromatic characteristics of visual elements in images.

\subsection{Contextual Features}
In the realm of medical image analysis, the extraction and utilization of both contextual and non-contextual image features play a pivotal role in the accurate interpretation and diagnosis of medical conditions. Contextual features in medical image analysis pertain to the spatial relationships, interdependencies, and relative positioning of image elements or regions within the overall image \cite{vega2006contextual}. These features provide valuable information about how different elements or structures within the image relate to one another. They often include:
\begin{enumerate}
    \item Texture Analysis
    \item Shape and Contour Descriptors
    \item Spatial Arrangement
    \item Contextual Information from Multiple Modalities
\end{enumerate}

\subsection{Gray Level Co-Occurrence Features}

Gray-level co-occurrence matrix (GLCM) features are a set of texture descriptors widely used in image analysis, including medical image processing. These features capture spatial relationships between pixel intensities in an image, providing information about the texture and patterns present. The GLCM is a statistical representation of how often different pairs of pixel intensities occur in a given spatial relationship (e.g., distance and angle) within an image. From the GLCM, various texture features can be derived, such as energy, entropy, contrast, and homogeneity, among others. For medical image analysis, GLCM features are valuable in characterizing the fine details and structural nuances within images, aiding tasks such as tumor identification or tissue classification where textural information is crucial. They enable a quantitative and robust description of the image texture, contributing to the overall understanding and analysis of complex medical images.

\subsubsection{Haralick Texture Features}
Haralick et al. computed a texture feature for image classification for the images of three different types including satellite images, aerial images, and the images of microscopic objects \cite{haralick1973textural}. The image is been taken as a two-dimensional array of numbers. The representation can be written as in equation \ref{eq:haralick_image}. The features of the images are the statistical functions on the gray-tone spatial dependence (GTSD) matrices.

\begin{equation}
\label{eq:haralick_image}
haralick_{feature}(Image) \leftarrow stats(GT(Image))
\end{equation}

The Haralick features operate under the assumption that the texture information within any image is encapsulated within the Gray-Level Co-occurrence Matrices (GLCM). These matrices, denoted as GTSD matrices, encompass information regarding the frequency of occurrence of various color contrasts in horizontal, vertical, upper diagonal, and lower diagonal neighboring directions. Here, the neighbors are defined as pixels at a distance 'd'. The computation of GTSD can be expressed through equation \ref{eq:gtsd}.
\begin{equation}
\label{eq:gtsd}
    GTSD(i,j,d,horizontal) \leftarrow count(|gt.loc(i)-gt.loc(j)|==d)
\end{equation}

The Haralick features are the texture features using the image in different rotations. The features are very lightly rotation variant as the features are of 4 different angles of the image.

The image region of interest (ROI) was categorized into two types of regions, the convex polyhedron (a polygon in 3D) and the rectangular parallel pipeds (A polygon with 6 parallel faces). The two types of ROIs had different rules to decide the class of ROI or image by using the rules of pair-wise linear discriminant and min-max, similar to the leave one out, for polyhedron and parallel pipes respectively.

The procedure of the Haralick features computation can be presented as algorithm \ref{algo:haralick}.

\begin{algorithm}
    \caption{Haralick Feature Extraction}
    \label{algo:haralick}
    \begin{algorithmic}
    	\Require $Image_{RGB}$
    	\Ensure $Features_{Haralick}$
    	\Procedure{Feature Extraction}{$Image$}
    	\State $Image_{gray} \gets Gray-Converter(Image_{RGB})$    	
    	\State $GrayToneValues \gets Analyse-Image(Image_{gray})$
    	\State $GTSD_{matrices} \gets GTSD-Computation(GrayToneValues,Image_{gray})$
    	\State $Features_{Haralick} \gets Statistical-Functions(GTSD_{matrices})$
    	\State $a\gets Recursion(a)$ \Comment{Call Recursion again}
    	\State \textbf{return} $Features_{Haralick}$
    	\EndProcedure
    \end{algorithmic}
\end{algorithm}

The features were used for the classification on three different types of datasets. Those datasets are as follows:
\begin{enumerate}
  \item The dataset of Earth Resources Technology Satellite (ERTS) images had the images of seven classes. These features produced $83\%$ accuracy on the ERTS dataset.
  \item Panchromatic (1:20 000) aerial photographs of the land use having eights classes. The testing accuracy on that dataset was $82\%$.
  \item The dataset of photomicrographs of sandstones has five classes. The Haralick features provided $89\%$ accuracy on the testing part of the dataset.
\end{enumerate}
The Haralick features proved a good accuracy on three different datasets and the features are easy, with respect to computational time, to compute. These features gave a good accuracy even on some other datasets later as well.

An analysis of the various approaches using Texture Features is shown in Table~\ref{table:annalysis_texture}.

\footnotesize
\setlength{\arrayrulewidth}{0.5pt}
\begin{longtable}{|p{0.1cm}|p{1.5cm}|p{2cm}|p{1.5cm}|p{1cm}|p{0.1cm}|p{0.1cm}|}
\caption{Analysis of the Approaches using texture features}
\label{table:annalysis_texture}
    \\ \hline
    \textbf{Sr.} & \textbf{Paper} & \textbf{Technique Used} & \textbf{Features Used} & \textbf{Dataset} & \textbf{Acc.} & \textbf{F1} \\
    \hline
    \endfirsthead
    
    \hline
    \multicolumn{7}{|c|}{\textbf{Texture Features Approaches Continued}} \\
    \hline
    \textbf{Sr.} & \textbf{Paper} & \textbf{Technique Used} & \textbf{Features Used} & \textbf{Dataset} & \textbf{Acc.} & \textbf{F1} \\
    \hline
    \endhead
    
    % Your table content goes here
    1 & Z. Guo's LBP \cite{guo2010completed} & LDSMT, CLBPC, CLBP-s and CLBP-m & LBP \cite{ojala2002multiresolution} & Clinically-Obtained & X & X \\
    \hline
    2 & Polyp-alert \cite{wang2015polyp} & Multi derivative Edge \cite{wang2015polyp} & Texture  & Self-Captured Videos & 0.98 & \\
    \hline
    3 & Normal and cancerous colonic mucosa detection \cite{esgiar1998microscopic} & Section Profile Calculation  & Texture  &  Camera Images & 0.90 & \\
    \hline
    4 & Neural network-based colonoscopic diagnosis \cite{magoulas2004neural} &  Step-wise discriminant Analysis  &  Texture  &  N/A  &  N/A  &  N/A \\
    \hline
    5 & Texture and NN based Classification \cite{wang2001classification} & Neural Nets & Texture &  Clinically-Obtained  &  N/A  &  N/A \\
    \hline
    6 & Cold \cite{maroulis2003cold} & CoLD  & Texture  & Clinically-Obtained  & 0.95 &  N/A \\
    \hline
    % Add more rows as needed
    
\end{longtable}
\normalsize

\subsubsection{Neighborhood Preserving Embedding (NPE) on texture features}
Neighborhood Preserving Embedding is a technique that aims to learn a lower-dimensional representation of texture features while preserving the local neighborhood relationships. It is often used for dimensionality reduction in texture analysis \cite{nanni2012survey}.
Simplified Objective Function for NPE) can be expressed mathematically as Equation~\ref{eq:NPE}.
\begin{equation}
    \label{eq:NPE}
    \min_{W} J(W) = \left\| X - XW \right\|_F^2
\end{equation}
Where $X$ is the feature matrix, $W$ is the transformation matrix, and $F$ represents the Frobenius norm.

\subsection{Deep Features}
\label{sec:nnf}

The field of medical image analysis has witnessed remarkable progress over the years, largely due to the integration of cutting-edge technologies such as neural networks. The crucial role of feature extraction in the process of medical image analysis cannot be overstated. This task involves identifying and isolating significant patterns, structures, and anomalies within medical images. A multitude of neural network architectures has been explored to excel in this domain, each tailored to serve specific purposes. This article delves into some of the most prominent neural network architectures used for feature extraction in medical image analysis, highlighting their effectiveness and the diverse criteria upon which they are selected.

\subsubsection{ResNet: A Stalwart in Medical Image Feature Extraction}
The ResNet (Residual Network) architecture stands out as a pioneering force in applying deep learning to medical image analysis. Its revolutionary concept, residual connections, tackles the vanishing gradient problem, enabling the training of exceptionally deep neural networks. This multi-layered structure empowers ResNet to excel at capturing subtle and complex features within medical images. Consequently, it demonstrates remarkable performance in detecting abnormalities like tumors and fractures. Furthermore, leveraging ResNet as a feature extractor delivers impressive results and establishes a benchmark for evaluating other architectures.
ResNet is a groundbreaking architecture that introduced the concept of residual learning. Its key features include:
\begin{itemize}
    \item \textbf{Deep Residual Blocks:} ResNet uses deep residual blocks, allowing for the training of extremely deep neural networks. These blocks contain shortcut connections that bypass one or more layers, helping to mitigate the vanishing gradient problem.
    \item \textbf{Bottleneck Architecture:} ResNet also employs bottleneck architectures, which reduce the number of parameters and computational cost while maintaining model performance.
    \item \textbf{State-of-the-Art Accuracy:} ResNet models have consistently achieved top-tier accuracy in image classification tasks, making them a strong choice for disease detection.
    \item \textbf{Medical Image Analysis:} ResNet deep architecture enable the extraction of intricate features from medical images, enhancing the model's ability to detect subtle signs of diseases.
\end{itemize}

\subsubsection{DenseNet: Maximizing Feature Reuse}
In addition to ResNet, DenseNet, short for Densely Connected Convolutional Networks, has made a significant impact on medical image analysis. DenseNet's innovative approach to connectivity involves dense connections between layers, ensuring that each layer receives the feature maps from all preceding layers. This maximizes feature reuse and facilitates information flow throughout the network. By enabling the network to access features from multiple depths simultaneously, DenseNet has proven to be highly effective in extracting discriminative features from medical images, thus improving the accuracy of abnormality detection.
DenseNet is known for its dense connectivity pattern, wherein each layer receives input from all previous layers. Key features of DenseNet include:

\begin{itemize}
    \item \textbf{Dense Connectivity:} DenseNet's densely connected layers enhance feature reuse and information flow, which results in a more compact model and improved gradient flow.
    \item \textbf{Feature Concatenation:} Instead of simple addition as in ResNet, DenseNet employs feature concatenation, which encourages feature diversity and boosts the network's representational power.
    \item \textbf{Reduction in Vanishing Gradient:} Dense connectivity combats the vanishing gradient problem, facilitating the training of very deep networks.
    \item \textbf{Medical Image Analysis:} DenseNet's deep architecture enables the extraction of intricate features from medical images, enhancing the model's ability to detect subtle signs of diseases.
\end{itemize}

\subsubsection{VGG: The Elegance of Simplicity}
The VGG (Visual Geometry Group) architecture is characterized by its simplicity, with a consistent 3x3 kernel size and the exclusive use of 2x2 max-pooling layers. Although it may not be as deep as ResNet or as densely connected as DenseNet, VGG's straightforward architecture has demonstrated robust performance in medical image analysis. The uniformity of VGG's structure simplifies network design and training, making it an excellent choice for feature extraction in medical image analysis. Its elegant simplicity is especially valuable when computational resources are constrained, or real-time processing is required.

VGG is characterized by its uniform architecture, using small convolutional filters with a consistent 3x3 kernel size. Key features of VGG include:

\begin{itemize}
    \item \textbf{Simplicity and Uniformity:} VGG's straightforward architecture employs repeated 3x3 convolutions and 2x2 max-pooling layers, making it easy to understand and implement.
    \item \textbf{Multiple Variants:} The VGG family includes variations such as VGG16 and VGG19, which differ in the number of layers, providing a trade-off between model size and accuracy.
    \item \textbf{Strong Baseline:} VGG networks serve as robust baseline architectures, with reliable performance in various computer vision tasks.
    \item \textbf{Robestness:} VGG provides a robust baseline and can be a reliable choice for disease classification tasks, especially when computational resources are not a constraint.
\end{itemize}

\subsubsection{MobileNet}
In some medical imaging scenarios, real-time detection of abnormalities is of paramount importance. MobileNet and MobileNet V2 have emerged as solutions for this requirement. These architectures are tailored to operate efficiently on resource-constrained platforms, such as mobile devices and embedded systems. MobileNet reduces computational complexity through depth-wise separable convolutions, resulting in lightweight yet effective networks. MobileNet V2 builds upon its predecessor by introducing inverted residuals and linear bottlenecks. These enhancements lead to better performance without a substantial increase in computational cost. Real-time detection of abnormalities becomes a reality through the implementation of these architectures.

MobileNet is designed for mobile and embedded devices, emphasizing model efficiency. Key features of MobileNet include:

\begin{itemize}
    \item Depthwise Separable Convolutions: MobileNet relies on depthwise separable convolutions, which reduce the computational cost significantly while preserving accuracy.
    \item Lightweight and Fast: MobileNet models are lightweight and suitable for real-time applications on resource-constrained devices.
    \item Squeeze-and-Excitation Blocks: MobileNet V2 incorporates squeeze-and-excitation blocks, which adaptively recalibrate channel-wise feature responses.
\end{itemize}

\subsection{Fine-Tuning of Neural Networks for Feature Extraction}
In the context of medical image analysis, neural networks are often fine-tuned to adapt to the specific characteristics of a dataset. Fine-tuning entails initializing a pre-trained network, such as ResNet, DenseNet, or VGG, with weights from a general image dataset like ImageNet. The network is then adjusted to the target medical image dataset, with learning rates and data augmentation techniques tailored to the task. This process allows the network to leverage the knowledge acquired from general image features while fine-tuning its parameters to the intricacies of medical images. As a result, the network becomes highly adept at extracting relevant features for abnormality detection.

\subsubsection{Neural Network Architectures (NNA) for Image Classification}
\label{sec:nna}
A lot of work has been done on image classification by using representation learning/deep learning. The basic idea behind the neural network is the use of multi-layer perceptron (MLP). The MLPs are small functional units to compute the dot product of the input and weight and addition of the base shown in equation \ref{eq:mlp}. Then the MLP applies an activation function, i.e. sigmoid, on the input as in equation \ref{eq:mlp_sigmoid}.
\begin{equation}
    \label{eq:mlp}
    Y=function(W^TX+b)
\end{equation}
\begin{equation}
    \label{eq:mlp_sigmoid}
    \begin{split}
    Y=\omega(W^TX+b) \\
    \omega(x)=\frac{e^x}{1+e^x}
    \end{split}
\end{equation}
There are some well-known datasets available, for general image classification, including MNIST \cite{deng2012mnist}, MS-COCO \cite{lin2014microsoft}, imageNet \cite{deng2009imagenet,russakovsky2015imagenet}, VisualQA \cite{agrawal2017vqa} and Fashion-MNIST \cite{xiao2017fashion}. The models have been built and trained on these available datasets. There are various models for different types of image classification problems and datasets \cite{lecun1998gradient,lecun1998gradient,krizhevsky2012imagenet,simonyan2014very,chen2017rethinking}. The commonly used Neural Network Architectures for medical image classification are as below.

\subsubsection {LeNet}
\label{sec:lenet}
LeCun et al. proposed a neural network for image classification \cite{lecun1998gradient}. The network is a convolution neural network (CNN). CNN is different from standard MLP in terms of weight-sharing and pooling operations. The weight sharing allows the CNN to detect a similar object in multiple regions of an image at once and the pooling layer shrinks the image in such a way that every pixel of the resultant image contains the information of the neighboring pixels of the original image as well. The final part of a CNN is similar to the standard MLP where a set of fully connected layers are been used for the prediction of the output labels. LeNet is a network of two convolution Layers. The kernel size is maximum at the input corner and reduces towards output layers ($size_{kernal} \propto 1/layer_{number}$). The network uses hyperbolic tangent as an activation unit that can be computed as in the Equation~\ref {eq:tanh}.
\begin{equation}
    \label{eq:tanh}
    tanh(x)=\frac{e^z-e^{-z}}{e^z+e^{-z}}
\end{equation}

The network is used by various researchers for medical image analysis and specialized tasks of gastro-intestinal abnormalities detection.

\subsubsection {AlexNet}
\label{sec:alexnet}
AlexNet is a CNN with five convolution layers \cite{krizhevsky2012imagenet}. The kernel size is maximum at the input corner and reduces towards output layers ($size_{kernal} \propto 1/layer_{number}$). The Alexnet uses a rectified linear unit (ReLU) as an activation unit that can be computed as in Equation~\eqref{eq:relu} \cite{nair2010rectified}.
\begin{equation}
    \label{eq:relu}
    relu(x)=max(0,x)
\end{equation}

\subsubsection {VGG or OxfordNet}
\label{sec:vgg}
The VGG19 or the Oxfordnet is a much deeper network than LeNet and AlexNet \cite{simonyan2014very}. The network has 19 layers including Convolution and Pooling layers. The network won the image classification challenge of the ImageNet in 2014 \cite{deng2009imagenet}.

\section{Binary Classification Techniques (BCT)}
Binary classification pertains to the task of categorizing data instances into one of two distinct classes. In the context of medical diagnostics, this problem concerns the determination of the presence or absence of anomalies, such as abnormal sessions or polyps, within a given dataset. The objective is to develop a robust and accurate model that can effectively discern between these two classes, facilitating timely and accurate identification of pathological conditions, thereby aiding in informed medical decision-making. This classification task holds significant relevance in healthcare and medical research, where the early detection of anomalies can be crucial for patient outcomes and the advancement of diagnostic procedures.

There exists a substantial body of literature dedicated to the binary classification of medical image datasets. This extensive research encompasses a wide array of image modalities, including but not limited to radiological images, histopathological slides, and various forms of medical scans. These studies aim to harness advanced machine learning and deep learning techniques to discern between normal and abnormal patterns, thus contributing to the early detection and diagnosis of medical conditions. The significance of such work is underscored by its potential to enhance diagnostic accuracy, reduce human error, and ultimately improve patient care. Researchers and practitioners have explored a plethora of methodologies, algorithms, and evaluation metrics tailored to the unique challenges posed by medical image classification, further advancing the field and paving the way for the development of more effective and precise diagnostic tools.

A. Kumar et al. \cite{kumar2016ensemble} presented an ensemble of the well-known convolutional neural network (CNN) for the feature extraction of the images. The methodology used by the authors is based on the combination of the features from AlexNet \cite{krizhevsky2012imagenet} and Google LeNet\cite{szegedy2015going} with the imageNet weights and the finetuned both networks. These combined features have been used as input for the ensemble network with a set of multiclass SVM classifiers. The approach is tested on the dataset of ImageClef2016 \cite{de2016overview}, containing 6776 training and 4166 test images, and resulted in top 1 accuracy of $82.48\%$ and top five accuracy of $96.59\%$.

N. Tajbakhsh et al. \cite{tajbakhsh2015comprehensive} proposed a two-stage computer-aided polyps detection system. The stage one is for the shape feature extraction and the stage two is to classify with minimization of the false positive. In stage one canny edges have been extracted and minimizes the edges i.e. removes the non-polyps boundary edges using three class random forest classifier having 100 fully grown trees. Stage two uses texture, color, shape, etc. features in a CNN to learn the weights of these features and then classify them. The experiments have been executed on a set of 40 colonoscopy videos with 18900 training and testing frames with different class distributions. The analysis resulted in $50\%$ sensitivity at the frame rate of 0.002 FPs/frame.

Wang et al. (2013) present a groundbreaking method for automated polyp detection in colonoscopy images using part-based multiderivative edge cross-sectional profiles (ECSP) \cite{wang2013part}. This approach builds upon traditional ECSP by incorporating derivative functions and segmenting profiles into distinct parts. This allows the method to capture intricate features specific to each polyp region, including shape, texture, and surface variations.

The researchers compared their technique against existing methods using a diverse dataset of 42 polyps with varying protrusion levels, viewing angles, lighting conditions, and scales. Their method achieved superior performance in both accuracy and processing speed. It demonstrated a higher area under the ROC curve, signifying its effectiveness.

For example, at a true positive rate of 81.4\% for polyp detection, the proposed method identified significantly fewer false positives (0.32 per image) compared to the best existing method (1.8 per image). Additionally, it precisely marked polyp boundaries, providing valuable visual feedback for clinicians. These results suggest that part-based multiderivative ECSP has the potential to become a valuable tool for reliable polyp detection and visual guidance in colonoscopy.

This study by Alexandre et al. (2008) explores two approaches for polyp detection in endoscopic video: texture-based methods and a simpler method combining color and position features \cite{alexandre2008color}. While texture-based methods have shown success elsewhere, the authors compare them to this more straightforward approach. Even though the color and position method generates a larger number of features (8,000 per image), a Support Vector Machine (SVM) with a Radial Basis Function (RBF) kernel effectively handles this high dimensionality and outperforms the texture methods on a dataset of 4,620 endoscopic images.

The proposed method works by dividing each image into smaller 40x40 pixel squares, analyzing each for polyps. If a square is identified as a polyp, the entire image is flagged. This cautious approach prioritizes minimizing missed polyps (false negatives) at the potential cost of some false positives. Notably, the study focuses on accurate sub-image classification and doesn't extensively analyze false positives/negatives.

Despite the high number of features due to solely relying on color and pixel position, the SVM classifier successfully handles this complexity. The results are promising, achieving an Area Under the Curve (AUC) of 94.87 on the 4,620 image dataset using 10-fold cross-validation. The authors propose future work applying this method to video capsule endoscopy images instead of colonoscopy videos.

In their work, Hwang et al. (2007) propose a method for polyp detection in colonoscopy videos, leveraging elliptical shapes as a key feature \cite{hwang2007polyp}. This approach facilitates the identification of potential polyps, ultimately aiming to improve the accuracy of polyp detection in colonoscopy procedures.

The research paper, "Towards Automatic Polyp Detection with a Polyp Appearance Model" by Bernal et al. (2012) tackles the challenge of automatically detecting polyps in colonoscopy videos \cite{bernal2012towards}. Their method tackles this in three stages: segmentation, description, and classification. Segmentation isolates potentially relevant regions, ensuring any polyps are fully contained within a single area while excluding irrelevant background.  A novel descriptor, SA-DOVA (Sector Accumulation-Depth of Valleys Accumulation), is introduced to describe these regions. While SA-DOVA identifies promising regions, it doesn't guarantee the presence of a polyp. Finally, classification differentiates between regions using the SA-DOVA descriptor's peak values. Initial results show promising performance, particularly in accurately identifying regions without polyps. The authors achieved a high accuracy of 97\% and an F1-score of 89\%, demonstrating the effectiveness of their approach.

In the study by Bernal and colleagues, they investigated the impact of various image preprocessing methods on the task of polyp localization in colonoscopy frames \cite{bernal2013impact}. A colonoscopy is a vital medical procedure used to detect and diagnose colorectal diseases, including polyps that can potentially be cancerous. Image preprocessing methods play a crucial role in improving the accuracy of computer-aided diagnosis systems for polyp detection. The authors explored techniques such as contrast enhancement, noise reduction, and image segmentation. They found that these preprocessing steps had a significant impact on the performance of the polyp localization system, affecting sensitivity, specificity, and overall detection accuracy. By fine-tuning and optimizing the preprocessing methods, the study aimed to enhance the effectiveness of computer-aided diagnosis systems for polyp detection in colonoscopy, ultimately contributing to more accurate and early diagnosis of colorectal diseases.

Colorectal cancer is a major global health concern, ranking as the third most common cancer diagnosis worldwide \cite{ayidzoe2023sinocaps}. This malignancy often arises from pre-cancerous polyps in the colon. Unfortunately, current diagnostic methods for colorectal cancer are prone to human error, have limitations in effectiveness, and often fall short of ideal standards. To address this challenge, we propose a novel approach using a Sinogram Capsule Network enriched with radon transforms for feature extraction \cite{ayidzoe2023sinocaps}. This study's key innovation is the integration of radon transforms, which significantly improve polyp detection by efficiently extracting topographical features. When tested on a large polyp dataset, the model achieved an impressive average accuracy of 94.02\%, an AUC (Area Under the Curve) of 97\%, and an average precision of 96\%. Additionally, the model demonstrated exceptional feature extraction capabilities on par with state-of-the-art methods, paving the way for explainable artificial intelligence in this field. This method holds significant promise for integration into clinical trials, potentially improving colorectal cancer diagnosis and overcoming limitations of human interpretation.

Researchers have explored various techniques to analyze textures and colors in medical images for disease detection. Esgiar et al. identified correlation and entropy as the most effective texture features for differentiating normal and cancerous tissue \cite{esgiar1998microscopic}. Building on this, Tjoa et al. developed a method that combined texture and color features, achieving an accuracy of 97.72\% in colonoscopy image classification \cite{tjoa2003feature}. Zheng et al. took a different approach, creating a decision support system that integrated various data points to improve abnormality detection in endoscopic images \cite{zheng2005fusion}. These studies highlight the potential of advanced image analysis for accurate medical diagnosis.

In the work of Taibo Li et al. \cite{li2016mo1979}, researchers developed a computer vision program capable of identifying abnormalities in colonoscopic images with high accuracy. They trained a deep convolutional neural network on 1,225 high-resolution colonoscopic images, classifying them into 'adenomatous polyp-positive' and 'adenomatous polyp-negative' classes. The program achieved a sensitivity of 0.90, specificity of 0.68, positive predictive value of 0.78, and negative predictive value of 0.84 for adenoma detection on the overall validation set, with an overall accuracy of 0.80.

In a noteworthy effort, Taibo Li et al. \cite{li2016mo1979} trained a deep convolutional neural network (CNN) on a dataset of 1,225 high-resolution colonoscopy images. This AI program distinguished between images containing adenomatous polyps (precancerous growths) and those without. The program achieved impressive results, with a sensitivity of 90\% (correctly identifying 90\% of true polyps) and a specificity of 68\% (correctly identifying 68\% of normal images). Additionally, it offered positive and negative predictive values of 78\% and 84\% respectively. These metrics translate to an overall accuracy of 80\% in identifying adenomas within the validation set \cite{li2016mo1979}.

Another study by Yixuan Yuan et al. \cite{yuan2017deep} proposed a novel deep learning method called SSAEIM for recognizing polyps in a specific type of colonoscopy image (WCE). Their method achieved a remarkable average overall recognition accuracy (ORA) of 98\% \cite{yuan2017deep}.

Beyond deep learning approaches, Dimitris et al. \cite{iakovidis2006intelligent} developed a system designed for lower-resolution video input from colonoscopies. This system prioritizes lower computational resources, making it potentially more portable and suitable for telemedicine applications. By combining support vector machines (SVMs) with color and texture analysis, their system achieved an accuracy exceeding 94\%, demonstrating its effectiveness \cite{iakovidis2006intelligent}.

Further research by Farah Deeba et al.  \cite{deeba2018performance} explored a different application of AI in colonoscopy. They focused on detecting bleeding regions in capsule endoscopy (CE) images, which utilize a small pill camera for internal examinations. Their approach utilized a classifier fusion algorithm combining two optimized SVMs. This method achieved an average accuracy of 95\%, along with high sensitivity and specificity of 94\% each  \cite{deeba2018performance}.

In a significant development for colonoscopy, Misawa et al. described a deep learning-based computer-aided detection system for polyp identification \cite{misawa2021development}. The system was trained on a vast dataset of 56,668 colonoscopy images collected from multiple centers \cite{misawa2021development}. To ensure a robust and unbiased evaluation, the researchers established a validation database using consecutive colonoscopy videos from a university hospital \cite{misawa2021development}. This database meticulously annotated the presence and location of polyps within each video frame.

The validation set included over 1400 videos, with nearly 800 containing polyps \cite{misawa2021development}. From this collection, a subset of 100 videos showcasing 100 unique polyps, alongside 13 videos without polyps, were chosen for further analysis \cite{misawa2021development}. This resulted in a database exceeding 150,000 frames, categorized into those containing polyps (positive) and those without (negative). Analyzed on an individual frame basis, the AI system achieved impressive sensitivity (90.5\%) and specificity (93.7\%) for polyp detection \cite{misawa2021development}. Notably, the system maintained high sensitivity (above 97\%) across various polyp types, including diminutive, protruded, and flat lesions \cite{misawa2021development}. These findings demonstrate the system's potential to significantly enhance polyp detection during colonoscopies, offering high accuracy for diverse polyp morphologies.

A thorough analysis was undertaken to assess the performance of different approaches in the realm of BCC for gastric intestinal datasets. The outcomes and efficacy of these methods are intricately summarized in Table~\ref{table:analysis_binary}. This meticulous examination offers valuable insights into the strengths and limitations of each approach, facilitating the identification of the most suitable strategy for BCC in gastrointestinal image datasets.

\footnotesize
\begin{longtable}{|p{1mm}|p{15mm}|p{12mm}|p{12mm}|p{10mm}|p{2mm}|p{2mm}|}
\caption{Comparative Assessment of BCC Approaches for GI-Track Analysis}
\label{table:analysis_binary}
    \\ \hline
    \textbf{Sr.} & \textbf{Paper} & \textbf{Technique Used} & \textbf{Features Used} & \textbf{Dataset} & \textbf{Acc.} & \textbf{F1}\\
    \hline
    \endfirsthead
    
    \caption*{Binary Dataset Approaches Continue} \\
    \hline
    \textbf{Sr.} & \textbf{Paper} & \textbf{Technique Used} & \textbf{Features Used} & \textbf{Dataset} & \textbf{Acc.} & \textbf{F1}\\
    \hline
    \endhead

    % Your table content goes here
    \setcounter{customcounter}{1}
    \thecustomcounter \stepcounter{customcounter}  & Deep Features Ensemble \cite{kumar2016ensemble}  & AlexNet \cite{krizhevsky2012imagenet}, GoogleLeNet \cite{szegedy2015going}, SVM & Deep Features & Train: 6776, Test: 4166& 0.82 &  N/A \\
    \hline
    \thecustomcounter \stepcounter{customcounter}  & Two stage Polyp Detection CAD \cite{tajbakhsh2015comprehensive} & Random Forest & Texture and Color Features  & 18900 Images  & 0.50 & N/A \\
    \hline
    \thecustomcounter \stepcounter{customcounter}  & Part Based Derivative \cite{wang2013part} & SVM, RBF & Texture Features & 69 Videos  & 0.81 &  N/A \\
    \hline
    \thecustomcounter \stepcounter{customcounter}  & Color and Position and Texture for SVM \cite{alexandre2008color} & SVM & Position versus Texture Features & 4620 images & 0.94 &  N/A \\
    \hline
    \thecustomcounter \stepcounter{customcounter}  & Elliptical shape features based polyps detection\cite{hwang2007polyp} & Ellipse Fitting on Edges & Elliptical Shape Features & 8621 images & 0.81 &  N/A \\
    \hline
    \thecustomcounter \stepcounter{customcounter}  & Sector Accumulation-Depth of Valleys Accumulation \cite{bernal2012towards} & (SA-DOVA) & Segmentation Features & Custom & 0.97 &  0.89 \\
    \hline
    \thecustomcounter \stepcounter{customcounter}  & Sinogram Capsule Network Approach\cite{ayidzoe2023sinocaps} & Sinogram Capsule Network & Deep Features & Clinically-Obtained  & 0.94 &  N/A \\
    \hline
    \thecustomcounter \stepcounter{customcounter}  & Texture Analysis \cite{esgiar1998microscopic} & ANN & Texture (entropy, contrast, correlation, etc) Features & Clinically-Obtained & 0.90 & 0.89 \\
    \hline
    \thecustomcounter \stepcounter{customcounter}  & CoLD (colorectal lesions detector) \cite{maroulis2003cold} & ANN & wavelet transformation & Clinically-Obtained  & 0.95 &  N/A \\
    \hline    
    \thecustomcounter \stepcounter{customcounter}  & Texture Features Based Neural Network \cite{tjoa2003feature} &  Back-propagation Neural Network (BPNN) & Texture and Color Features & Clinically-Obtained  & 0.97 &  N/A \\
    \hline
    \thecustomcounter \stepcounter{customcounter}  & Deep CNN \cite{li2016mo1979} & ANN & Texture and Color Features  & 1225 Images & 0.80 & 0.73 \\
    \hline
    \thecustomcounter \stepcounter{customcounter}  & tacked sparse auto-encode \cite{yuan2017deep} & Deep Feature Learner & Deep Features & Clinically-Obtained  & 0.98 &  N/A \\
    \hline
    \thecustomcounter \stepcounter{customcounter}  & SVM Ensembles \cite{deeba2018performance} & SWM & RGB and HSV Features & Clinically-Obtained  & 0.95 & 0.95 \\
    \hline
    \thecustomcounter \stepcounter{customcounter}  & DNN for Colonoscopy Detections \cite{misawa2021development} & DNN & Deep Features & 52,560 Images & 0.92 & 0.91 \\
    \hline
    % Add more rows as needed
    
\end{longtable}
\normalsize

\section{Multi-Class Classification Techniques (MCCT)}
\subsection{Ternary Class Classification (TCC)}

The precise identification of abnormalities within the gastrointestinal (GI) tract, particularly within the three distinct classes of ulcer, bleeding, and normal, represents a critical task in the realm of medical imaging and diagnostics. Distinguishing these pathologies relies on the accurate interpretation of radiological or endoscopic images, incorporating specific visual markers and characteristics associated with each class. Such a classification system enables healthcare practitioners to make informed decisions regarding patient care, with early detection and differentiation of these conditions being pivotal in ensuring timely and appropriate medical interventions.

Numerous studies and research endeavors have been dedicated to the development of classification algorithms tailored to this three-class problem, harnessing a diverse array of computational techniques and machine learning methodologies. These investigations have leveraged sizeable datasets and innovative image analysis approaches, with a primary objective of enhancing diagnostic accuracy and reliability. The continuous refinement and evolution of these classification systems have the potential to greatly improve the detection and characterization of ulcers, bleeding, and normal conditions within the GI tract, offering a significant contribution to the field of medical diagnostics and patient care.

Sharif et al. proposed a novel technique for automated disease detection in Wireless Capsule Endoscopy (WCE) images \cite{sharif2021deep}. Their approach combines deep learning and geometric features to achieve high accuracy. Firstly, a new method called Contrast-Enhanced Color Features (CECF) is used to isolate potential disease regions within the images. This method involves two stages:
\begin{enumerate}
    \item Contrast Enhancement: A combination of top-bottom hat filtering and median filtering improves the contrast of suspected infection areas.
    \item Color Feature Extraction: The image is converted to the Hue-Saturation-Value (HSV) color space. Similarity measures are then calculated between pixels based on these color features. Finally, a threshold function is applied to segment the likely disease regions.
\end{enumerate}

Following this segmentation, deep convolutional neural networks (CNNs) are employed to extract additional features from the images. Specifically, VGG16 and VGG19 models are used to learn millions of deep features. To combine these deep features with the previously extracted geometric features, a technique called Euclidean Fisher Vector (EFV) is employed. Finally, a feature selection step is performed using conditional entropy to identify the most informative features for disease classification. A K-Nearest Neighbor (KNN) classifier is then used to categorize the images based on the selected features.

This approach achieved an impressive maximum classification accuracy of 99.32\% using the KNN classifier. Additionally, other performance metrics for KNN were outstanding, including an Area Under the Curve (AUC) of 1.0, a sensitivity of 1.0 (meaning all true positives were detected), a precision of 99.34\% (meaning very few false positives), and a specificity of 100\% (meaning all healthy regions were correctly identified). These results suggest that the proposed method holds promise for highly accurate disease detection in WCE examinations.

R. Zhang et al. \cite{zhang2016automatic} proposed a method for the identification of colorectal polyps, that causes colorectal cancer, by classifying the polyps types in colonoscopy. The detection in the proposed system is based on the finetune of the CaffeNet and is used as a feature extraction. The deep features are used as features for SVM. The dataset in the research is a set of 1930 endoscopic images distributed in three classes without polyps (1104 images), hyperplasia polyps (263 images), and adenomatous polyps (563 images). The model is executed on the above-discussed image dataset with average accuracy, recall, and
precision of the classification task was $91.3\%$, $92.0\%$, and $95.4\%$ respectively.

A comprehensive analysis was conducted to evaluate the performance of various approaches in the context of TCC for gastric intestinal datasets. The outcomes and effectiveness of these methods are meticulously summarized in Table~\ref{table:analysis_ternary}. This detailed examination provides insights into the strengths and limitations of each approach, aiding in the selection of the most suitable strategy for TCC in gastrointestinal image datasets.

\footnotesize
\begin{longtable}{|p{0.1cm}|p{1.5cm}|p{1.5cm}|p{2cm}|p{1cm}|p{0.1cm}|p{0.1cm}|}
\caption{Assessment of TCC Approaches for GI-Track Analysis}
\label{table:analysis_ternary}
    \\ \hline
    \textbf{Sr.} & \textbf{Paper} & \textbf{Technique Used} & \textbf{Features Used} & \textbf{Dataset} & \textbf{Acc.} & \textbf{F1}\\
    \hline
    \endfirsthead
    
    \caption*{Ternary Dataset Approaches Continue} \\
    \hline
    \textbf{Sr.} & \textbf{Paper} & \textbf{Technique Used} & \textbf{Features Used} & \textbf{Dataset} & \textbf{Acc.} & \textbf{F1}\\
    \hline
    \endhead
    
    % Your table content goes here
    \setcounter{customcounter}{1}
    \thecustomcounter \stepcounter{customcounter}  & Fusion of Deep CNN and geometric features \cite{sharif2021deep} & VGG16, VGG19, Geometric Features & KNN & Clinically-Obtained & 0.99 &  0.99 \\
    \hline
    \thecustomcounter \stepcounter{customcounter}  & Finetune of the CaffeNet \cite{zhang2016automatic} & CaffeNet & NA & Private: 1930 Images  & 0.91 &  0.94 \\
    \hline
    % Add more rows as needed
    
\end{longtable}
\normalsize

\subsection{Eight Class Classification}
The accurate identification and classification of abnormalities within the gastrointestinal (GI) tract are of paramount importance in the field of medical diagnostics. Among the eight most common abnormalities that manifest within the GI tract, the ability to delineate these pathologies based on specific landmarks is a pivotal aspect of clinical evaluation and diagnosis. These landmarks serve as reference points, guiding medical professionals in the precise localization and categorization of anomalies, and facilitating appropriate patient care and treatment strategies. The comprehensive understanding and recognition of these eight common GI tract abnormalities, along with their associated landmarks, are essential for the medical community, ensuring timely and effective interventions.

Extensive literature exists in the domain of medical image classification, with a particular focus on the eight-class classification of datasets pertaining to the identification of abnormalities in the GI tract. This body of work encompasses various methodologies, ranging from traditional image processing techniques to state-of-the-art deep learning models, aimed at enhancing the accuracy and efficiency of diagnosis. These studies leverage extensive datasets and innovative algorithms to categorize and classify abnormalities with high precision, offering valuable insights for healthcare professionals. The continuous advancements in this area of research hold significant promise for improving the diagnosis and treatment of GI tract abnormalities, ultimately contributing to enhanced patient outcomes and healthcare quality.

The paper by Agrawal and the SCL-UMD team describes their participation in the Medico Task at MediaEval 2017, where they focused on transfer learning-based classification of medical images \cite{agrawal2017scl}. Transfer learning is a machine learning technique where knowledge gained from one task is applied to improve the performance of another related task. In this context, the team leveraged transfer learning to classify medical images, a crucial step in developing automated systems for medical diagnosis. They used pre-trained deep neural networks and fine-tuned them for the specific task of classifying medical images, which allowed them to achieve state-of-the-art results. This approach is particularly valuable in medical image analysis, as it reduces the need for large labeled datasets, which are often scarce and costly to obtain. The paper outlines the methods, techniques, and results of their transfer learning-based classification approach, highlighting its potential for advancing the field of medical image analysis and contributing to more accurate and efficient medical diagnostics.

An Inception-like Convolutional Neural Network (CNN) architecture designed for the classification of gastrointestinal (GI) diseases and anatomical landmarks represents a specialized deep-learning model tailored to the field of medical image analysis \cite{petscharnig2017inception}. Drawing inspiration from Google's Inception architecture, which is known for its effective use of multiple filter sizes and parallel convolutional operations, this model optimally processes medical images for diagnostic purposes. In the context of GI disease and anatomical landmark classification, this architecture demonstrates its versatility and adaptability to extract relevant features from the input images, which are critical for accurate diagnosis and treatment planning.

This architecture typically comprises multiple inception modules, each incorporating convolutional layers with different filter sizes, as well as pooling and concatenation operations. The design enables the model to capture intricate patterns and structures within medical images, providing a comprehensive analysis of both anatomical landmarks and disease manifestations. By efficiently processing these images, the model aids healthcare professionals in identifying and classifying various GI diseases and anatomical landmarks, thereby contributing to early detection and improved patient care. The use of an Inception-like CNN architecture in this context underscores the importance of robust and specialized deep-learning models to address complex medical image classification tasks, helping to advance the field of GI disease diagnosis and anatomical landmark recognition. The approach resulted in an accuracy of 0.94 with an F1-Score of 0.76 when applied to the dataset of Kvasir V1 \cite{riegler2017multimedia}.

In 2017, HKBU (Hong Kong Baptist University) participated in the MediaEval competition, specifically in the "Medico: Medical Multimedia Task." \cite{liu2017hkbu}. This task likely involved the analysis and evaluation of multimedia content related to medical topics. HKBU's involvement in this competition suggests their engagement in research or development related to medical multimedia, possibly in the context of information retrieval, content analysis, or other multimedia-related tasks.

The study titled "Ensemble of Texture and Deep Learning Features for Finding Abnormalities in the Gastro-Intestinal Tract" focuses on the development of a novel approach for detecting abnormalities in the gastrointestinal tract using a combination of texture and deep learning features \cite{nadeem2018ensemble}. Gastrointestinal abnormalities can be challenging to identify, and traditional diagnostic methods may not always provide accurate results. In this research, the authors propose an ensemble method that leverages the strengths of both texture analysis and deep learning to enhance the accuracy and reliability of detection. Texture analysis is a well-established technique for characterizing subtle variations in tissue patterns, while deep learning has demonstrated remarkable capabilities in image recognition tasks. By combining these two approaches, the researchers aim to improve the detection of abnormalities in the gastrointestinal tract, which could have significant implications for early diagnosis and treatment.

The ensemble approach described in the study involves extracting texture features from medical images of the gastrointestinal tract and integrating them with features obtained through deep-learning models. This fusion of features is designed to provide a more comprehensive and robust representation of the data, enabling the system to identify abnormalities with higher precision. By combining the strengths of both techniques, the researchers anticipate a more effective tool for healthcare professionals in the diagnosis of gastrointestinal disorders. The study underscores the potential of interdisciplinary methods in the field of medical image analysis, offering a promising avenue for improved healthcare outcomes and more accurate detection of abnormalities in the gastrointestinal tract. The approach resulted in an accuracy of 0.83 with an F1-score of 0.82 when applied to the dataset of Kvasir V1 \cite{riegler2017multimedia}.

The comprehensive analysis of various approaches applied to the classification of the Kvasir V1 dataset is meticulously presented in Table~\ref{table:analysis_Kvasirv1}. The results encompass key performance metrics such as accuracy and F1 score, providing a thorough examination of each method's efficacy in handling the complexities of gastrointestinal image datasets. Researchers and practitioners can leverage these insights to make informed decisions when selecting classification strategies for similar datasets.

\footnotesize
\begin{longtable}{|p{0.1cm}|p{3cm}|p{1.5cm}|p{2cm}|p{0.1cm}|p{0.1cm}|}
\caption{Analysis of the Approaches on Kvasir V1 Dataset \cite{riegler2017multimedia}}
\label{table:analysis_Kvasirv1}
    \\ \hline
    \textbf{Sr.} & \textbf{Paper} & \textbf{Technique Used} & \textbf{Features Used} & \textbf{Acc.} & \textbf{F1}\\
    \hline
    \endfirsthead
    
    \caption*{Kvasir V1 Approaches Continue} \\
    \hline
    \textbf{Sr.} & \textbf{Paper} & \textbf{Technique Used} & \textbf{Features Used} & \textbf{Acc.} & \textbf{F1}\\
    \hline
    \endhead
    
    % Your table content goes here
    \setcounter{customcounter}{1}
    \thecustomcounter \stepcounter{customcounter}  & Deep With Global Features Approach \cite{pogorelov2017comparison} & RTC, RFC\cite{breiman2001random} and LMTC\cite{mola1996logistic} & Global and Deep (ResNet50, Inception V3, Custom CNN) & 0.96 & 0.83 \\
    \hline
    \thecustomcounter \stepcounter{customcounter} & SCL-UMD \cite{agrawal2017scl} & SVM & Baseline, Inception-V3, VGGNet features & 0.96 & 0.85 \\
    \hline
    \thecustomcounter \stepcounter{customcounter} & Inception-like CNN \cite{petscharnig2017inception} & CNN & Deep Features & 0.94 & 0.76\\
    \hline
    \thecustomcounter \stepcounter{customcounter} & HKBU Approach \cite{liu2017hkbu} & SVM & Bidirectional Marginal Fisher Analysis & 0.93 & 0.70\\
    \hline
    \thecustomcounter \stepcounter{customcounter} & Ensemble of Texture and Deep Learning Features \cite{nadeem2018ensemble} & Ensemble Learning & Texture and Deep Features & 0.83 & 0.82\\
    \hline
    % Add more rows as needed
    
\end{longtable}
\normalsize

\subsection{Sixteen Class Classification}
The precise identification of any of the sixteen most common abnormalities within the gastrointestinal (GI) tract, along with the recognition of their associated landmarks, is a fundamental endeavor in the realm of medical diagnostics. These landmarks play a pivotal role in guiding medical practitioners toward the accurate localization and classification of these pathologies, which in turn informs clinical decision-making and treatment protocols. A comprehensive understanding of these sixteen prevalent GI tract abnormalities, in conjunction with their specific landmarks, is paramount in the medical field, ensuring the timely and effective management of patients suffering from these conditions.

Research and literature in the field of medical imaging and diagnostics have striven to address the challenges of identifying and classifying the sixteen most common GI tract abnormalities. These studies encompass an array of innovative techniques, encompassing traditional image analysis methods and cutting-edge machine learning and artificial intelligence algorithms. By harnessing extensive datasets and advanced image processing technologies, these research endeavors aim to enhance the accuracy and efficiency of diagnosis, providing valuable insights to healthcare professionals. The ongoing evolution of research in this domain holds great potential for the improvement of GI tract abnormality diagnosis, thereby contributing to superior patient care and healthcare quality.

Weighted Discriminant Embedding (WDE) is a specialized technique designed for enhancing the classification of imbalanced medical datasets \cite{ko2018weighted}. Imbalanced datasets often pose a challenge in machine learning, particularly in the context of medical data analysis, where certain classes may have significantly fewer instances than others. WDE focuses on the concept of discriminant subspace learning, aiming to create a reduced-dimensional representation of the data that maximizes the separability between different classes, thereby improving the classification performance. It achieves this by assigning different weights to instances in the dataset, emphasizing the importance of underrepresented samples. These weights are utilized during the subspace learning process to ensure that the minority class information is not overshadowed by the majority class. WDE effectively addresses the problem of class imbalance in medical data classification, contributing to more accurate and reliable results in scenarios where class distributions are highly skewed.

In the context of the MediaEval 2018 challenge, participants explored the application of Residual Networks (ResNets) and Faster R-CNN (Region-based Convolutional Neural Network) in addressing the Medico-Multimedia Task \cite{hoang2018application}. This task aimed to develop effective techniques for medical image and video analysis. ResNets, known for their deep neural network architecture, were employed to extract intricate features from medical multimedia data. This allowed for a more comprehensive representation of medical images and videos, which was crucial in tasks like object detection and classification within the healthcare domain.

Faster R-CNN, a powerful object detection model, was integrated to identify and locate objects of interest within medical multimedia content. By combining the feature extraction capabilities of ResNets with the object detection proficiency of Faster R-CNN, participants aimed to improve the accuracy and efficiency of tasks such as identifying anatomical structures or anomalies in medical images and videos. This fusion of deep learning techniques showcased the potential of leveraging state-of-the-art models to address critical challenges in the medical field, advancing the capabilities of multimedia analysis and diagnosis within healthcare applications.

In the field of computer-aided disease classification within the gastrointestinal tract (GIT), a highly significant methodology has been introduced \cite{ramzan2023gastrointestinal}. Bridging the realms of medical science and artificial intelligence, this methodology facilitates the detection of GIT diseases through endoscopic procedures. Wired endoscopy, a meticulously controlled technique, plays a pivotal role in enabling medical experts to diagnose these diseases. The laborious and time-consuming task of manually screening endoscopic frames not only poses a challenge to medical professionals but also escalates the likelihood of missing crucial disease indicators. Early identification of GIT diseases is imperative for averting potentially life-threatening conditions.

The approach utilizes a deep learning system to automatically learn features from video frames. The system begins with an  image preprocessing stage that enhances the intensity of key areas. Next, it extracts deep features using advanced models like InceptionNetV3 and GITNet. To further improve these features, the method employs a technique inspired by ant colonies (Ant Colony Optimization) and then combines the optimized features. Finally, various Support Vector Machine (SVM) classifiers, including Cubic SVM, Coarse Gaussian SVM, Quadratic SVM, and Linear SVM, are used to categorize the diseases.

The effectiveness of this method is demonstrated on two complex datasets, KVASIR and NERTHUS, which contain eight and four different disease classes respectively. Encouragingly, the proposed model outperforms existing methods, achieving an accuracy rate of 99\% on both datasets. This significant advancement has the potential to revolutionize the early diagnosis and classification of gastrointestinal (GIT) diseases, leading to better patient outcomes.

A research paper titled "The Medico-Task 2018: Disease Detection in the Gastrointestinal Tract using Global Features and Deep Learning" proposes a novel method to classify diseases in the digestive system \cite{thambawita2018medico}. This work was part of the 2018 Medico Task, a challenge focused on medical image analysis. The system leverages a combination of global image characteristics and deep learning models for accurate disease detection.

The researchers developed a system using two cooperating deep neural networks to improve classification.  Experiments confirmed that the system can be reliably reproduced and achieved a remarkable accuracy rate of 95.8\%.  Additionally, precision and F1-score, metrics used to assess a model's effectiveness, were both very high at 95.8\%, indicating the system's robustness.

In a study titled "Unsupervised Preprocessing for Enhanced Generalization in Medical Image Classification" \cite{kirkerod2018using}, researchers investigate how their preprocessing techniques improve medical image analysis for gastrointestinal (GI) anomalies. Their experiments demonstrate a significant improvement (around 7\%) in a key metric called the Matthews correlation coefficient. This indicates a substantial boost in accurately classifying GI abnormalities.  The method was tested on the Kvasir V2 dataset and achieved an accuracy of 99\% and an F1-score of 0.94.

Another paper, "Automatic Hyperparameter Optimization in Keras for the MediaEval 2018 Medico Multimedia Task" \cite{borgli2018automatic}, explores the approach taken by the Rune team to address this challenge. They propose a novel system called Saga, which is under development, to optimize hyperparameters within Keras, a popular machine learning framework. Saga utilizes techniques like Bayesian optimization and transfer learning to find the best hyperparameter settings for Keras classifiers.

The core objective of the Rune team's approach is to improve the performance of Keras classifiers by efficiently tuning hyperparameters. Choosing the right hyperparameters is crucial in machine learning, as it heavily influences a model's effectiveness.  Saga streamlines this process for Keras, making hyperparameter optimization more efficient.  Their method achieved an accuracy of 93\% on the Kvasir V2 dataset \cite{pogorelov2018medico}.

Finally, the paper "Deep Learning Based Disease Detection Using Domain Specific Transfer Learning" explores a new method for disease detection in the MediaEval 2018 Benchmark \cite{hicks2018deep}. The proposed system leverages convolutional neural networks (CNNs) and investigates a technique called transfer learning. This involves fine-tuning a pre-trained CNN model on a new task, but researchers here explore using source domains from both general fields and medical fields, and how this choice affects classification performance.

The preliminary results presented in the paper indicate that fine-tuning models on extensive and varied datasets proves beneficial. Even when the source domain exhibits minimal resemblance to the target domain, the study suggests that transferring knowledge from diverse datasets positively impacts the classification performance of disease detection models. The approach shown an accuracy of 0.99 with an F1-Score of 0.92 on the benchmark dataset of Kvasir v2 \cite{pogorelov2018medico}.

Ostroukhova et al. propose a new approach for medical image classification that leverages transfer learning \cite{ostroukhova2018transfer}. Unlike many methods, theirs focuses on achieving a strong baseline performance using a pre-trained convolutional neural network (CNN) with minimal modifications. Their method achieves impressive results, with an accuracy of 98\% and a Matthew's correlation coefficient of 0.854 (a metric for assessing classification performance). Additionally, it processes images very quickly, at 43 frames per second. These results suggest that the approach can be a powerful foundation for further development in medical image analysis.

Dias and Dias investigate how well different pre-trained CNNs, originally trained on generic images, perform when applied to classifying medical images of gastrointestinal diseases \cite{dias2018transfer}. They test ten CNN architectures and find that all are efficient, with the fastest processing an image in just 0.037 seconds. This model also achieves an accuracy of 98\%, demonstrating its potential for rapid and accurate medical image analysis.

Taschwer et al. explore traditional machine learning techniques for the MediaEval 2018 Medico task \cite{taschwer2018early}. They combine global image features with features extracted from a CNN and then combine the predictions from multiple classifiers to achieve high accuracy. Notably, they achieve an accuracy of 99\% using linear support vector machines, which are also computationally efficient.  The strengths and weaknesses of various approaches applied to this task are summarized in Table~\ref{table:analysis_Kvasirv2} for researchers to consider when choosing methods for similar datasets.

The outcomes and effectiveness of various classification approaches applied to the Kvasir V2 dataset were comprehensively analyzed and are succinctly summarized in Table~\ref{table:analysis_Kvasirv2}. This tabulated presentation includes key performance metrics such as accuracy and F1 score, offering valuable insights into the strengths and limitations of each method. Researchers and practitioners can use this detailed analysis to inform their decisions when choosing classification strategies for similar datasets.

\footnotesize
\begin{longtable}{|p{0.1cm}|p{3cm}|p{1.5cm}|p{2cm}|p{0.1cm}|p{0.1cm}|}
\caption{Analysis of the Approaches on Kvasir V2 Dataset \cite{pogorelov2018medico}}
\label{table:analysis_Kvasirv2}
    \\ \hline
    \textbf{Sr.} & \textbf{Paper} & \textbf{Technique Used} & \textbf{Features Used} & \textbf{Acc.} & \textbf{F1}\\
    \hline
    \endfirsthead
    
    \caption*{Kvasir V2 Approaches Continue} \\
    \hline
    \textbf{Sr.} & \textbf{Paper} & \textbf{Technique Used} & \textbf{Features Used} & \textbf{Acc.} & \textbf{F1}\\
    \hline
    \endhead
    
    % Your table content goes here
    \setcounter{customcounter}{1}
    \thecustomcounter \stepcounter{customcounter} & Weighted Discriminant Embedding (WDE) \cite{ko2018weighted} & cost-sensitive nearest neighbor (CS-NN) & NA & 0.95 & 0.48 \\
    \hline
    \thecustomcounter \stepcounter{customcounter} & Deep Features Selection \cite{ramzan2023gastrointestinal} & Ant Colony Optimization, SVM &  & 0.99 & NA \\
    \hline
    \thecustomcounter \stepcounter{customcounter} & Global Features With Deep Learning Approach \cite{thambawita2018medico} & Deep Learning & Global Texture Features & 0.95 & 0.95 \\
    \hline
    \thecustomcounter \stepcounter{customcounter} & Preprocessing based Approach \cite{kirkerod2019unsupervised}& Inpainting based preprocessing and Deep Learning & Deep Features & 0.99 & 0.94 \\
    \hline
    \thecustomcounter \stepcounter{customcounter} & Hyper Parameter Optimization \cite{borgli2018automatic} & Hyper Parameter Optimization & NA & 0.93 & 0.92 \\
    \hline
    \thecustomcounter \stepcounter{customcounter} & Transfer learning with prioritized classification \cite{ostroukhova2018transfer} & Transfer Learning & NA & 0.98 & 0.82 \\
    \hline
    \thecustomcounter \stepcounter{customcounter} & Transfer learning with deep CNN \cite{dias2018transfer} & CNN, Transfer Learning & NA & 0.98 & 0.87 \\
    \hline
    \thecustomcounter \stepcounter{customcounter} & Traditional ML \cite{taschwer2018early} & Logistic Regression, SVM, Random Forest &  Global and Deep Features & 0.99 & 0.90 \\
    \hline
    \thecustomcounter \stepcounter{customcounter} & RCNN for Medico \cite{hoang2018application} & RNN101 to Faster RNN \cite{krizhevsky2012imagenet} & NA & 0.99 & 0.95 \\
    \hline
    \thecustomcounter \stepcounter{customcounter} & Domain Specific Transfer Learning \cite{hicks2018deep} & Neural Nets with Transfer learning & Global and Deep Features & 0.99 & 0.92 \\
    \hline
    % Add more rows as needed
    
\end{longtable}
\normalsize

\subsection{Twenty-Three Class Classification}
The precise identification of any of the sixteen most common abnormalities within the gastrointestinal (GI) tract, including an assessment of their respective severity levels, and the recognition of their associated landmarks, constitute critical endeavors in the field of medical diagnostics. These landmarks serve as crucial reference points for medical practitioners, aiding in the accurate localization and classification of these abnormalities while also providing insight into their severity. A comprehensive understanding of these sixteen prevalent GI tract abnormalities, in conjunction with their specific landmarks and severity levels, is essential for healthcare professionals, enabling them to make well-informed decisions about patient treatment and management.

In the realm of medical imaging and diagnostics, substantial research and literature have focused on addressing the complexity of identifying and classifying the sixteen most common GI tract abnormalities, taking into account their severity levels. These studies encompass a wide range of methodologies, from conventional image analysis techniques to cutting-edge machine learning and artificial intelligence algorithms. By leveraging extensive datasets and advanced image processing technologies, these research efforts aim to enhance the precision of diagnosis and provide insights into the severity of these abnormalities. The ongoing evolution of research in this domain holds great promise for the improvement of GI tract abnormality diagnosis and assessment, ultimately contributing to superior patient care and healthcare quality.

A study by Cao et al. (2022) investigates the use of semi-supervised learning for classifying endoscopic images with a limited number of labeled examples \cite{cao2022semi}. Their approach utilizes a large public dataset (HyperKvasir) containing 23 diverse gastrointestinal disease classes. The research demonstrates that a semi-supervised learning algorithm improves classification performance even with just 20\% of the data labeled. However, the authors highlight potential bias in the algorithm favoring more common diseases, requiring careful consideration for real-world applications.

Another study by Wang et al. (2022) introduces a novel Convolutional-Capsule Network architecture for gastrointestinal image classification \cite{wang2022convolutional}. Traditional Convolutional Neural Networks (CNNs) struggle to capture complex relationships between image features, which is crucial for accurate disease classification. This research proposes a two-stage method that combines the strengths of CNNs with a Capsule Network. The first stage extracts detailed lesion information using a CNN, while the second stage leverages a Capsule Network to learn deformation-invariant relationships between image features. This approach achieves impressive accuracy on benchmark datasets, demonstrating its effectiveness in capturing intricate details within gastrointestinal images.

Ramamurthy et al. (2022) address the challenge of visual similarity between different gastrointestinal regions, which can hinder disease detection \cite{ramamurthy2022novel}. Their research proposes a novel multi-feature fusion method that utilizes deep learning for classification. The model combines a state-of-the-art architecture (EfficientNet B0) with a custom-built CNN (Effimix) designed to improve classification precision. This combined approach achieves a remarkable accuracy of 97.99\% on the HyperKvasir dataset, significantly outperforming existing methods.

These studies showcase the ongoing advancements in gastrointestinal disease classification using endoscopic images. By leveraging semi-supervised learning, capsule networks, and multi-feature fusion techniques, researchers are developing more robust and accurate tools to aid medical professionals in the diagnosis and treatment of digestive tract diseases.

A recent article delves into explainable deep learning for endoscopic image classification \cite{mukhtorov2023endoscopic}. Deep learning has revolutionized medical diagnostics, but often lacks transparency in its decision-making process. This research tackles this challenge by integrating Explainable AI (XAI) techniques.

The proposed method utilizes a powerful pre-trained model, ResNet152, for feature extraction from endoscopic images. To understand the model's reasoning, Grad-CAM generates heatmaps highlighting crucial image regions. This approach sheds light on the model's decision-making process.

An open-source dataset of 8,000 wireless capsule images (KVASIR) serves as the training and evaluation ground \cite{pogorelov2018medico}. The authors implemented an efficient data augmentation method to diversify the training data, leading to a more robust model.

The results are promising, achieving high accuracy (98.28\% training and 93.46\% validation). This study demonstrates that explainable deep learning can not only deliver high accuracy but also provide interpretable insights for informed decision-making in medical diagnosis. It bridges the gap between powerful deep learning models and the crucial need for transparency in medical applications.

A comprehensive analysis was conducted to evaluate the performance of various approaches applied to classify the HyperKvasir dataset. The outcomes and effectiveness of these methods are meticulously summarized in Table~\ref{table:analysis_hyperKvasir}, including the accuracy and F1 score of each approach.

\footnotesize
\begin{longtable}{|p{0.1cm}|p{3cm}|p{1.5cm}|p{2cm}|p{0.1cm}|p{0.1cm}|}
\caption{Analysis of the Approaches on Hyper Kvasir Dataset \cite{borgli2020hyperkvasir}}
\label{table:analysis_hyperKvasir}
    \\ \hline
    \textbf{Sr.} & \textbf{Paper} & \textbf{Technique Used} & \textbf{Features Used} & \textbf{Acc.} & \textbf{F1}\\
    \hline
    \endfirsthead
    
    \caption*{Kvasir V2 Approaches Continue} \\
    \hline
    \textbf{Sr.} & \textbf{Paper} & \textbf{Technique Used} & \textbf{Features Used} & \textbf{Acc.} & \textbf{F1}\\
    \hline
    \endhead
    
    % Your table content goes here
    \setcounter{customcounter}{1}
    \thecustomcounter \stepcounter{customcounter} & FixMatch Semisupervised Learning Approach \cite{cao2022semi} &  FixMatch semi-supervised learning algorithm & NA & 0.91 & NA \\
    \hline
    \thecustomcounter \stepcounter{customcounter} & Convolutional-capsule Network \cite{wang2022convolutional} & CNN & NA & 0.86 & NA \\
    \hline
    \thecustomcounter \stepcounter{customcounter} & Multi-Feature Fusion \cite{ramamurthy2022novel} & Fusion of Features & Visual Features & 0.98 & 0.97 \\
    \hline
    \thecustomcounter \stepcounter{customcounter} & Explainable Deep Learning \cite{mukhtorov2023endoscopic} & ResNet152 & NA & 0.93 & NA \\
    \hline
    
    % Add more rows as needed
    
\end{longtable}
\normalsize

\section{Segmentation Techniques}
The segmentation of the abnormal region within the gastrointestinal (GI) tract is a crucial step in the process of further treatment for identified abnormalities. Segmentation involves the delineation and isolation of the specific area where the abnormality is located, which is essential for targeted medical interventions. By accurately defining the boundaries of the abnormal region, healthcare professionals can plan and execute treatments with greater precision, minimizing the impact on healthy tissues and enhancing the overall effectiveness of therapeutic measures.

This segmentation process often relies on advanced medical imaging techniques, such as magnetic resonance imaging (MRI), computed tomography (CT), or endoscopic imaging, which provide detailed visualization of the GI tract. Medical professionals use these images to identify and isolate the abnormal region, creating a clear and well-defined area for treatment. The successful segmentation of the abnormal region not only aids in planning surgery or other therapeutic procedures but also allows for real-time monitoring during treatments, ensuring that healthcare providers can precisely target and address the abnormality while preserving healthy tissue. Overall, segmentation is a critical step in the continuum of care for GI tract abnormalities, facilitating optimized treatment strategies and better patient outcomes.

Colon polyp detection in colonoscopy images is crucial for early colorectal cancer diagnosis. Bernal et al. investigated how image preprocessing techniques like contrast enhancement and noise reduction impact polyp localization systems \cite{bernal2013impact}. They found these methods significantly affect sensitivity, specificity, and overall accuracy. Optimizing preprocessing can improve computer-aided diagnosis systems and lead to earlier disease detection.

Precise radiation therapy is essential for treating GI cancers. Sharma et al. proposed a new method for segmenting organs like the stomach and intestines in GI scans \cite{sharma2023u}. This segmentation helps radiation oncologists target tumors more accurately while minimizing damage to healthy tissues. Their method utilizes a U-Net model trained on multiple pre-trained models (Inception V3, SeResNet50, etc.) to effectively segment small organs. Evaluation metrics show impressive results, suggesting this approach can improve the precision of GI cancer treatment.

Wang et al. introduced a polyp detection system called Polyp-Alert \cite{wang2015polyp}. This system uses edge-cross-section visual features and a rule-based classifier to identify polyp edges. In their experiments, the software achieved a high detection rate (97.7\%) on colonoscopy video footage.

Magoulas et al. explored the concept of online neural network training for computer-assisted colonoscopy diagnosis \cite{magoulas2004neural}. Their approach focuses on continuously training the network as it analyzes images, allowing it to adapt to changing patterns within the colonoscopy video. This method shows promise in detecting potentially malignant regions with good accuracy.

Karkanis et al. proposed a tumor detection approach based on feature extraction and neural networks \cite{karkanis2001detection}. Their method extracts features from images using wavelet transformation and then feeds them into a multi-layered neural network for tumor detection.

The neural network is trained using features extracted from both normal and tumor regions. The classification results obtained from this approach demonstrate promise, as the system exhibits the capability to effectively classify and pinpoint regions corresponding to lesions. The reported success rates range from 0.94 to 0.99 in accurately identifying such regions within a sequence of video frames. This suggests a high level of efficacy in the proposed system's ability to discern and locate tumor-related regions in magnetic resonance imaging of the breast.

In the work conducted by Gloria Fernandez et al. \cite{fernandez2016exploring}, the concept of polyp boundaries is elucidated. According to the authors, polyp boundaries are characterized as valleys within the context of image intensity. The information pertaining to these valleys is amalgamated into energy maps, which serve as graphical representations denoting the probability or likelihood of the existence of a polyp in the given context. In essence, the study introduces and defines polyp boundaries as features discerned from the image intensity landscape, and subsequently employs energy maps as a means to convey the probability of polyp presence based on the identified valleys in the image data.

\chapter{Data Collections}
\label{sec:data_collection}

Several publicly available datasets exist for training and evaluating algorithms designed to detect abnormalities in the gastrointestinal (GI) tract. These datasets are compiled from images collected at various hospitals and meticulously labeled by medical experts. Notably, many datasets are constructed with a balanced class distribution, ensuring a roughly equal number of patients and images for each type of abnormality. This balanced approach helps machine learning models perform better during training and evaluation. The datasets used in this research are of two types based on the usability of the datasets.

\section{Abnormality Detection}
The dataset of the GI-Tract abnormality detection. There are several datasets for the detection of various abnormalities in the endoscopy or colonoscopy video frames. Some of the datasets for the abnormailty detection used in the research are as follows:

\subsection{Kvasir V1}
The Kvasir dataset is a significant resource for developing computer-aided disease detection in gastrointestinal (GI) tract analysis using medical images.  Limited access to medical image datasets has previously slowed research progress in this critical field.  The Kvasir dataset addresses this challenge by providing a platform for researchers to compare and improve disease detection methods, ultimately aiming to improve global healthcare \cite{pogorelov2017kvasir,riegler2017multimedia}.

Kvasir V1 offers 4,000 high-resolution GI tract images annotated by medical experts \cite{riegler2017multimedia}.  The dataset is segmented into eight classes, each containing 500 images.  These classes include anatomical landmarks (Z-line, pylorus, cecum), pathological findings (esophagitis, polyps, ulcerative colitis), and polyp removal stages (dyed and lifted polyps, dyed resection margins). Image sizes vary from $720 \times 756$ to $1920 \times 1072$ pixels.

\subsection{Kvasir V2}
Kvasir V2 is a dataset of GI-tract images annotated by medical experts \cite{pogorelov2018medico}. The dataset consists of 14033 images, distributed in two parts 5293 for training and 8740 for testing. There are 16 classes of images and each class consists of a different number of training and testing images. The classes are of five categories including anatomical landmarks, pathological and normal findings, endoscopic procedures, and not for diagnosis.

\begin{figure}
	\includegraphics[width=1.0\linewidth]{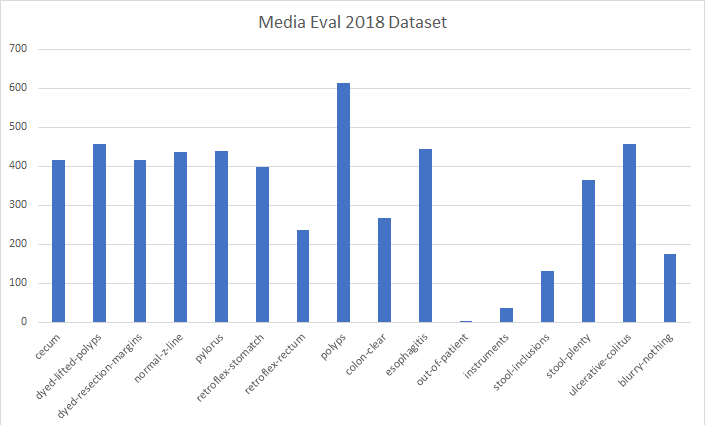}
	\caption{Class Distribution of Kvasir 2018 dataset}
	\label{fig:me2018_classes}
\end{figure}

\begin{enumerate}
	\item The anatomical landmarks are normal-z-line, normal-pylorus, normal-cecum, retroflex-rectum, and retroflex-stomach.
	\item The pathological findings include esophagitis, polyps, and ulcerative colitis.
	\item Pathological and normal findings are the pre, while, and post-surgical findings having classes of the dyed-lifted-polyps, the dyed-resection-margins and the instruments.
	\item Endoscopic procedures contain the classes of the normal tissue with or without stool contamination, which are colon-clear, stool-inclusions, and stool-plenty.
	\item Not for diagnosis consists of classes blurry-nothing and the out-of-patient.
\end{enumerate}
The images in the dataset are of different sizes varying from $720 \times 756$ to $1920 \times 1072$ pixels. The details of the class distribution are shown in the figure \ref{fig:me2018_classes}.

\subsection{Kvasir 2017 (10 Versions)}

Kvasir 2017 is a dataset of GI-tract images annotated by medical experts \cite{pogorelov2017kvasir} for the various GI-tract abnormalities. The dataset consists of 80000 images, distributed in two parts 72000 for training and 8000 for testing. There are 8 classes of the images and each class consists of 9000 training and 1000 testing images. The classes of the images are anatomical landmarks including Z-line, pylorus, and cecum, the pathological findings include esophagitis, polyps, and ulcerative colitis and removal of polyps including the dyed and lifted polyp and the dyed resection margins. The images in the dataset are of different sizes varying from 720*756 to 1920*1072 pixels.

\subsection{ICPR EndoTech 2020}
\begin{figure}
	\centering
	\subfloat[ulcerative-colitis-grade-1]{\label{sfig:a}\includegraphics[width=.30\textwidth,height=2cm]{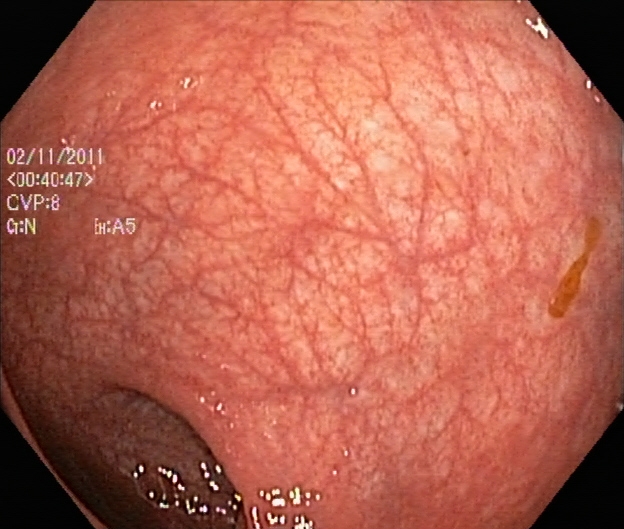}}\hfill
	\subfloat[ulcerative-colitis-2-3]{\label{sfig:b}\includegraphics[width=.30\textwidth,height=2cm]{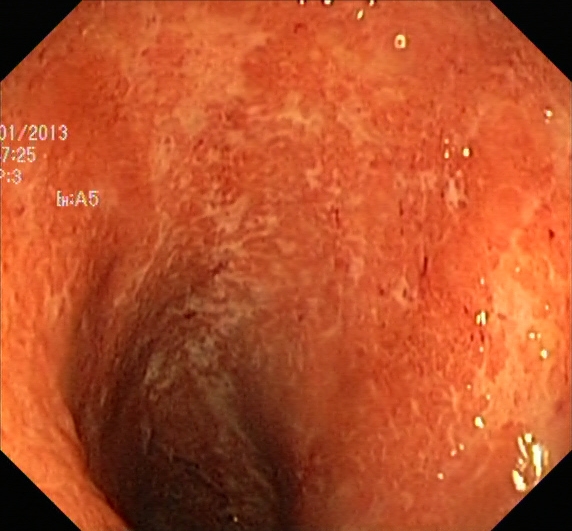}}\hfill
	\subfloat[retroflex-stomach]{\label{sfig:c}\includegraphics[width=.30\textwidth,height=2cm]{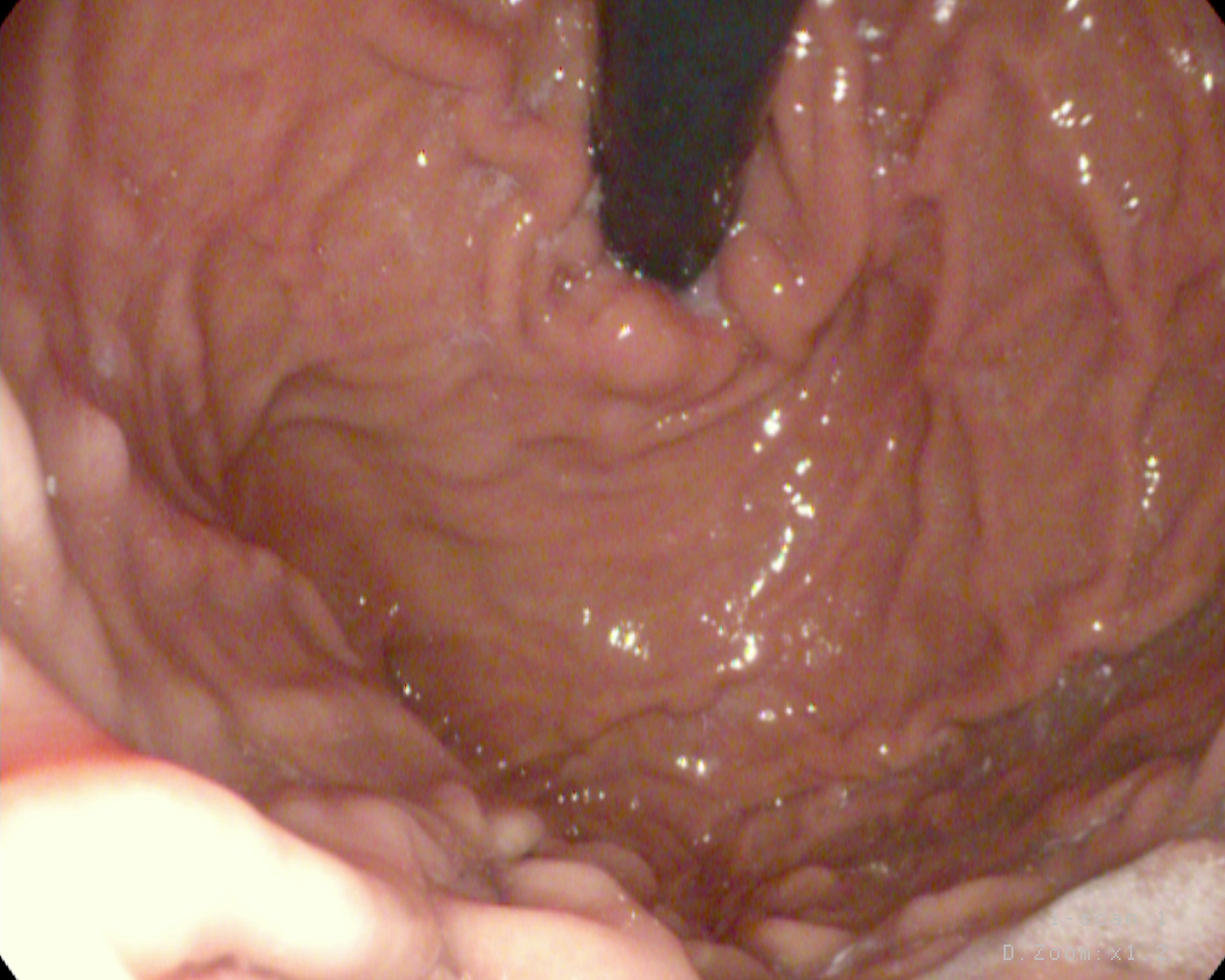}}\hfill
	\subfloat[polyps]{\label{sfig:d}\includegraphics[width=.30\textwidth,height=2cm]{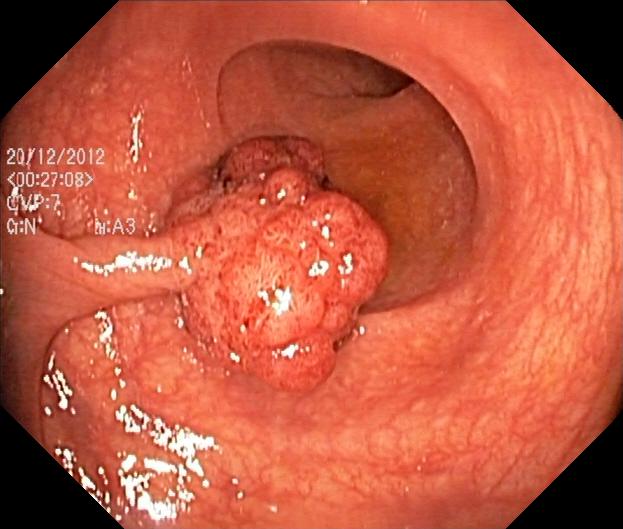}}\hfill
	\subfloat[ileum]{\label{sfig:e}\includegraphics[width=.30\textwidth,height=2cm]{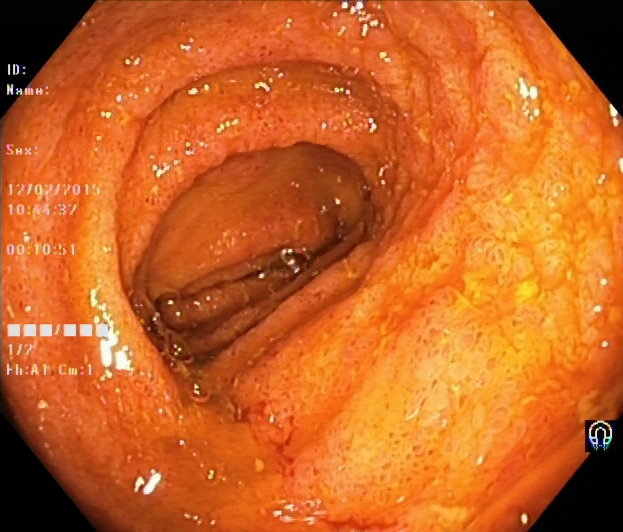}}\hfill
	\subfloat[esophagitis-b-d]{\label{sfig:f}\includegraphics[width=.30\textwidth,height=2cm]{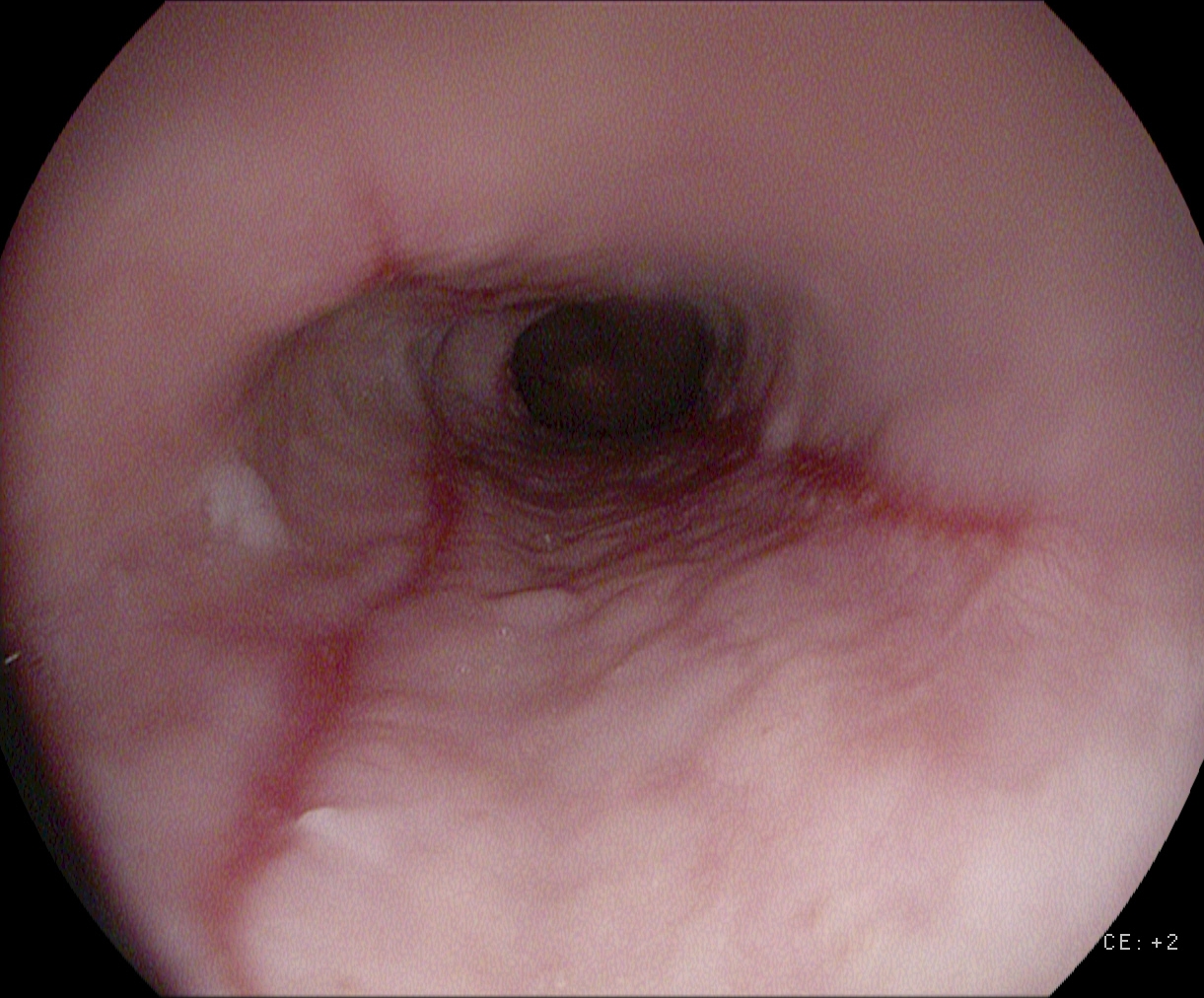}}\hfill
	\\
	\caption{Some images from ICPR 2020 Dataset}
	\label{fig:icpr2020}
\end{figure}

The ICPR Endo 2020 dataset contains images of the gastrointestinal (GI) tract annotated by medical professionals \cite{borgli2020hyperkvasir}. It includes 110,800 JPEG images showcasing various GI tract abnormalities in varying sizes as shown in the Figure~\ref{fig:icpr2020}. The dataset is divided into three parts: a labeled training set (10,662 images across 23 classes), an unlabeled set (99,417 images), and a test set (721 images). Image sizes range from $352 \times 332$ to $1079 \times 1920$ pixels, all with 3 color channels. Notably, the dataset exhibits significant class imbalance, with the number of images per class ranging from just 6 (hemorrhoids) to 1148 (bbps-2-3 class). The image categories encompass anatomical landmarks, pathological and normal findings, and endoscopic procedures.

\begin{enumerate}
	\item The anatomical landmarks are normal-z-line, normal-pylorus, normal-cecum, retroflex-rectum, and retroflex-stomach.
	\item The pathological findings include esophagitis-a,esophagitis-b-d, polyps, ulcerative-colitis-0-1, ulcerative-colitis-1-2, ulcerative-colitis-2-3, ulcerative-colitis-grade-1, ulcerative-colitis-grade-2, ulcerative-colitis-grade-3, bbps-0-1, bbps-2-3 and hemorrhoids.
	\item Pathological and normal findings are the pre, while, and post-surgical findings having classes of the dyed-lifted-polyps, the dyed-resection-margins, and the ileum.
	\item Endoscopic Procedures like impacted-stool.
\end{enumerate}

\begin{figure}
	\includegraphics[width=1.0\linewidth]{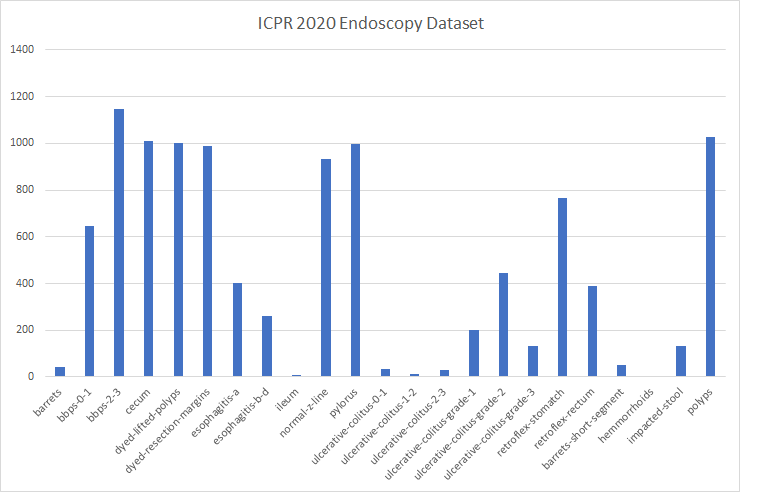}
	\caption{Class Distribution of ICPR Endo Tech 2020 dataset}
	\label{fig:icpr2020_classes}
\end{figure}

The details of the dataset with the class distribution are shown in Table~\ref{dataset} and Figure~\ref{fig:icpr2020_classes}.

\begin{table}
	\caption{Description of Hyper Kvasir Dataset \cite{borgli2020hyperkvasir}}
	\label{dataset}
	\begin{tabular}{|p{3 cm}|p{2 cm}|p{3.5 cm}|p{2 cm}|}
		\hline
		\textbf{Class Label} & \textbf{Number of Images} & \textbf{Class Label} & \textbf{Number of Images}\\
		\hline
		barretts & 41 & barretts-short-segment & 53 \\
		bbps-0-1 & 646 & bbps-2-3 & 1148 \\
		cecum  & 1009 & dyed-lifted-polyps & 1002 \\
		dyed-resection-margins & 989 & esophagitis-a & 403\\
		esophagitis-b-d & 260 & hemorrhoids & 6 \\
		ileum & 9 & impacted-stool & 131 \\
		normal-z-line & 932 & polyps & 1028 \\
		pylorus & 999 & retroflex-rectum & 391 \\
		ulcerative-colitis-0-1 & 35 & ulcerative-colitis-1-2 & 11 \\
		ulcerative-colitis-2-3 & 28 & ulcerative-colitis-grade-1 & 201 \\
		ulcerative-colitis-grade-2 & 443 & ulcerative-colitis-grade-3 & 133 \\
		retroflex-stomach & 764 &  &  \\
		\hline
	\end{tabular}
\end{table}

\subsection{EAD 2019}
The Endoscopy Artifact Detection (EAD) 2019 dataset is a collection of images used to train and test algorithms for detecting artifacts in endoscopic videos \cite{EAD2019endoscopyDataset}.  The dataset includes images from six different medical centers around the world, including Italy, Switzerland, the United Kingdom, France, and Russia \cite{EAD2019endoscopyDataset}. It contains a total of 2,342 images categorized into seven different classes: specularity, saturation, artifact, blur, contrast, bubbles, and instruments \cite{EAD2019endoscopyDataset}. These classes represent various visual issues that can hinder the quality and analysis of endoscopic videos.

\subsection{EAD 2020}
The EAD 2020 is the extension of the dataset of EAD 2019 \cite{EAD2019endoscopyDataset} consisting of 2200 training and 99 testing images of 8 classes \cite{ali2020objective}. The classes in this version are all the 7 classes of the previous version with the addition of a class blood in it.

\subsection{DowPK}
The DowPK is data gathered from the GI department of the DOW University hospital \cite{zeshan2023}. The data is unlabeled and will be verified after the labeling is done by the machine/algorithm. There are 844 images of the various sizes for different GI-tract findings.

\subsection{Comparison}
The datasets above have a different number of classes/diseases/ abnormalities and a different number of images for disease detection. A compact analysis of these datasets is provided in table \ref{table:datasets-main}.

\begin{table}
	\centering
 \caption{Abstract Analysis of the GI-Tract image datasets}
	\label{table:datasets-main}
	%\tiny
	\begin{tabular}{|p{2.5cm}|p{1cm}|p{1cm}|p{1cm}|p{2.5cm}|}
		\hline
		Dataset & Classes & Training Samples & Testing Samples & Class Distribution\\
		\hline
		Kvasir V1 \cite{riegler2017multimedia}&8&4000&4000&Equal\\
		\hline
		Kvasir V2 \cite{pogorelov2018medico}&16&5293&8740&Un-equal\\
		\hline
		Hyper Kvasir \cite{borgli2020hyperkvasir}&23&10662&721&Un-equal\\
		\hline
		Kvasir 2017 (10 Versions) \cite{pogorelov2017kvasir}&8&72000&8000&Equal\\
		\hline
		EAD 2019 \cite{EAD2019endoscopyDataset}&7&2147&195&Un-equal\\
		\hline
		EAD 2020 \cite{ali2020objective}&8&2200&99&Un-equal\\
		\hline
		RLE 2018 \cite{coelho2018deep}&2&3295&600&Un-equal\\
		\hline
		DowPk~\cite{zeshan2023}&8&NA&844&Un-equal\\
		\hline
	\end{tabular}
\end{table}

\section{Abnormality Region Identification}
There is another category of the dataset used in the reserach based on identifcation of the abnormality locating in the GI-Tract videos and images.

\subsection{RLE 2018}
Red Lesion Endoscopy (RLE2018) Dataset is a dataset for the GI-tract images consisting of 3295 training and 600 testing images of two classes i.e. the Lesion or no lesion with the bounding box of the lesion if that exists \cite{coelho2018deep}.

\subsection{CVC-ClinicDB}
CVC-ClinicDB is a collection of images extracted from colonoscopy videos, specifically designed for polyp detection research \cite{bernal2012towards}. These images include various polyp examples along with corresponding ground truth masks, which accurately highlight the polyp regions within each image. The database is built from 25 different video studies, ensuring a diverse range of polyp appearances. Notably, at least one sequence containing a polyp was extracted from each study, resulting in a total of 29 unique sequences included in CVC-ClinicDB. Each data point within the dataset consists of two separate TIFF image files: an RGB image representing the colonoscopy frame and a ground truth mask identifying the polyp area.

\subsection{EndoVisSRI 2015}
The Endoscopy Vision sub-challenge of Instrument segmentation (EndoVisSRI 2015) dataset is a collection of 160 labeled and 60 unlabeled collected from 6 different laparoscopic colorectal surgeries \cite{EndoVisSRI:2020}. The dataset contains the location of the instrument detected in the endoscopy image.

\subsection{EndoTech 2020 Segmentation}
This dataset contains 1200 endoscopic images for polyp detection and localization.  1000 of the images are labeled, providing information on the presence and location of polyps. The remaining 200 images are unlabeled and can be used for various tasks such as model training validation.  The image sizes vary, ranging from $332 \times 487$ pixels to $1920 \times 1072$ pixels, and all images use the RGB color format.

\subsection{INESC-TEC Lesion Endoscopy}
INESC-TEC C-BER Bioimaging Lab's Red Lesion Endoscopy Dataset is the collection of 3895 images of the endoscopy with the Red Lesion \cite{INESC-TEC:2018,coelho2018deep}. The 3295 images are labeled while the 600 are part of test set with all images of $320 \times 320$ pixels. The data is collected clinically and labeled by the medical experts.

\subsection{Vocal Folds Dataset}
The vocal folds dataset consists of 536 images from 8 endoscopic analysis of two patients \cite{laves2019dataset}. The dataset consists of 7 types of the regions in every image including vocal folds, other tissues, glottal space, pathology, surgical tool, intubation and void. The regions of the different types is divided by using different colors for the above mentioned regions.

\chapter{Contributions}
\label{sec:approaches}
This chapter delves into advanced methodologies for medical image analysis, particularly those geared towards detecting abnormalities in the GI-tract. It highlights the novel contributions made by the author in this domain.

One key area of focus is feature selection techniques. By carefully selecting the most informative features from the image data, the author aims to improve the accuracy and efficiency of the classification process. The chapter explores both homogeneous and heterogeneous classifiers, evaluating their effectiveness in identifying GI-tract abnormalities.

Furthermore, the chapter investigates the application of lightweight networks for real-time detection. This approach prioritizes speed while maintaining a reasonable level of accuracy, making it suitable for practical clinical applications. Finally, the author explores the utilization of genetic algorithms for optimizing threshold computation. This technique helps determine the optimal point at which to distinguish abnormal from normal tissue within the images, leading to more robust and reliable classifications.

\section{Majority Voting of Heterogeneous Classifiers}
\label{approach-hetero}

The foundation of the approach involves creating a heterogeneous ensemble of classifiers, each contributing to the detection process by extracting features related to the texture of GI tract images. Texture features, including patterns, gradients, and structures, play a crucial role in distinguishing between normal and abnormal conditions. The ensemble employs Majority Voting as a decision-making mechanism, combining predictions from each heterogeneous classifier to mitigate biases and errors, resulting in a more reliable abnormalities detection process \cite{khan2018majority}.

Feature Engineering constitutes a pivotal and intricate facet within the realm of Machine Learning, standing as a paramount and formidable challenge to address. As delineated in Figure~\ref{fig:p1-methodology}, denoting the illustrative depiction of our proposed model, the discernment of salient features emerges as an essential prerequisite for the effective approximation of mathematical functions. The organizers of the task thoughtfully furnished a compendium of six pre-computed visual attributes for every image in the dataset. These attributes encompass JCD (Joint Composite Descriptor), Tamura, Color Layout, Edge Histogram, Auto Color Correlogram, and PHOG (Pyramid Histogram of Oriented Gradients). In conjunction with these pre-determined visual characteristics, our approach further leverages deep learning features to uncover and extract deep feature for enrichment of the understand-ability of the images for classification.

In view of the inherent limitations posed by the relatively modest scale of our training dataset, consisting of a mere 5293 images, we embarked on a judicious strategy. We harnessed the power of VGG19, a pre-trained convolutional neural network model introduced by Simonyan and Zisserman \cite{simonyan2014very}. VGG19 is known for its impressive depth, consisting of 19 weight layers, including 16 convolutional layers and 3 fully-connected layers. Originally designed for large-scale image classification tasks, VGG19 has achieved significant success on various datasets.

\begin{figure}
    \centering
    \includegraphics[width=15cm]{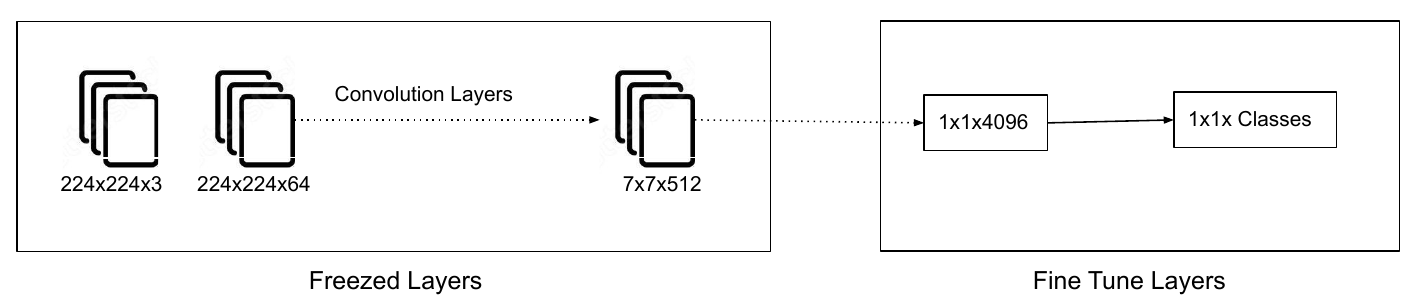}
    \caption{Proposed Model for VGG19 Finetune For Deep Features Extraction}
    \label{fig:p1-methodology-homo}
\end{figure}

\begin{figure}
    \centering
    \includegraphics[width=15cm]{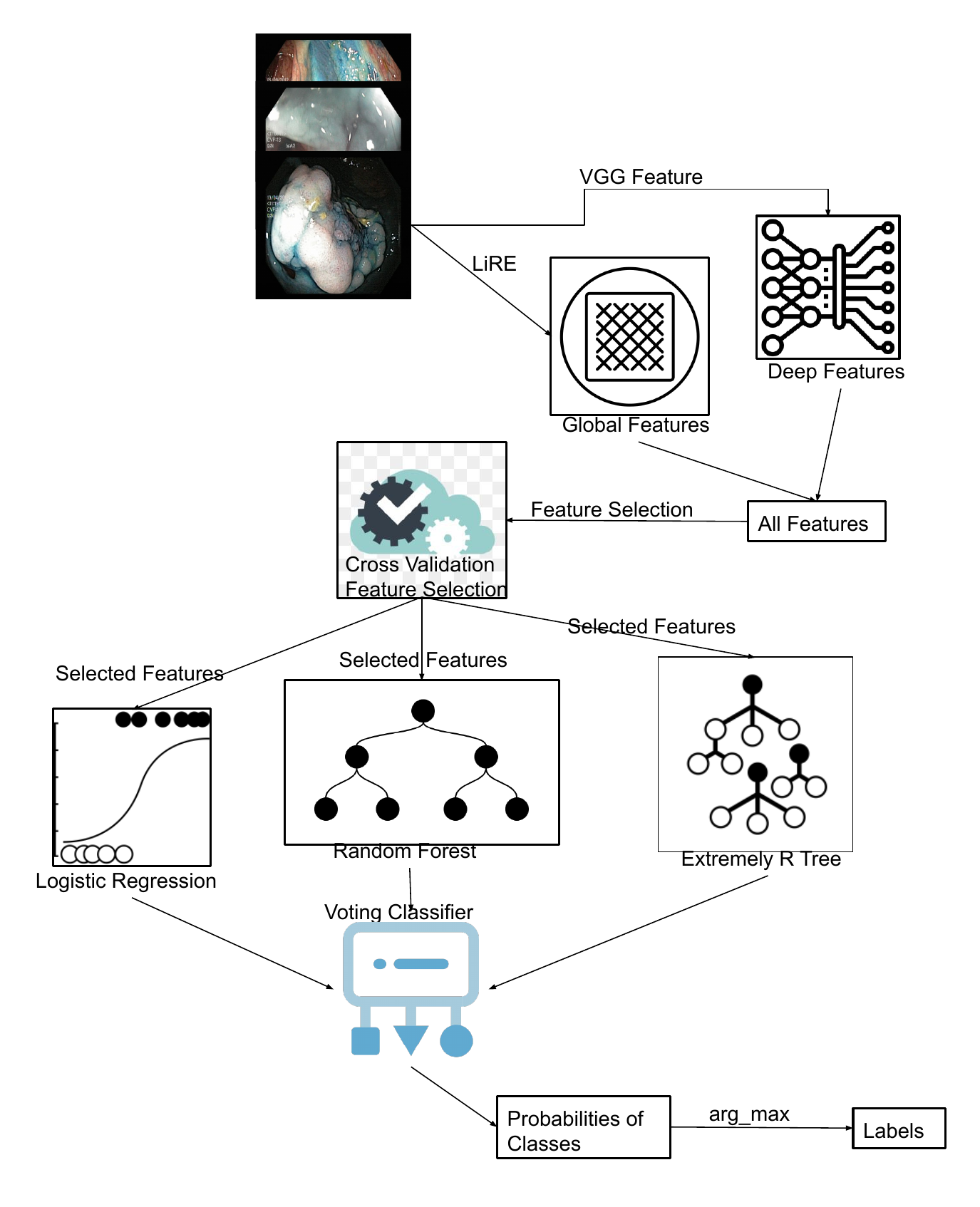}
    \caption{Proposed Model for Heterogeneous Approach for GI-Tract Abnormalities Detection}
    \label{fig:p1-methodology}
\end{figure}

Our approach, as shown in Figure~\ref{fig:p1-methodology-homo}, entails the use of a pre-trained VGG19 model, which was originally fine-tuned on a massive corpus of data hailing from the ImageNet challenge. Following this pre-training, the final two layers of the VGG19 model underwent a process of retraining, calibrated to the nuances of our specific dataset of medical images. The outcome of this strategic amalgamation of pre-trained and retrained layers is a repository of abundant visual concepts ready for deployment in our classification endeavors.

To translate these visual features into actionable insights and robust classification models, we employed a range of classifiers, including logistic regression \cite{bishop2006pattern}, random forest \cite{pal2005random}, and the esteemed extremely random trees classifier \cite{geurts2006extremely}. These classifiers served as our trusty companions in navigating the intricate terrain of feature space. It is imperative to emphasize that we were presented with two distinct categories of features: pre-computed texture features and VGG features, the latter representing those derived from our VGG19 pre-trained model. Our ensemble model, which emerged as the final arbiter of classification, is predicated on the principle of weighted majority voting across the spectrum of independent models, each uniquely trained on a distinct set of features. The weighting of this ensemble reflects the accuracy achieved by each of the independent classifiers.

In the pursuit of optimal performance, it is worth noting that we diligently explored an array of advanced machine learning techniques. However, the outcome of this rigorous experimentation unequivocally affirmed the supremacy of logistic regression, random forest, and extremely random trees classifiers. Consequently, these exemplary classifiers comprise the focal point of our present report, encapsulating the state-of-the-art results obtained in our research.

This endeavor unfolded against a backdrop characterized by the convergence of two noteworthy challenges. First and foremost, we grappled with the paucity of training data, a constraint that invariably hampers the training of sophisticated machine learning models. Additionally, the vexing issue of class imbalance introduced further complexity to our task. In order to mitigate these challenges and restore balance to our dataset, we harnessed the technique of random resampling, deftly generating additional data for each class. The new samples were generated in such a proportion that can lead samples of all classes equal to the 110\% of the majority class.

For example, let's say the Kvasir V3 dataset has a majority class with 1148 samples and a minority class with only 6 samples.  This imbalance would make it difficult for the model to learn the characteristics of the minority class.  Through random resampling, we increased the size of each class to 110\% of the majority class.  In this example, each class would now have 1263 samples (1148 x 1.1).  This process resulted in a total dataset size of 29049 samples (23 classes x 1263 samples/class).

This deliberate resampling strategy not only fostered the creation of supplementary features for each class but also rectified the class imbalance conundrum. Ultimately, it engendered an equitable distribution of instances across the 16 available classes, thereby enabling the expansion of our training dataset and affording us the opportunity to validate a diverse array of models.

Figure ~\ref{fig:p1-methodology}, which serves as the architectural representation of our proposed model having phases of the feature extraction and feature selection followed by voting classifier.

\section{Ensemble of Homogeneous Classifiers}
\label{approach-homo}

An ensemble methodology is introduced that utilizes data bagging to fine-tune multiple instances of deep neural networks for GI-Tract abnormality detection. This method enhances the robustness and adaptability of the classification process by employing an ensemble of neural networks as feature extractors. Data bagging involves creating diverse versions of the dataset to mitigate biases in the fine-tuning process, resulting in neural network instances capturing various facets of the data distribution \cite{khan2021medical}.

A 169-layer DenseNet architecture serves as the foundational deep neural network for fine-tuning, with eleven instances trained on different bags of the dataset. Features extracted from each instance are aggregated using Majority Voting and a neural network-based approach. Majority Voting selects the majority class from predictions of all eleven models, while the neural network-based approach employs a specially designed neural network to provide class probabilities. The fusion of data bagging, fine-tuning, and ensemble-based feature amalgamation creates a comprehensive abnormality detection framework, addressing challenges in GI-Tract image analysis.

Bagging, a technique introduced by Breiman \cite{breiman1996bagging}, involves generating multiple versions of the dataset, either with or without repetition, to obtain diverse predictions. The aggregation of these predictions serves to mitigate biases and enhance the overall robustness of the model. The predictive outcome in bagging, as expressed in Equation \ref{eq:bagging}, involves summing the probabilities of each class across different data versions. Here, $C(X|D)$ represents the prediction of class $X$ by classifier $C$ at input $D$, and $P(C(X, D)==j)$ signifies the probability of class $j$ at input $X$ using classifier $C$ with a data version of $D$.

\begin{equation}
Y(j|X) = \sum_{D \in \text{Data-Versions}} P(C(X, D) == j)
\label{eq:bagging}
\end{equation}

The approach utilized a combination of bagging and neural network fine-tuning to predict abnormalities in GI-tract images.  Here's how it functioned: the data was divided into various chunks, ranging from 5 to 20. Each chunk contained 70\% of the data and was evaluated with repetition. The researchers measured both the individual accuracy of each data chunk and the overall diversity of the results. While an increase in the number of chunks led to greater diversity, it also came at the cost of increased computation time.

The model achieved its highest combined accuracy when using 11 data bags. Interestingly, the accuracy remained the same for 14 and 15 bags. However, using more or less than this range (between 5 and 20) resulted in lower accuracy during testing. Considering the trade-off between accuracy, computation time, and the number of bags, 11 data bags were chosen as the optimal solution. This selection allowed for a good balance between achieving high accuracy and maintaining reasonable computation time.

 To address the neural network architecture, the research looked at various options documented in existing literature. Studies have shown that ResNet and DenseNet architectures achieve strong results on similar datasets. Here, different versions of both architectures were evaluated on 50\% of the data without any preprocessing. The best performing model utilized pre-trained weights from a 169-layer DenseNet trained on ImageNet.
 
The final fully connected (FC) layer of the DenseNet169 was replaced with a new FC layer containing 23 output neurons, employing the softmax activation function. Subsequently, eleven instances of the DenseNet169 were fine-tuned independently for 500 epochs, each on a distinct data bag or chunk of the dataset. This fine-tuning process resulted in unique weights for each instance, effectively creating eleven individualized instances of the DenseNet169.

The training and testing datasets were input to each of the eleven instances of the DenseNet169, generating a 23-dimensional vector as the output class probabilities for each image. These eleven vectors, each representing the results from a different instance of the DenseNet169, were concatenated to form a single 253-dimensional feature vector for each image. This composite feature vector was then utilized for classification through two distinct approaches. 

\subsection{Hard Majority Voting}
The Hard Majority Voting approach constitutes a classification strategy applied to the class probabilities derived from all eleven fine-tuned models. For each individual image, the majority class among the predictions of all eleven models was determined. Subsequently, the class label with the highest frequency among the eleven retrieved labels was chosen as the final classification for that particular image. This method relies on the collective decision-making power of multiple models to enhance the overall robustness and accuracy of the abnormality detection process.

\subsection{Neural Network for Class Prediction}
In the Neural Network for Class Prediction approach, a specific neural network architecture was formulated, consisting of five fully connected layers. These layers incorporated various activation functions, namely Rectified Linear Unit (ReLU), Sigmoid, and Softmax. The designed neural network underwent training for 500 epochs, utilizing the 253-dimensional input vector. The primary objective of this training was to generate an output comprising class probabilities for the 23 different classes. Subsequently, the trained neural network was employed to predict the class associated with the highest probability for each image. This approach aimed to leverage the learned features from the fine-tuned DenseNet169 instances to enhance the accuracy of abnormality classification in GI-Tract images.

\section{Real time abnormalities detection}
\label{approach-realtime}

This section provides a comprehensive overview of the core methodology employed in realyt=time detection. The system is designed with a structured approach, comprising seven main components as shown in the Figure~\ref{fig:p4-methodology_abs} \cite{khan2023real,zeshan_realtime}. Each of these components is meticulously explored and implemented using state-of-the-art techniques, and they are discussed in detail as follows:

\subsection{Data Preprocessing}
\label{sec:preprocess}

\begin{figure}[ht]
    \centering
    \begin{overpic}[width=\textwidth]{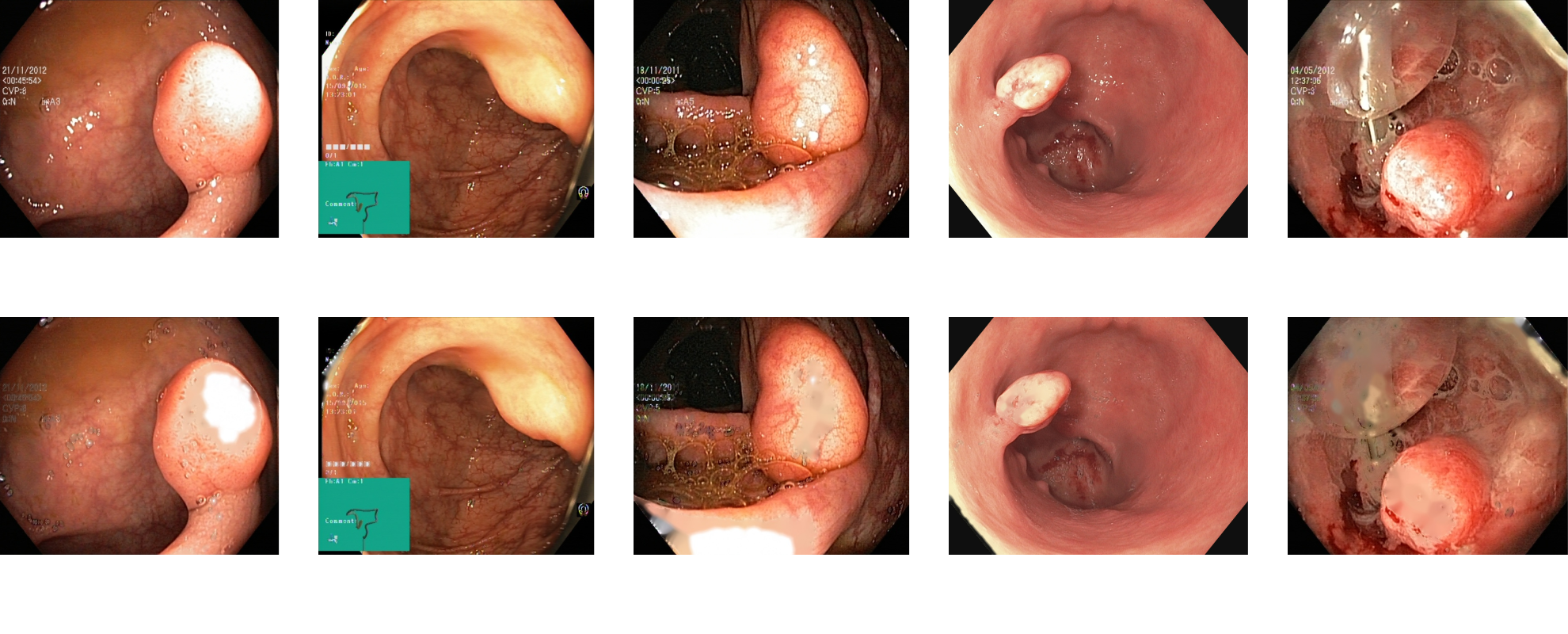}
        \put(0,21){(A) Images with Reflections}
        \put(0,0){(B) Images after removal of Reflections}
    \end{overpic}
  \caption{Reflections on the images of the Kvasir V2 \cite{pogorelov2018medico} polyps category.}
  \label{fig:reflection}
\end{figure}

The presence of reflections in endoscopic images captured during video recording represents a critical issue that necessitates immediate attention. It is imperative to eliminate reflections because they have the potential to significantly distort the image and adversely impact the accuracy of subsequent abnormality and anatomical landmark detection processes. The detrimental effects of these reflections are visually evident in a sample set of polyp images, as illustrated in Figure~\ref{fig:reflection}. Machine learning and computer vision (CV) based approaches heavily rely on color differentiation for their analysis. Specifically, they distinguish between different objects based on variations in color. In the context of polyp detection, the primary distinction often relies on color differences. Notably, the difference between a regular polyp and a dyed lifted polyp hinges significantly on these color variations. However, the presence of reflections can lead to the alteration of color, effectively transforming red or blue regions into an indistinguishable white. This phenomenon presents a formidable challenge for CV-based approaches, rendering them incapable of accurate classification. The significance of preprocessing, specifically reflection removal techniques, is underscored by instances where certain image samples, drawn from the Kvasir V2 dataset \cite{pogorelov2018medico}, were initially misclassified but were subsequently correctly categorized after the application of reflection removal techniques, as exemplified in Figure~\ref{fig:reflection}.

\begin{figure}[ht]
    \centering
	\includegraphics[width=14cm]{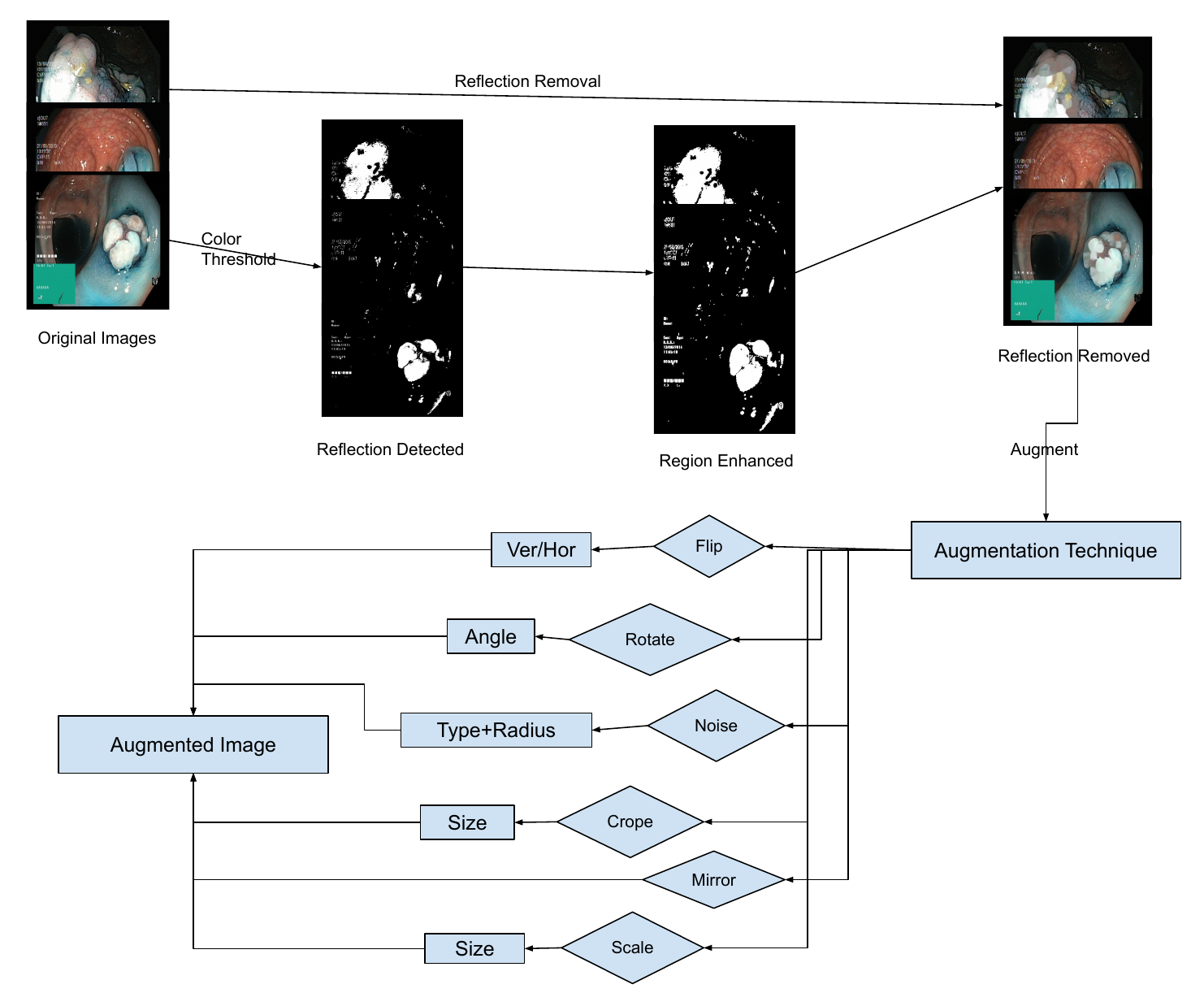}
	\caption{Data preprocessing for the reflection removal and data augmentation for the class imbalance problem.}
	\label{fig:p4-methodology_pre}
\end{figure}

\begin{figure}
    \centering
	\includegraphics[height=20cm]{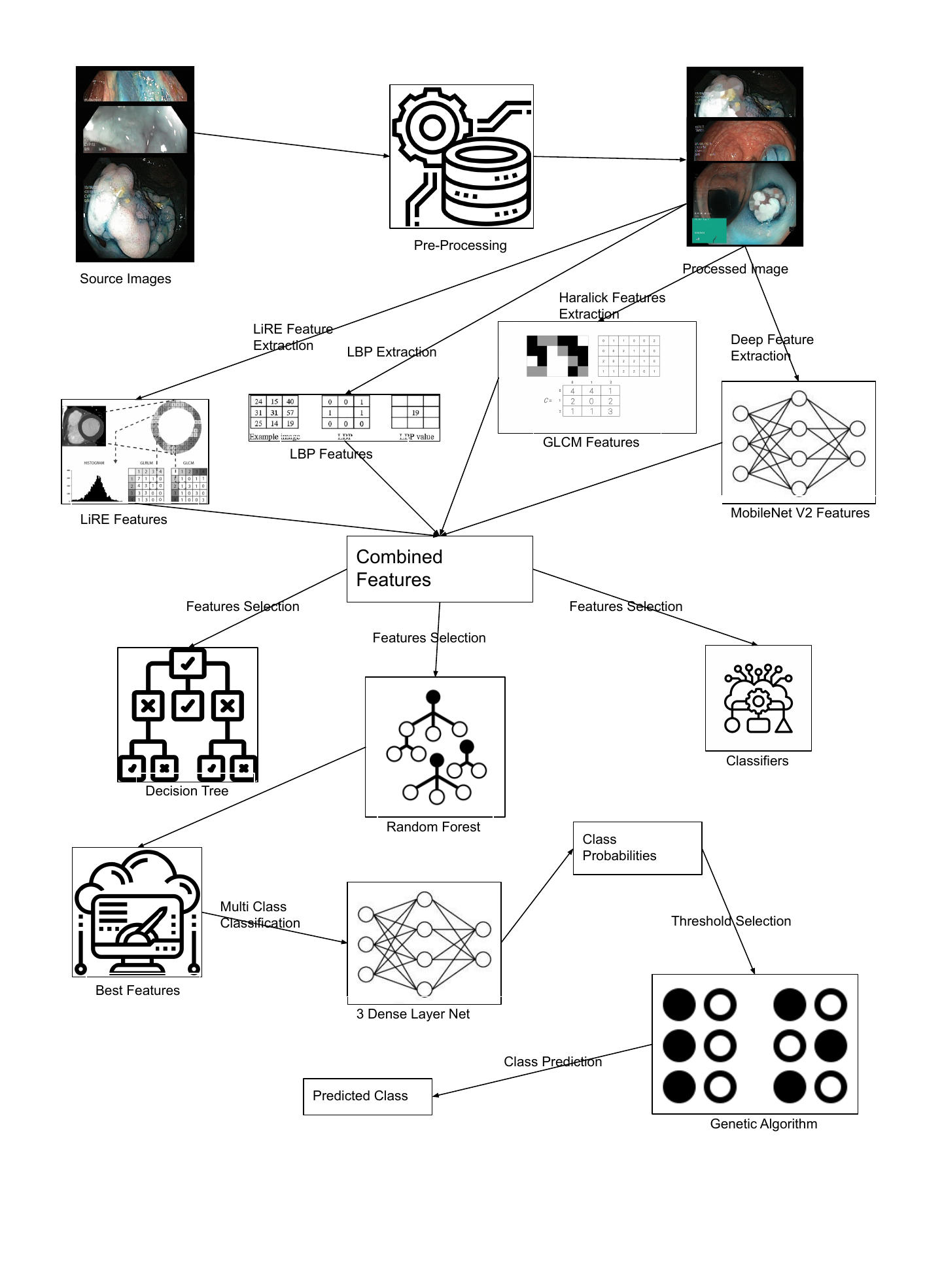}
	\caption{Methodology of the Realtime Detection of GI-Tract Abnormalities.}
	\label{fig:p4-methodology_abs}
\end{figure}

Several methodologies and procedures exist for effectively removing reflections, both with and without human intervention. Notably, some of the standard reflection detection and removal procedures applied within the scope of this research are as follows:

\begin{itemize}
    \item \textbf{Image Crop:} A simple way to remove reflections is to crop the image, cutting out the reflective area. While this removes the reflection and is easy to implement, it also reduces the image size. This can accidentally crop out abnormalities, losing important information for diagnosis. Cropping is only ideal when reflections solely affect the background.  Researchers applied this method to the Kvasir datasets and achieved an average F1-score of 0.72 on Kvasir V2 using various features and a decision tree classifier. However, it's important to note that cropping other datasets resulted in lower accuracy compared to using the whole image.
    \item \textbf{Supervised Detection:} One approach to removing reflections relies on supervised learning. This method trains a model on a large amount of data where human experts have already painstakingly removed reflections from the images. While this technique has shown some success on specific datasets, it requires a significant investment in time and effort to annotate enough data. Additionally, applying this method to new datasets yielded only minor improvements in accuracy.
    \item \textbf{Unsupervised Detection:} One approach to detecting reflections in images, without relying on pre-trained models, focuses on color layout differences. This method assumes that reflected areas appear brighter. In endoscopy images, experiments have shown that pixels with a grayscale value exceeding 180 are likely affected by reflection.  In a typical color image, each pixel has a value for each color channel (red, green, blue). The unsupervised approach removes pixels where any of these channels surpasses the 180 threshold, assuming they represent unwanted reflections. This detection process can be formulated mathematically, as expressed in Equation ~\eqref{eq:reflection}.
    \begin{equation}
		\label{eq:reflection}
		I_{x,y} \leftarrow 
		\begin{cases}
			255,& \text{if } I_{x,y} > 180 \\
			0,	& \text{otherwise}\\
		\end{cases}
	\end{equation}

    The approach described above primarily focuses on detecting areas of solid reflection, where a strong reflection causes the grayscale value to reach 180. However, there are instances of weak reflections where the pixel's grayscale value falls within the range of 150 to 180. These weaker reflections can sometimes be challenging to distinguish from regular, lighter image pixels. To enhance the detection of weak reflections and minimize the possibility of false positives, a refinement technique is applied. This enhancement methodology is founded on the concept of expanding the reflection region.
    
    In essence, the approach for detecting weak reflections is predicated on the condition that, when a pixel exhibits a weak reflection value (between 150 and 180), and its adjacent pixel is identified as a reflection, that particular pixel is classified as part of the reflection region. This condition can be formally represented by Equation ~\eqref{eq:reflectionen} and Equation ~\eqref{eq:reflectionadj}.
    
	\begin{equation}
		\label{eq:reflectionen}
		I_{x,y} \leftarrow 
		\begin{cases}
			255,& \text{if } I_{x,y} > 130 \wedge \exists Adj(I_{x,y}) > 180 \\
			0,	& \text{otherwise}\\
		\end{cases}
	\end{equation}
	
	\begin{equation}
		\label{eq:reflectionadj}
		Adj(I_{x,y}) \leftarrow I_{x-i,y-j} \forall (i \in {1,0,-1}) \forall (j \in {1,0,-1})
	\end{equation}
     
    The reflection removal procedure, along with the enhancement approach for weak reflection detection, is visually illustrated in Figure~\ref{fig:p4-methodology_pre}. This technique plays a critical role in refining the process of removing reflections, ultimately leading to improved image quality and more accurate abnormality and anatomical landmark detection.
    
    Unsupervised removal approaches for reflections have been implemented in openCV, utilizing techniques like the Fast Marching Method by Telea \cite{telea2004image} and the concept of Navier-Stokes fluid dynamics \cite{bertalmio2001navier}. Within the scope of this research, the unsupervised detection and removal method proposed by Telea yielded the most favorable results in terms of accuracy, detection precision, and processing speed.

    This method, based on the principles introduced by Telea, achieved notable success. When applied in the research context, it resulted in an average F1-score of 0.84. This promising score was obtained through the utilization of various features and a decision tree classifier, outperforming the Image Crop method. The Telea-based unsupervised detection and removal method thus emerges as a robust and effective approach, offering superior accuracy and speed in comparison to alternative techniques like image cropping. This achievement underscores its potential significance in the domain of endoscopic image analysis, particularly in the context of reflection removal and enhancing the quality of diagnostic images.

\end{itemize}

\subsection{Data Augmentation}
\label{sec:augmentation}
Endoscopic image datasets often suffer from class imbalance, where some diseases are rare and consequently poorly represented. This imbalance can significantly impact the performance of machine learning (ML) and deep learning (DL) models. For instance, the Kvasir V2 dataset has a major class with 613 samples and a minor class with only 4. This disparity becomes even more pronounced in Kvasir V3, where hemorrhoids (minor class) have just 6 samples compared to 1148 for BBPS 2-3 (major class).

These imbalanced datasets, coupled with limited overall data availability, pose a challenge for ML and DL models. Detection accuracy for major classes is typically good, but minor classes often suffer.

One way to address class imbalance and data scarcity is data augmentation. This technique creates new training data by applying transformations to existing images. In endoscopic image analysis, common augmentation methods include generative models (like Generative Adversarial Networks, GANs) and simpler image manipulation techniques.

While GAN-based augmentation has been explored, it can be time-consuming for image generation. Conversely, image manipulation methods have shown promise in both accuracy and efficiency. This research compared GANs and image manipulation on the Kvasir V2 dataset. Interestingly, image manipulation techniques emerged as the better choice, achieving superior accuracy and faster training times.

The list of techniques used for the image manipulation, in the research are as follows:
\begin{enumerate}
    \item Image Rotation (Random Angles)
    \item Image Horizontal and Vertical Flip (Random Angles)
    \item Image Crop (Random size from original to $256 \times 256$)
    \item Image Resize (Random size from original to $256 \times 256$)
    \item Noise injection with various ratios
\end{enumerate}

The image manipulations are carried out as a post-step after reflection removal and a pre-step before feature extraction and fine-tuning of the neural network. This ensures that all classes have a balanced number of image samples to rectify the class imbalance issue. The augmented dataset is available at \cite{kvasirv1_aug,kvasirv2_aug,kvasirv3_aug} The number of images generated for each image in a class is determined by the Equation~\eqref{eq:augmentation}:

\begin{equation}
\label{eq:augmentation}
N(C,I) = \frac{N(C) \times (Max\_Images-1)}{L(C)} 
\end{equation}

Where:

\begin{itemize}
    \item $N(C,I)$ represents the number of images generated for image $I$ from class $C$.
    \item $N(C)$ is the initial number of images in class $C$.
    \item $Max_Images$ is the maximum desired number of images for any class.
    \item $L(C)$ denotes the number of images in the class with the largest sample size.
\end{itemize}

The application of image manipulation methods in conjunction with a decision tree classifier across various sets of features has yielded notable results in terms of detection accuracy. The average F1-score achieved using this approach consistently surpasses the 0.84 threshold. This high F1-score demonstrates the effectiveness of the image manipulation techniques in enhancing the classification performance for abnormalities in endoscopic images. It signifies that the balanced and augmented dataset, along with the decision tree classifier, is well-suited for the accurate detection of abnormalities, ultimately contributing to the reliability of the diagnostic process.

\subsection{Texture Feature Extraction}

Endoscopic image analysis relies on a variety of state-of-the-art image features that have proven to be effective in detecting abnormalities and localizing and segmenting anatomical landmarks. Some of these features, each with its unique characteristics, have been instrumental in achieving accurate classification results. Here is an overview of these features:
\begin{itemize}
    \item Color Layout: Color layout describes how colors are arranged within an image. It's a special kind of color descriptor that focuses on the distribution of colors, not the image size. This allows researchers to compare images based on their color patterns, regardless of their resolution \cite{lux2008lire,sikora2001mpeg,huang1997image}.
    \item Edge Histogram: An edge histogram provides a map of how edges are distributed throughout an image. Edges often occur at transitions between different colors, so the information captured in an edge histogram is similar to what a color layout descriptor uses for decision-making \cite{lux2008lire,sikora2001mpeg}.
    \item Tamura Features: Tamura features are a powerful tool for image analysis, capturing six key aspects of texture: coarseness, contrast, directionality, line-likeness, regularity, and roughness. By analyzing these features together, researchers can gain a comprehensive understanding of an image's textural properties \cite{lux2008lire,tamura1978textural}.
    \item Color and Edge Directivity Descriptor (CEDD): CEDD (Color and Edge Directivity Descriptor) is a unique image feature that efficiently captures both color and textural information. Unlike other texture and deep learning-based features, CEDD requires minimal computational resources, making it a fast and efficient choice for various image analysis tasks \cite{lux2008lire,chatzichristofis2008cedd}.
    \item Fuzzy Color and Texture Histogram (FCTH): FCTH (likely referring to a specific feature extraction technique) offers a simplified representation of both image textures and color distributions. This combined information makes it a powerful tool for image analysis tasks \cite{lux2008lire,chatzichristofis2008fcth}.
    \item Color Histograms (HSV and RGB): A color histogram is a graphical representation that shows how often each color appears in an image. It's like a bar chart where each bar represents a specific color, and the height of the bar indicates how many pixels in the image have that color \cite{lux2008lire}.
    \item Gabor Features: Gabor filters are image processing tools designed to capture specific features like edges and textures. This makes them particularly useful for object detection tasks where identifying the edges of an object is crucial for distinguishing it from the background \cite{lux2008lire}.
    \item Auto Color Correlation: In image analysis, Auto Color Correlation is a technique that measures how often specific color differences appear between neighboring pixels within an image \cite{lux2008lire}.
    \item Auto Color Correlogram: The Auto Color Correlogram defines the color distribution of an image by considering the probability of color differences between pixels at various distances \cite{lux2008lire}.
    \item Speeded Up Robust Features (SURF): SURF is a keypoint descriptor known for its scale and rotation invariance, enabling efficient extraction of robust image features \cite{lux2008lire}.
    \item Pyramid Histogram of Oriented Gradients (PHOG): PHOG is constructed from grids of Histograms of Oriented Gradient descriptors, which are particularly effective for human identification and endoscopic image classification \cite{lux2008lire}.
    \item Local Binary Patterns (LBP): LBP is a texture representation that characterizes the relationship between a pixel and its neighbors. It is calculated by summing the values of the circular neighborhood of each pixel \cite{ojala1996comparative}.
    \item Local Ternary Patterns (LTP): Similar to LBP, LTP is a texture representation that uses three states to describe the relationship between a pixel and its neighbors.
    \item Gray-Level Co-Occurrence Matrix (GLCM): GLCM, or Gray Level Co-occurrence Matrix, is a technique used in image analysis. It analyzes how often pairs of pixels with specific gray levels appear next to each other in an image. This can reveal patterns in the texture of the image, such as smoothness, roughness, or regularity \cite{haralick1973textural,ojala1996comparative}.
    \item Haralick Features (Statistics of GLCM): Haralick features encompass 14 texture statistics calculated from GLCM matrices, providing insights into texture characteristics \cite{haralick1973textural}.
\end{itemize}

These features are commonly used in the detection of abnormalities and anatomical landmarks within endoscopic images. They have been instrumental in achieving accurate results in various classification tasks related to endoscopic image analysis.

\subsection{Deep Feature Extraction}

In the realm of endoscopic image analysis, various Convolutional Neural Network (CNN)-based approaches have been employed to detect abnormalities in endoscopy and colonoscopy images. Notable CNN architectures such as DenseNet \cite{huang2017densely,hicks2018deep,jha2021comprehensive}, ResNet \cite{he2016deep,hoang2018application,kirkerod2018using,jha2021comprehensive}, and MobileNet V2 \cite{howard2017mobilenets,sandler2018mobilenetv2,harzig2019automatic,jha2021comprehensive} have been leveraged to achieve high detection accuracies in this domain.

DenseNet \cite{huang2017densely}, characterized by its dense connectivity patterns, has shown promise in endoscopic image analysis tasks. ResNet \cite{he2016deep}, with its residual connections, has also demonstrated robust performance in detecting abnormalities in medical images. MobileNet V2, known for its lightweight architecture, offers a compelling balance between detection accuracy and processing speed, making it a favorable choice for real-time applications.

When these three approaches, DenseNet, ResNet, and MobileNet V2, undergo various preprocessing steps, their detection accuracies become comparable. However, MobileNet V2 distinguishes itself with its superior detection speed. In the context of the research at hand, MobileNet V2 is fine-tuned to serve as a deep feature extractor.

MobileNet V2 is designed as a lighter network with 105 layers. It exhibits efficient inference times, processing approximately 41 images per second on a CPU-based machine, with even faster performance on GPU machines, processing over six times the rate. When employed in its original form, MobileNet V2 outputs 1,000 classes with associated probabilities based on its training on the ImageNet dataset \cite{howard2017mobilenets,sandler2018mobilenetv2,deng2009imagenet}. After the Flatten operation, the network's first layer produces deep features characterized by a feature size of 1280.

This configuration positions MobileNet V2 as an effective choice for feature extraction, particularly when processing endoscopic images, thanks to its winning combination of accuracy and speed, which is crucial in real-time diagnostic scenarios.

\begin{figure}[ht]
    \centering
    \includegraphics[width=\linewidth]{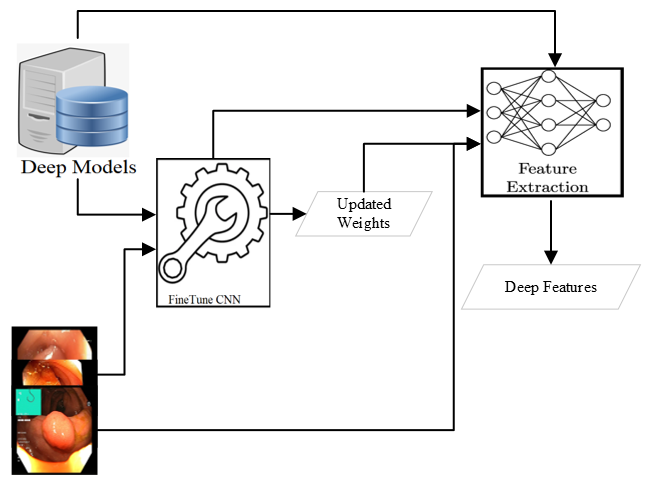}
	\caption{Deep feature extraction methodology.}
	\label{fig:p4-methodology_deep}
\end{figure}

\subsection{Feature Selection}

Feature selection is a crucial step in the classification of endoscopic images, as it determines the most informative and relevant features for accurate detection. After experimenting with various feature combinations and employing a voting classifier, a final set of features has been identified, resulting in optimal detection accuracies. The selected features, which have proven to be most effective for classification, are as follows:

\begin{itemize}
    \item Auto Color Correlogram \cite{lux2008lire}
    \item Color Layout \cite{lux2008lire,sikora2001mpeg,huang1997image}
    \item Edge Histogram \cite{lux2008lire,sikora2001mpeg}
    \item Gabor Features \cite{lux2008lire}
    \item JCD (Joint Collection Distances) \cite{lux2008lire}
    \item Color and Edge Directivity Descriptor \cite{lux2008lire,chatzichristofis2008cedd}
    \item Fuzzy Color and Texture Histogram \cite{lux2008lire,chatzichristofis2008fcth}
    \item Tamura Features (Contrast, Direction etc.) \cite{lux2008lire,tamura1978textural}
    \item Local Binary Patterns (Radius from 1 to 5) \cite{haralick1973textural,ojala1996comparative}
    \item Deep Features \cite{howard2017mobilenets,sandler2018mobilenetv2,harzig2019automatic}
\end{itemize}

In the classification network of the research, neural networks are employed as powerful decision-making tools for image classification. This network consists of two main components: the feature extractor and the decision-making component.

In the previous steps of the approach, deep learning is used as the feature extractor, and a fine-tuned MobileNet V2 is employed to extract deep features. In this step, the decision-making component of the neural network processes the selected set of feature vectors, which have values ranging from 0 to 1. The decision network comprises three layers, aiming for a balance between generalization, detection accuracy, and classification speed.

To introduce non-linearity and improve generalization, the ReLU (Rectified Linear Unit) activation function is utilized in the first and second layers of the neural network. ReLU is defined by the Equation~\ref{eq:relup4}.

\begin{equation}
    \label{eq:relup4}
    Relu(x) = \max(0, x)
\end{equation}

The final output from the third layer is required to be the probability of various classes for each input feature vector. To achieve this, a sigmoid activation function is applied to convert the results into the range of 0 to 1. The sigmoid function is defined by the Equation~\ref{eq:sigmoid}

\begin{equation}
    \label{eq:sigmoid}
    f(x) = \sigma(x) = \frac{1}{1 + e^{-x}}
\end{equation}

The complete forward path of the neural network can be represented by the Equation~\ref{eq:network}.

\begin{equation}
    \label{eq:network}
    Y = \sigma(W_3^T \cdot \text{Relu}(W_2^T \cdot \text{Relu}(W_1^T \cdot X + B_1) + B_2) + B_3)
\end{equation}

In the evaluation of various optimizers for the neural network, the Nadam optimizer was selected due to its effectiveness in adjusting weights and rapid convergence. The Nadam optimizer is defined by the equation provided in the text.

The resulting configuration of the neural network is as follows:
\begin{enumerate}
    \item Layer 1: 64 Fully Connected Neurons with ReLU Activation Function.
    \item Layer 2: 64 Fully Connected Neurons with ReLU Activation Function.
    \item Layer 3: 64 Fully Connected Neurons with Sigmoid Activation Function.
\end{enumerate}

This neural network-based approach achieves an impressive accuracy of 0.99, an F1-score of 0.90, and a Matthews Correlation Coefficient (MCC) of 0.90, all while maintaining a fast detection time of 41 frames per second (FPS). The notable enhancements in F1-score and MCC signify the effectiveness of this approach in image classification tasks.

\subsection{Genetic Algorithm for Threshold Detection}

In your research, a Genetic Algorithm (GA-Boost) is employed for threshold detection to enhance the detection of abnormalities in different classes. The intra-class and inter-class differences are taken into account, recognizing that some classes have higher differences than others. This difference in class separability can impact the detection probabilities achieved by a neural network architecture.

The GA-Boost algorithm works by learning thresholds for each detected class using a genetic algorithm. Initially, random thresholds are chosen from a set of predefined values listed in Equation~\eqref{eq:valrangega}.
\begin{equation}
    \label{eq:valrangega}
    (0.0, 0.1, 0.2, 0.3, 0.4, 0.5, 0.6, 0.7, 0.8, 0.9)
\end{equation}

In GA-Boost, the crossover operator is based on addition with a modulus operation, ensuring that the resulting thresholds remain within the range of 0 to 1. For each pair of original chromosomes ($x$ and $y$), two newly generated chromosomes ($X_n$ and $Y_n$) are created using this crossover operator. Mathematically, the crossover operation is represented by the equation provided in your text.

The GA algorithm operates with a population size of 10 chromosomes, each initialized with random threshold values between 0 and 1. A mutation rate of 20\% is applied, and the algorithm executes for 20 iterations to optimize the threshold values. The F1-score, evaluating the accuracy of abnormality detection using the Decision Tree classifier on the combined feature vector of the Kvasir V2 dataset \cite{pogorelov2018medico}, serves as the evaluation measure for selecting thresholds in the crossover process.

The GA-Boost algorithm significantly enhances accuracy measures, resulting in an F1-score of 0.91. This indicates that optimizing thresholds for different classes through genetic algorithms leads to improved detection of abnormalities.

\section{A Layered Decision Ensembles (TreeNet)}
\label{approach-treenet}
The proposed methodology for medical image analysis is designed to effectively address challenges related to limited data, noisy images, and class imbalance. Drawing insights from a thorough analysis of existing approaches, the system aims to integrate the advantages of Neural Networks, Decision Trees, and ensemble learning. The architecture is meticulously crafted to leverage the strengths of each approach while mitigating their individual limitations.

\subsection{Neural Network Component}

The neural network component of the system is optimized for efficient learning with reduced computational costs. Departing from traditional backpropagation-heavy networks, this architecture adopts a forward-only relationship between layers, eliminating the need for extensive gradient computations. This design choice streamlines the learning process and minimizes computational overhead. Additionally, the system capitalizes on the progressive nature of neural networks by using each layer's results as inputs for subsequent layers, enhancing the model's ability to capture intricate features and relationships within medical images.

\subsection{Decision Trees Integration}

The methodology incorporates decision trees renowned for their effective decision-making capabilities. Acknowledging the bias of decision trees towards major classes, the system introduces tree forests to counteract bias by aggregating decisions from multiple trees. To further enhance decision diversity and counteract bias, ensemble learning is introduced at each layer. Multiple tree ensembles are employed, and their results are amalgamated, ensuring a balanced and robust decision-making process. This strategic combination of decision trees and ensemble learning contributes to the system's efficacy in handling class imbalances and conducting informed medical image analyses.

\subsection{System Architecture}

\begin{figure*}
	\includegraphics[width=1.0\linewidth]{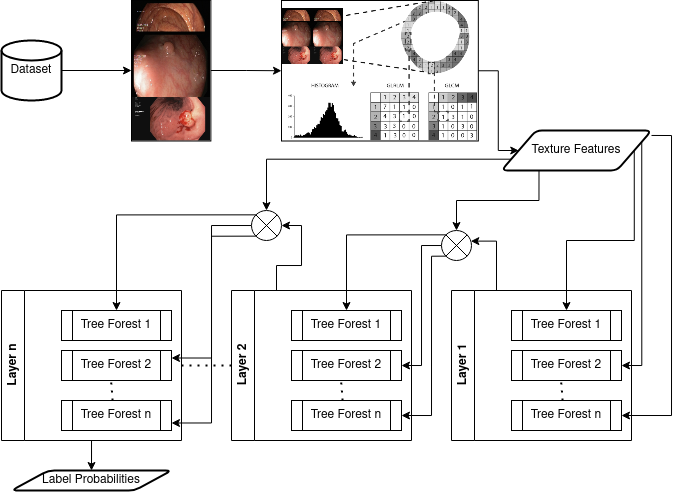}
	\caption{Methodology of TreeNet Architecture.}
	\label{fig:p8-methodology}
\end{figure*}

\begin{figure*}
	\includegraphics[width=1.0\linewidth]{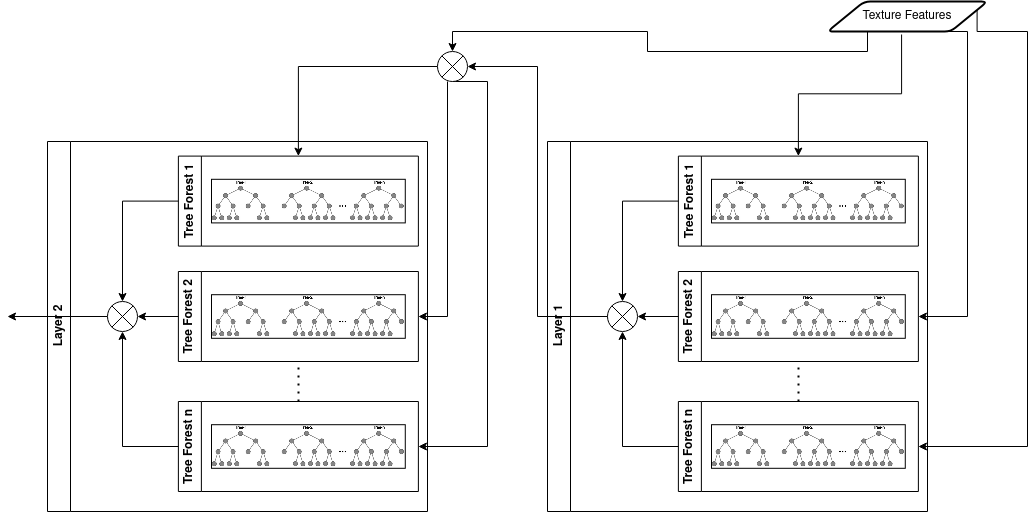}
	\caption{In-site of the layers of TreeNet Model.}
	\label{fig:methodology_layers}
\end{figure*}

The overall architecture of the proposed system is visually represented in Figure~\ref{fig:p8-methodology} and Figure~\ref{fig:methodology_layers}. These figures illustrate the integration of Neural Networks, Decision Trees, and ensemble learning, showcasing the system's unique blend of approaches. This innovative methodology positions the proposed system as a comprehensive solution for medical image analysis under challenging conditions. Through this approach, the system aims to enhance diagnostic accuracy, overcome data limitations, and mitigate the impact of noise, contributing to advancements in the field of medical image analysis.

The system proposed encompasses the features listed in Table~\ref{tab:comparison}.

\footnotesize
\begin{longtable}{|p{0.5cm}|p{1.8cm}|p{1cm}|p{1.2cm}|p{3.5cm}|}
    \caption{Selection of the various characteristics of machine learning deciders}
    \label{tab:comparison} \\
    \hline
    \textbf{Sr. No.} & \textbf{Characteristics} & \textbf{Decider Models} & \textbf{Strength/ Weakness} & \textbf{Details} \\
    \hline
    \endfirsthead

    \hline
    \multicolumn{5}{|c|}{\textbf{Characteristics Selection Continue}} \\
    \hline
    \textbf{Sr. No.} & \textbf{Characteristics} & \textbf{Decider Models} & \textbf{Strength/ Weakness} & \textbf{Details} \\
    \hline
    \endhead

    % Add your content rows here
    \setcounter{customcounter}{1}
    \thecustomcounter \stepcounter{customcounter} & Layered Processing & Neural Network Architecture & Strength & Feature Hierarchy and Abstraction \\  \hline
        \thecustomcounter \stepcounter{customcounter} & Feature Transformation & Neural Network Architecture & Strength &  Feature Hierarchy and Abstraction \\  \hline
        \thecustomcounter \stepcounter{customcounter} & Model Complexity & Neural Network Architecture & Weakness & Affects Inference and Training Time \\  \hline
        \thecustomcounter \stepcounter{customcounter} & Hyper-parameters & Neural Network Architecture & Weakness & High Training time and Error probability based on Hyper Parameters \\  \hline
        \thecustomcounter \stepcounter{customcounter} & Differential Mode (Weight Adjustment) & Neural Network Architecture & Weakness & High Training Time \\  \hline
        \thecustomcounter \stepcounter{customcounter} & Gradient Descent & Neural Network Architecture & Weakness & Local Minima and Plateaus \\  \hline
        \thecustomcounter \stepcounter{customcounter} & Generalization & Ensemble Learning Model & Strength & - \\  \hline
        \thecustomcounter \stepcounter{customcounter} & Diversity-driven Performance & Ensemble Learning Model & Strength & -  \\  \hline
        \thecustomcounter \stepcounter{customcounter} & Input Feature Manipulation & Ensemble Learning Model & Strength & Increases Generalization \\  \hline
        \thecustomcounter \stepcounter{customcounter} & Output Manipulation & Ensemble Learning Model & Strength & Increases Generalization \\  \hline
        \thecustomcounter \stepcounter{customcounter} & Tree-based Decisions & Decision Tree Deciders & Strength & Feature Importance \\  \hline
        \thecustomcounter \stepcounter{customcounter} & Feature Priority Organization & Decision Tree Deciders & Strength & Feature Importance \\  \hline
        \thecustomcounter \stepcounter{customcounter} & Time-Optimal Feature Ignorance & Decision Tree Deciders & Strength & Removal of misleading features \\  \hline
        \thecustomcounter \stepcounter{customcounter} & Interpretability & Decision Tree Deciders & Strength & Transparent Decision-Making \\  \hline
        \thecustomcounter \stepcounter{customcounter} & Non-parametric & Decision Tree Deciders & Strength & Feature Importance based Selection \\  \hline
        \thecustomcounter \stepcounter{customcounter} & Feature Importance based Decision & Decision Tree Deciders & Strength & -  \\  \hline
        \thecustomcounter \stepcounter{customcounter} & Handling Non-linear Relationships & Decision Tree Deciders & Strength & -  \\  \hline
        \thecustomcounter \stepcounter{customcounter} & Overfitting & Decision Tree Deciders & Weakness & A single tree decider is susceptible to overfitting due to its inherent inclination to capture noise and intricacies in the training data, potentially leading to a suboptimal generalization to new and unseen data.\\  \hline
        \thecustomcounter \stepcounter{customcounter} & Instability & Decision Tree Deciders & Weakness & Overfitting \\  \hline
        \thecustomcounter \stepcounter{customcounter} & Bias towards Dominant Classes & Decision Tree Deciders & Weakness &  The purity or information gain is higher when partitioning the major class compared to the minor class.\\  \hline
        \thecustomcounter \stepcounter{customcounter} & Greedy Nature & Decision Tree Deciders & Weakness & A solitary decision tree exhibits a greedy decision-making process that may result in reaching a local optimum. \\  \hline
        \thecustomcounter \stepcounter{customcounter} & Discretization & Decision Tree Deciders & Reduced with ensembles & Possible information Loss in decision trees \\  \hline
        \thecustomcounter \stepcounter{customcounter} & Limited Expressive Power & Decision Tree Deciders & Weakness & Inability to Capture Complex Relationships \\  \hline    

\end{longtable}
\normalsize

\subsection{Proposed Model Characteristics}

Table~\ref{tab:mode_proposed} succinctly outlines key characteristics of the proposed model, emphasizing its fusion of advantages derived from neural networks, ensemble learning, and decision trees. The model strategically incorporates neural network features, leveraging layered processing and feature transformation to optimize learning. Notably, it streamlines the learning process by omitting computationally intensive gradient calculations for weight updates, addressing a significant time-consuming aspect of traditional neural networks.

Moreover, the model harnesses ensemble learning benefits, capitalizing on its capacity for generalization and input-output manipulation. This is particularly valuable in scenarios with class imbalance, where ensemble methods excel in improving decision robustness. Integration of ensemble learning aims to provide more reliable and balanced outcomes, crucial in medical image analysis where class imbalances impact diagnostic accuracy.

The methodology integrates decision tree advantages, including feature prioritization-based decisions, interpretability, and identification of feature relationships. These characteristics enhance the overall decision-making process, allowing the system to interpret and extract meaningful information from intricate medical images. Decision tree features align with the need for transparent and interpretable decision-making in medical contexts, fostering trust among practitioners and facilitating the integration of advanced technologies into clinical workflows.

Table~\ref{tab:mode_proposed} offers a concise overview of the proposed model's characteristics, showcasing its ability to harness the strengths of neural networks, ensemble learning, and decision trees. By strategically combining these elements, the model aims to overcome challenges such as class imbalance, improve decision robustness, and enhance interpretability in the context of complex medical data.

\begin{table}
    \centering
    \caption{Characteristics of Proposed Model}
    \label{tab:mode_proposed}
    \begin{tabular}{|p{0.5cm}|p{5cm}|p{5cm}|}
        \hline
        \textbf{Sr. No.} & \textbf{Characteristics} & \textbf{Decider Models} \\ \hline
        \setcounter{customcounter}{1}
        \thecustomcounter \stepcounter{customcounter} & Layered Processing & Neural Network Architecture \\  \hline
        \thecustomcounter \stepcounter{customcounter} & Feature Transformation & Neural Network Architecture \\  \hline
        \thecustomcounter \stepcounter{customcounter} & Generalization & Ensemble Learning Model \\  \hline
        \thecustomcounter \stepcounter{customcounter} & Diversity-driven Performance & Ensemble Learning Model \\  \hline
        \thecustomcounter \stepcounter{customcounter} & Input Feature Manipulation & Ensemble Learning Model \\  \hline
        \thecustomcounter \stepcounter{customcounter} & Output Manipulation & Ensemble Learning Model \\  \hline
        \thecustomcounter \stepcounter{customcounter} & Tree-based Decisions & Decision Tree Deciders \\  \hline
        \thecustomcounter \stepcounter{customcounter} & Feature Priority Organization & Decision Tree Deciders \\  \hline
        \thecustomcounter \stepcounter{customcounter} & Time-Optimal Feature Ignorance & Decision Tree Deciders \\  \hline
        \thecustomcounter \stepcounter{customcounter} & Interpretability & Decision Tree Deciders \\  \hline
        \thecustomcounter \stepcounter{customcounter} & Non-parametric & Decision Tree Deciders \\  \hline
        \thecustomcounter \stepcounter{customcounter} & Feature Importance based Decision & Decision Tree Deciders \\  \hline
        \thecustomcounter \stepcounter{customcounter} & Handling Non-linear Relationships & Decision Tree Deciders \\  \hline
    \end{tabular}
\end{table}

\section{Segmentation using Depth-Wise Separable Convolution}

The approach employed in this study closely aligns with the methodology originally proposed by Jha et al. in their work, as detailed in the reference \cite{jha2019resunet++}. Central to this approach is the utilization of the ResUNet++ architecture, a sophisticated framework designed for semantic segmentation tasks. The ResUNet++ architecture is characterized by its incorporation of an encoder and decoder structure, effectively facilitating the segmentation process. To bridge the encoder and decoder blocks, a pyramid pooling mechanism was adopted, while the encoder block itself features residual units that leverage skip connections within the neural network. These skip connections play a crucial role in enabling the training of deep neural networks without any adverse impact on performance. In addition, the utilization of Squeeze and Excitation blocks, a technique introduced by Hu et al. \cite{hu2018squeeze}, further enhances the framework by ensuring equitable weighting of channel output features. Furthermore, an attention mechanism was introduced within the decoder block. This attention mechanism, inspired by its successful application in natural language processing (NLP), aims to create pixel-wise predictions. Just as attention is directed to each word in a sentence in NLP, in semantic segmentation, this mechanism is harnessed to allocate focus to every pixel in an image, enabling pixel-level predictions, as highlighted by Huang et al. \cite{huang2019ccnet}.

In the realm of semantic segmentation, the adaptation of the attention mechanism stands as a fundamental step towards achieving high-precision pixel-level predictions. By allocating attention to individual pixels within an image, akin to the attention granted to individual words in a sentence in natural language processing (NLP), this approach capitalizes on a well-established concept to enhance the accuracy of semantic segmentation. This intricate architecture, as proposed in reference \cite{jha2019resunet++}, has demonstrated its effectiveness in various computer vision tasks and continues to be a significant asset in the domain of image analysis. The inclusion of the ResUNet++ framework, pyramid pooling, residual units, Squeeze and Excitation blocks, and the strategic application of the attention mechanism collectively contribute to the robustness and precision of the segmentation process. The noteworthy synergy of these elements signifies the thoughtful integration of cutting-edge techniques to address the challenges inherent to semantic segmentation in computer vision \cite{hu2018squeeze,huang2019ccnet}.

\tikzset{every picture/.style={line width=1 pt}} %set default line width to 0.75pt

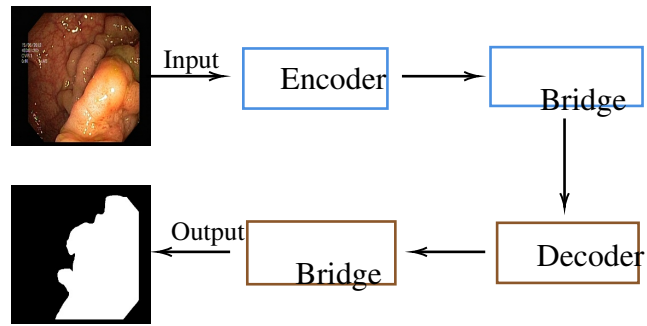
\begin{figure}[!h]
\centering

\begin{tikzpicture}[x=0.5pt,y=0.5pt,yscale=-1,xscale=1]
%uncomment if require: \path (0,300); %set diagram left start at 0, and has height of 300
\centering
%Image [id:dp136730669692412]
\draw (98,70) node  {\includegraphics[width=52.5pt,height=52.5pt]{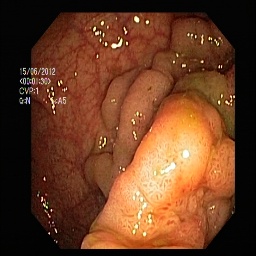}};
%Image [id:dp2492542067576411]
\draw (98,205) node  {\includegraphics[width=52.5pt,height=52.5pt]{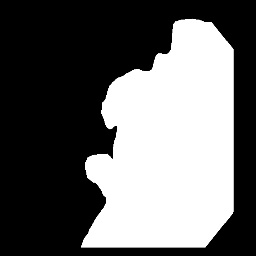}};
%Shape: Rectangle [id:dp5674526950445551]
\draw  [color={rgb, 255:red, 74; green, 144; blue, 226 }  ,draw opacity=1 ] (221,49) -- (330,49) -- (330,94.5) -- (221,94.5) -- cycle ;
%Shape: Rectangle [id:dp3201150951232319]
\draw  [color={rgb, 255:red, 74; green, 144; blue, 226 }  ,draw opacity=1 ] (408,48) -- (523,48) -- (523,92.5) -- (408,92.5) -- cycle ;
%Shape: Rectangle [id:dp13998836234493894]
\draw  [color={rgb, 255:red, 139; green, 87; blue, 42 }  ,draw opacity=1 ] (224,180) -- (336,180) -- (336,227.5) -- (224,227.5) -- cycle ;
%Shape: Rectangle [id:dp6655591393736568]
\draw  [color={rgb, 255:red, 139; green, 87; blue, 42 }  ,draw opacity=1 ] (413,181) -- (520,181) -- (520,226.5) -- (413,226.5) -- cycle ;
%Straight Lines [id:da46194281676248683]
\draw    (148,70) -- (204,70) ;
\draw [shift={(206,70)}, rotate = 180] [color={rgb, 255:red, 0; green, 0; blue, 0 }  ][line width=0.75]    (10.93,-3.29) .. controls (6.95,-1.4) and (3.31,-0.3) .. (0,0) .. controls (3.31,0.3) and (6.95,1.4) .. (10.93,3.29)   ;
%Straight Lines [id:da4119951631324619]
\draw    (341,70) -- (397,70) ;
\draw [shift={(399,70)}, rotate = 180] [color={rgb, 255:red, 0; green, 0; blue, 0 }  ][line width=0.75]    (10.93,-3.29) .. controls (6.95,-1.4) and (3.31,-0.3) .. (0,0) .. controls (3.31,0.3) and (6.95,1.4) .. (10.93,3.29)   ;
%Straight Lines [id:da2772802070552487]
\draw    (211,203) -- (158,203) ;
\draw [shift={(156,203)}, rotate = 360] [color={rgb, 255:red, 0; green, 0; blue, 0 }  ][line width=0.75]    (10.93,-3.29) .. controls (6.95,-1.4) and (3.31,-0.3) .. (0,0) .. controls (3.31,0.3) and (6.95,1.4) .. (10.93,3.29)   ;
%Straight Lines [id:da8115835906603084]
\draw    (404,203) -- (351,203) ;
\draw [shift={(349,203)}, rotate = 360] [color={rgb, 255:red, 0; green, 0; blue, 0 }  ][line width=0.75]    (10.93,-3.29) .. controls (6.95,-1.4) and (3.31,-0.3) .. (0,0) .. controls (3.31,0.3) and (6.95,1.4) .. (10.93,3.29)   ;
%Straight Lines [id:da27624544077878155]
\draw    (464,102) -- (464,167.5) ;
\draw [shift={(464,169.5)}, rotate = 270] [color={rgb, 255:red, 0; green, 0; blue, 0 }  ][line width=0.75]    (10.93,-3.29) .. controls (6.95,-1.4) and (3.31,-0.3) .. (0,0) .. controls (3.31,0.3) and (6.95,1.4) .. (10.93,3.29)   ;

% Text Node
\draw (246,61) node [anchor=north west][inner sep=0.75pt]   [align=left] {\begin{minipage}[lt]{40.715pt}\setlength\topsep{0pt}
\begin{center}
Encoder
\end{center}

\end{minipage}};
% Text Node
\draw (443,64) node [anchor=north west][inner sep=0.75pt]   [align=left] {\begin{minipage}[lt]{32.204392000000006pt}\setlength\topsep{0pt}
\begin{center}
Bridge
\end{center}

\end{minipage}};
% Text Node
\draw (258,196) node [anchor=north west][inner sep=0.75pt]   [align=left] {\begin{minipage}[lt]{32.204392000000006pt}\setlength\topsep{0pt}
\begin{center}
Bridge
\end{center}

\end{minipage}};
% Text Node
\draw (440,195) node [anchor=north west][inner sep=0.75pt]   [align=left] {\begin{minipage}[lt]{41.278108pt}\setlength\topsep{0pt}
\begin{center}
Decoder
\end{center}

\end{minipage}};
% Text Node
\draw (158,48) node [anchor=north west][inner sep=0.75pt]   [align=left] {\footnotesize Input};
% Text Node
\draw (164,178) node [anchor=north west][inner sep=0.75pt]   [align=left] {\footnotesize Output};

\end{tikzpicture}
\caption{Process Flow}\label{Process Flow}
\end{figure}

A pivotal architectural feature employed in this study lies in the incorporation of a pyramid pooling bridge, strategically positioned between the encoder and decoder blocks\cite{chen2018encoder} \cite{chen2017deeplab}. Notably, this bridge leverages atrous convolution, serving as a conduit for the transformation of encoder output across distinct receptive fields. The atrous convolution process entails the convolution of features with kernels of varying dilation rates, and subsequently culminates in a concatenation of the outcomes from all these convolutions. This innovative approach effectively captures contextual information within the feature set, spanning multiple scales.

In the context of the ResUNet++ framework, the Atrous Spatial Pyramid Pooling (ASPP) block takes center stage. It is worth noting that this ASPP block exhibits two alternative configurations, one incorporating depth-wise separable convolution, and the other substituting it with the Deep Atrous Spatial Pyramid Pooling (DASPP) module, as introduced by Emara in 2019. The depth-wise separable convolution, a key element, is executed by applying kernels at the channel level, followed by passage through pointwise convolution utilizing a 1x1 kernel, an approach derived from Chollet (2017) \cite{chollet2017xception}. This strategic use of depth-wise convolution reduces the computational load in terms of GFLOPs and minimizes the number of model parameters. A separate experiment was conducted, implementing DASPP to assess whether deeper network architecture yields performance enhancements in the domain of polyp segmentation. Three distinct architectural variants were introduced for this evaluation:
\begin{enumerate}
    \item $sepv-conv-resunet++$, wherein the ASPP module from ResUNet++ was replaced with depth-wise separable convolution;
    \item $dsapp-resunet++$, featuring the replacement of the ASPP module with the DASPP module from Emara (2019) \cite{emara2019liteseg}, marked as (2) in subsequent discussions
    \item $dsapp-relu-resunet++$, a refinement of (2) with the incorporation of ReLu activation functions \cite{emara2019liteseg}.
\end{enumerate}

Semantic segmentation, in contrast to object detection, can be construed as a pixel-wise classification problem, where the primary objective is to classify each pixel within an image. In the specific context of polyp segmentation, the output for a pixel assumes one of two values: 0 or 1, indicating the pixel's class identity. Evaluation metrics conventionally adopted for semantic segmentation encompass accuracy, precision, recall, mean Intersection over Union (mIoU), and dice coefficient. In this study, all these metrics, with the exception of accuracy, were systematically employed to gauge model performance. Furthermore, a custom loss function tailored for the mIoU metric was devised and used as a basis for training all architectural configurations under investigation.

\section{Voting Neural Network (VNN) for Endoscopic Image Segmentation}

The research methodology comprises two main components: data preparation and image segmentation. The initial phase, data preparation, involves addressing challenges inherent in endoscopic data, such as image quality issues due to light reflections and class imbalance arising from limited data availability for certain abnormalities. The subsequent phase focuses on image segmentation.
\subsection{Data Preprocessing}
\textbf{Image Quality Issues:}
Reflection Removal: Light reflections in endoscopic images can adversely affect the visibility of abnormalities, particularly polyps. The removal process involves two steps: reflection detection and subsequent removal. Various approaches, including deep learning-based methods, are available for reflection detection and removal. Notably, unsupervised exemplar-based texture synthesis proves effective in handling light reflections.
Reflection Detection: Sudden changes in color channel values are observed to detect light reflections, as depicted in Figure 1.

\textbf{Class Imbalance:}
Data Augmentation: Due to the scarcity of segmented images in endoscopic datasets, underfitting can occur in robust models. Data augmentation is employed to address this issue, involving techniques such as random rotations, flips, crops, resizing, and noise addition. This time-efficient approach has shown promising results in deep learning and transfer learning applications.

GAN-Based Augmentation: An alternative augmentation strategy involves using Generative Adversarial Networks (GANs) with attention. However, in this research, it is noted that GAN-based augmentation did not contribute to an improvement in segmentation accuracy within the context of transfer learning approaches.

\subsection{Image Segmentation}
The second major part of the methodology involves the actual image segmentation process, where the goal is to precisely identify and delineate regions of interest, particularly focusing on polyp segmentation.

Motivated by the success of the UNet architecture in pixel-wise semantic polyp segmentation, the conventional UNet architecture is improved by integrating features from the ResUNet++ architecture. The UNet architecture, characterized by an encoder-decoder design, yields impressive results but lacks certainty and accuracy in regions with small polyp areas due to missing low-level features during down-sampling.

To enhance the UNet model, Residual blocks from the ResUNet++ architecture are incorporated. ResUNet++ is a deeply-supervised encoder-decoder network connected through Atrous Spatial Pyramid Pooling (ASSP), serving as a bridge. This design overcomes degradation problems by using skip-connections to propagate information without loss, resulting in a lighter model with fewer trainable parameters and improved segmentation performance.
\subsubsection{Architectural Components}
\textbf{Residual Blocks:} \cite{aslanzadeh2017efficient}:
        Residual blocks are added to the backbone (ResUNet) architecture to address degradation problems and enhance model performance.
        
\textbf{Squeeze and Excitation Blocks:} \cite{jha2019resunet++}:
        These blocks increase the model's sensitivity by performing dynamic channel-wise calibration. The squeeze block compresses channels using global average pooling, while the excitation block captures channel dependencies. Stacking these blocks with the residual block enhances generalization across different data and improves overall network performance.

\textbf{ASPP Module:} \cite{jha2019resunet++}:
        Applied in semantic segmentation tasks, the Atrous Spatial Pyramid Pooling (ASPP) module resamples features at multiple scales. This module captures features and useful image context at various scales, acting as a bridge in the ResUNet++ architecture between the encoder and decoder. It helps capture multi-scale information for better segmentation.

\textbf{Attention Mechanism:} \cite{jha2019resunet++}:
        Inspired by Natural Language Processing (NLP), attention mechanisms are introduced in the decoder part of the architecture. These mechanisms enable the network to dynamically highlight relevant features in input data, guiding the network to focus on specific areas. This enhances the quality of features in the neural network and improves overall results.

\subsubsection{Ensemble Learning}

Recognizing that different neural network architectures may excel on specific image sets, an ensemble learning technique is employed. The outputs of the top three neural networks are combined using majority voting. This technique involves pixel-wise voting, expressed in a grayscale image where a white (255) value represents segmentation.

\subsubsection{Research Approach}

The research leverages these architectural enhancements and ensemble learning techniques for improved polyp segmentation. The combination of UNet and ResUNet++ features, along with attention mechanisms and ensemble learning, contributes to the development of a robust model for pixel-wise semantic polyp segmentation.

\chapter{Results and Discussions}
\label{sec:resultsall}
This chapter delves into the results obtained by applying the methodologies discussed in the previous chapter. It presents a detailed analysis of the evaluation methodologies employed and the corresponding outcomes. The chapter fosters a deeper understanding by not only presenting the results but also delving into the potential explanations and justifications behind them. This critical exploration allows for a more nuanced interpretation of the findings and their significance.

The breakdown of the chapter focuses on two key areas of Evaluation Methodologies and Results and Discussion.

The Evaluation Methodologies section meticulously outlines the specific techniques and metrics used to assess the performance of the proposed methodologies. These evaluation methods provide a standardized framework for measuring the effectiveness of the approaches in detecting GI-tract abnormalities.

The section of Results and Discussion explains the actual results achieved. A comprehensive discussion then ensues, exploring the potential reasons behind the observed outcomes. This discussion might consider factors such as the strengths and weaknesses of the employed methodologies, any unexpected findings, and how the results align with existing knowledge in the field. By critically examining the results, the chapter aims to provide valuable insights into the efficacy of the proposed approaches for GI-tract image analysis.

\section{Majority Voting of Heterogeneous Classifiers}
\subsection{Experimental Setup}
\label{sec:experimental-setup-p12}
The methodologies expounded upon in the antecedent sections are subjected to evaluation utilizing the Kvasir V1 dataset. The heterogeneous module necessitated the adoption of diverse feature classification approaches.

The examination delves into the application of various machine learning models, notably linear regression, extremely randomized trees, and random forest, within the Python ecosystem's scikit-learn package. The primary objective is to leverage these models for the training of logistic regression, random forests, and extremely random trees, incorporating both deep and global features.

The inaugural phase of our inquiry entails the scrutiny of model performance on the training data. To ensure a robust evaluation, a 10-fold cross-validation strategy is employed. The preliminary findings are noteworthy, revealing an accuracy rate of 0.97, an F1 score of 0.90, and a Matthews Correlation Coefficient (MCC) of 0.81 attained through the 10-fold cross-validation of the training dataset. These metrics collectively attest to the promising efficacy of the proposed model.

The diverse configurations of the executions manifest certain advantages across various evaluation measures. For the definitive selection of the optimal setup, all six evaluation configurations, comprising three runs designed for expeditious result generation concerning speed and an additional three runs tailored explicitly for accuracy evaluation, undergo scrutiny on the test dataset. The enumerated evaluation setups are as follows:

\subsubsection{Ensemble on 60 Images}
In this initial iteration, we investigate the potential of an ensemble model incorporating seven distinct features, namely JCD, Tamura, Color Layout, Edge Histogram, Auto Color Correlogram, PHOG, and VGG features. These features are employed for training on a subset of 60 images. The classification process involves a voting mechanism that amalgamates the predictive capabilities of logistic regression, random forest, and extremely randomized trees algorithms.

\subsubsection{Ensemble on 300 Images}
Building upon the groundwork laid in Run 1, this iteration maintains the same ensemble of seven features but expands the training dataset to 300 images. Analogous to Run 1, logistic regression, random forest, and extremely randomized trees continue to contribute to the classification process, guided by the ensemble's voting mechanism.

\subsubsection{Ensemble on 5293 Images}
This iteration extends our exploration to a more extensive training dataset comprising 5293 images. As in the preceding runs, the ensemble utilizes the seven features and relies on logistic regression, random forest, and extremely randomized trees algorithms. The primary objective is to evaluate the model's performance under the influence of this comprehensive dataset.

\subsubsection{Selected Features 60 Images}
In contrast to the preceding iterations, Run 4 focuses on a refined ensemble model comprising six features—JCD, Tamura, Color Layout, Edge Histogram, Auto Color Correlogram, and PHOG. These features form the basis of a classification system employing logistic regression, random forest, and extremely randomized trees, guided by a voting mechanism. The training dataset remains limited to 60 images.

\subsubsection{Selected Features 300 Images}
Parallel to Run 4, this iteration maintains the same ensemble of six features but shifts its focus to a larger training dataset, encompassing 300 images. The use of logistic regression, random forest, and extremely randomized trees, along with the ensemble's voting strategy, remains consistent with our methodology.

\subsubsection{Selected Features All Images}
The concluding iteration mirrors Run 4 in terms of its selection of ensemble features, confined to the six aforementioned descriptors. However, it is distinguished by a significant increase in the training dataset, comprising 5293 images. The classification algorithms—logistic regression, random forest, and extremely randomized trees—remain consistent, with the ensemble's voting mechanism providing a uniform framework for classification.

These subsequent configurations have offered a more profound insight into the performance of the chosen machine learning models across diverse scenarios. This contributes significantly to a comprehensive evaluation of their capabilities and suitability for the designated task. Our overarching objective is to derive valuable insights into both the strengths and limitations of these models, facilitating well-informed decisions regarding their application within our specific context.

\subsection{Results and Analysis}
\label{sec:resultsp1}

Each setup was meticulously crafted to provide comprehensive insights into the performance of our ensemble models, systematically exploring the impact of varying dataset sizes, classifiers, and feature selections on the predictive accuracy and efficiency of logistic regression, random forest, and extremely randomized trees algorithms. Through the systematic evaluation of these configurations, our objective is to derive informed conclusions regarding the most suitable model setup for our specific research objectives.

A succinct summary of the evaluation criteria for the best-performing setup is presented in Table~\ref{tab:runs}. Notably, this setup exhibits an impressive accuracy of 0.979, complemented by an F-score of 0.75 and a Matthews Correlation Coefficient (MCC) of 0.76. The intriguing aspect of this observation is that the most successful setup exclusively utilizes global features, eschewing the incorporation of deep learning features. This noteworthy finding prompts further investigation into why deep features have demonstrated suboptimal performance, a matter we plan to delve into in our forthcoming research endeavors.

Our preliminary analysis has identified a noteworthy misclassification trend, specifically the misclassification of a substantial number of samples originally belonging to the "ulcerative-colitis" class as "esophagitis" when deep features are employed. This puzzling trend is likely to be a key area of focus in our upcoming research investigations.

The best-performing setup, denoted as \textbf{Selected Features 5293 Images}, utilizes all 5293 available images. This approach revolves around an ensemble of six distinct features, namely JCD, Tamara, Edge Histograms, Color Layout, Auto Color Correlogram, and PHOG. In the classification process, logistic regression, random forest, and extremely randomized trees serve as classifiers, guided by a weighted majority voting scheme in their decision-making.

\begin{table}
  \caption{Accuracy, F1, and MCC on different execution setups of the testing}
  \label{tab:runs}
  \centering
  \begin{tabular}{|c|c|c|l|}
  \hline
    &Accuracy&F1&MCC\\ \hline
    Ensemble on 60 Images&0.956&0.625&0.614\\ \hline
    Ensemble on 300 Images&0.957&0.587&0.603\\ \hline
    Ensemble on 5293 Images&0.954&0.549&0.572\\ \hline
    Selected Features 60 Images&0.961&0.611&0.597\\ \hline
    Selected Features 300 Images&0.976&0.745&0.741\\ \hline
    Selected Features 5293 Images&\textbf{0.979}&\textbf{0.752}&\textbf{0.756}\\ \hline
    Homogenous Approach&0.96&0.60&0.57\\ \hline
\end{tabular}
\end{table}

\section{Ensemble of Homogeneous Classifiers}
The researchers implemented a uniform strategy on the Kvasir dataset, dividing it into 70\% for training and 30\% for testing. To evaluate the performance during training, they calculated F1-scores, which measure a model's balance between precision and recall. The F1-score achieved by the majority voting approach was 0.88, while the neural network approach achieved a higher score of 0.92. Additionally, a combined approach yielded an accuracy exceeding 0.92, suggesting promising performance on the training data.

However, when applying this same methodology to the unseen test data provided by the competition organizers, the results were less impressive. The best run achieved an F1-score of 0.60, indicating a significant drop in performance compared to the training data. The Matthews Correlation Coefficient (MCC), another metric that considers true positives, negatives, and false classifications, was also calculated and reached a value of 0.57. These results are summarized in Table~\ref{tbl:accuracy}.

\begin{table}
\centering
\caption{Accuracy Measures on Test Data}
\label{tbl:accuracy}
\begin{tabular}{|p{3cm}|p{5cm}|p{3cm}|}
\hline
\textbf{Sr. No.} & \textbf{Measures} & \textbf{Value} \\ \hline
1 & True Positives & 435 \\ \hline
2 & True Negatives & 15576 \\ \hline
3 & False Positives & 286 \\ \hline
4 & False Negatives & 286 \\ \hline
5 & Precision & 0.60 \\ \hline
6 & Recall/Sensitivity & 0.60 \\ \hline
7 & Specificity & 0.98 \\ \hline
8 & F1-Score & 0.60 \\ \hline
9 & MCC & 0.57 \\ \hline
\end{tabular}
\end{table}

Table~\ref{tab:confall} provides a comprehensive overview of the confusion matrix across various classes. A thorough analysis of this matrix reveals approximately 1469 samples that have been misclassified. A closer examination indicates that the misclassification primarily arises from two categories: "dyed-lifted-polyps" and "dyed-resection-margins," with approximately 500 samples erroneously categorized (as referenced in Tables~\ref{tab:conf_dyed} and~\ref{tab:conf_ndyed}). These findings underscore the importance of addressing the challenges associated with these categories and devising strategies to improve the overall performance.

\begin{table}
  \caption{Confusion matrix of all classes. There are total 16 classes and summary of all classes is shown}
  \label{tab:confall}
  \centering
  \begin{tabular}{|c|c|l|}  \hline
    Predicted Actual&All&non-All\\ \hline
    ALL&7271&1469\\ \hline
    non-All&1469&129631\\ \hline
\end{tabular}
\end{table}

\begin{table}
  \caption{Confusion matrix for class dyed-lifted-polyps versus non dyed-lifted-polyps. df=dyed-lifted.}
  \label{tab:conf_dyed}
  \centering
  \begin{tabular}{|c|c|l|}  \hline
    Predicted Actual&df-polyps &non df-polyps\\ \hline
    dyed-lifted-polyps&339&236\\ \hline
    non-dyed-lifted-polyps&217&7948\\ \hline
\end{tabular}
\end{table}

\begin{table}
  \caption{Confusion matrix for class dyed-resection-polyps versus non dyed-resection-polyps. dr=dyed-resection.}
  \label{tab:conf_ndyed}
  \centering
  \begin{tabular}{|c|c|l|} \hline
    Predicted class Actual class& dr-margins & non-dr-margins \\ \hline
    dyed-resection-margins &387&232\\ \hline
    non-dyed-resection-margins &177&7944\\ \hline
\end{tabular}
\end{table}

Our forthcoming research endeavors are poised to explore local features to enhance the system's performance, particularly in these challenging classes. We believe that such efforts will not only rectify the existing misclassification issues but also contribute to the overall refinement and robustness of the classification system.

\section{Real time abnormalities detection}
\label{p4-results}

\subsection{Datasets Used}
\label{datasets}

In the realm of medical image analysis, particularly in the domain of endoscopic or colonoscopic examinations, there exists a wealth of datasets meticulously curated and annotated for the precise identification of anomalies and irregularities. A primary objective in this sphere of research is the accurate detection of conditions such as polyps or lesions during endoscopy. This endeavor revolves around the utilization of specialized datasets designed for detection tasks, which are expounded upon below.

The Kvasir V1 dataset \cite{pogorelov2017kvasir}, a foundational resource in this context, encompasses a total of 8,000 images, evenly distributed into 4,000 for training and 4,000 for testing purposes. These images exhibit varying dimensions, ranging from $720 \times 576$ to $1920 \times 1072$ pixels. The dataset comprises eight distinct classes representing gastrointestinal (GI) tract abnormalities, including anatomical landmarks, pathological and normal findings, as well as endoscopic procedures. Notably, each class is represented by an equal number of image samples, ensuring balanced representation. The specific classes include dyed-lifted-polyps, dyed-resection-margins, esophagitis, normal-cecum, normal-pylorus, normal-z-line, polyps, and ulcerative-colitis.

In a similar vein, the Kvasir V2 dataset \cite{pogorelov2018medico} emerges as a more extensive repository, encompassing a grand total of 14,033 images, divided into 5,293 training images and 8,741 testing images. The images within this dataset mirror the size variability found in Kvasir V1, spanning dimensions from $720 \times 576$ to $1920 \times 1072$ pixels. Kvasir V2 goes a step further by incorporating 16 distinct classes representing a gamut of GI tract abnormalities and landmarks. However, this dataset exhibits class imbalance, with some categories having as few as nine samples, while others boast 2,331 samples. The complexity and diversity of classes within Kvasir V2 are further elucidated in the associated literature.

The Hyper Kvasir Classification dataset \cite{borgli2020hyperkvasir} assumes prominence with its inclusion of 110,079 images and 374 videos, thereby enhancing the depth and breadth of available resources for research and experimentation. A subset of this extensive dataset, intended for challenge usage, features 10,662 training images spanning 23 diverse classes. Moreover, it introduces an additional 721 labeled images, thereby augmenting the repository for anatomical landmarks, pathological findings, normal findings, and endoscopic procedures. This dataset stands as an illustrative example of the rich resource pool available to researchers engaged in the challenging endeavor of automated medical image analysis.

Furthermore, the DowPK dataset \cite{zeshan2023dowpk}, hailing from Dow Hospital in Pakistan, contributes 841 images depicting a spectrum of abnormalities discernible in endoscopic and colonoscopic examinations. These images originate from various patients, with an average of four images procured from each individual. The dataset benefits from the expert annotations of a group of 8 MBBS final year students, with discrepancies in annotations resolved through majority voting. Of particular note is the inclusion of images depicting different classes, as determined by the amalgamated consensus of medical experts. This dataset is exemplified by the visual representation of select images, further accentuating its diagnostic utility. The class distribution of the DowPK dataset is thoughtfully laid out, aligning with its objective of supporting medical image analysis endeavors.

\subsection{Evaluation Measures}
\label{evaluation}

The assessment of image segmentation methodologies necessitates the utilization of diverse evaluation measures to gauge their effectiveness and precision. In this context, a comprehensive array of evaluation metrics were employed to appraise the performance of the segmentation methodology. These evaluation measures include but are not limited to Accuracy, F1-Score, Matthews Correlation Coefficient (MCC), Sensitivity, Specificity, and the Area under the receiver operating characteristic curve (AUC-ROC).

Accuracy, a fundamental evaluation metric, quantifies the ratio of correctly classified instances to the total number of instances within the given task. The methodology under consideration demonstrated an impressive accuracy of 0.99 across various datasets, indicative of its prowess in making accurate classifications. The F1-Score, a measure that encompasses both precision and recall, was adeptly employed to assess the proposed approach's performance on the Kvasir v2 and DowPK datasets, yielding F1-Score values of 0.91 and 0.93, respectively.

The Matthews Correlation Coefficient (MCC) served as a robust alternative to evaluate classification accuracy, addressing certain limitations inherent in traditional measures. By considering the difference between true and false predictions while accounting for the ratio of true predictions to all values, MCC effectively gauges the model's classification performance. Furthermore, Sensitivity, also known as recall or the true positive rate, played a crucial role in assessing the model's capability to correctly identify positive instances. The DowPK dataset showcased an impressive sensitivity value of 0.93, indicative of its remarkable precision in identifying positive outcomes. Specificity, the counterpart to Sensitivity, underscored the model's proficiency in recognizing negative instances within the DowPK dataset, boasting a specificity value of 0.93.

Lastly, the AUC-ROC metric was strategically employed to evaluate the model's ability to discriminate between positive and negative instances. The Kvasir V2 and Hyper Kvasir datasets demonstrated remarkable AUC-ROC values of 0.99, emphasizing the model's proficiency in accurately distinguishing between classes in these datasets. This comprehensive set of evaluation metrics ensures a thorough assessment of the model's performance across various aspects of classification accuracy.

\subsection{Compared Methods}
In the quest for robust and effective solutions within the domain of medical image analysis, it is imperative to conduct an in-depth evaluation and comparison of various methodologies and techniques. This section is dedicated to the systematic examination and critical assessment of the methods that have been employed in the context of endoscopic and colonoscopic abnormalities detection. By rigorously scrutinizing these compared methods, researchers and practitioners can discern the strengths, limitations, and nuances inherent in each approach. A comparison of the various approaches is shown in Tables~\ref{tbl:apprachesv1},~\ref{tbl:apprachesv2}, and ~\ref{tbl:apprachesv3}.

\begin{table}[ht]
	\centering
	\caption{Analysis of some of the approaches for classification of abnormalities and landmarks detection on the Kvasir V1 dataset \cite{pogorelov2017kvasir}.}
	\label{tbl:apprachesv1}
	\begin{tabular}
		{|p{0.5cm}|p{2.5cm}|p{1cm}|p{1.5cm}|p{0.8cm}|p{0.7cm}|}
		\hline
		\textbf{Sr. No.} & \textbf{Approach} & \textbf{Acc.} & \textbf{F1-score} & \textbf{MCC} & \textbf{FPS}\\ \hline
		 1 & HKBU17 \cite{liu2017hkbu,jha2021comprehensive}  & 0.93 & 0.70 & 0.66 & 2 \\ \hline
		 2 & Ensemble17 \cite{naqvi2017ensemble,jha2021comprehensive}  & 0.94 & 0.77 & 0.74 & 2 \\ \hline
		 3 & Inception17 \cite{petscharnig2017inception}  & 0.94 & 0.76 & 0.72 & 1 \\ \hline
		 4 & SCL-UMD17 \cite{agrawal2017scl}  & 0.96 & 0.85 & 0.83 & 1 \\ \hline
		 5 & DLGF17 \cite{pogorelov2017comparison,jha2021comprehensive}  & 0.96 & 0.83 & 0.79 & 46 \\ \hline
		 6 & Lire-CNN & 0.99 & 0.90 & 0.90 & 41 \\ \hline
	\end{tabular}
\end{table}

\begin{table}[ht]
	\centering
	\caption{Analysis of various approaches for classification of abnormalities and landmarks detection on the Kvasir V2 dataset \cite{pogorelov2018medico}.}
	\label{tbl:apprachesv2}
	\begin{tabular}
		{|p{0.5cm}|p{2.5cm}|p{1cm}|p{1.5cm}|p{0.8cm}|p{0.7cm}|}
		\hline
		\textbf{Sr. No.} & \textbf{Approach} & \textbf{Acc.} & \textbf{F1-score} & \textbf{MCC} & \textbf{FPS}\\ \hline
		 1 & EEIC19 \cite{hoang2019enhancing}  & 0.99 & 0.95 & 0.94 & 23 \\ \hline
		 1 & EEIC19 Fast \cite{hoang2019enhancing}  & 0.99 & 0.88 & 0.89 & 3 \\ \hline
		 2 & TLCSS18 \cite{dias2018transfer,jha2021comprehensive}  & 0.98 & 0.87 & 0.89 & 3 \\ \hline
		 2 & TLCSS18 Fast \cite{dias2018transfer,jha2021comprehensive}  & 0.99 & 0.91 & 0.90 & 27 \\ \hline
		 3 & FRCSS18 \cite{hoang2018application,jha2021comprehensive}  & 0.99 & 0.94 & 0.94 & 23 \\ \hline
		 4 & DSTL18 \cite{hicks2018deep,jha2021comprehensive}  & 0.99 & 0.89 & 0.89 & 1015 \\ \hline
		 5 & WDE18 \cite{ko2018weighted}  & 0.95 & 0.48 & 0.54 & 3744 \\ \hline
		 6 & MVHC18 \cite{khan2018majority}  & 0.98 & 0.75 & 0.81 & 43328 \\ \hline
		 7 & Lire-CNN & 0.99 & 0.90 & 0.90 & 41 \\ \hline
	\end{tabular}
\end{table}

\begin{table}[ht]
	\centering
	\caption{Analysis of several approaches for classification of abnormalities and landmarks detection on the Kvasir V3 (Hyper Kvasir) dataset \cite{borgli2020hyperkvasir}.}
	\label{tbl:apprachesv3}
	\begin{tabular}
		{|p{0.5cm}|p{2.5cm}|p{1cm}|p{1.5cm}|p{0.8cm}|p{0.7cm}|}
		\hline
		\textbf{Sr. No.} & \textbf{Approach} & \textbf{Acc.} & \textbf{F1-score} & \textbf{MCC} & \textbf{FPS}\\ \hline
		 1 & HMTA \cite{galdran2021hierarchical}  & 0.99 & 0.87 & 0.86 & 20 \\ \hline
		 2 & HLNN \cite{he2021hybrid}  & 0.98 & 0.90 & 0.89  & 129 \\ \hline
		 3 & EDTDE \cite{dutta2021efficient}  & 0.93 & 0.76 & 0.76 & 49 \\ \hline
		 4 & Lire-CNN & 0.99 & 0.91 & 0.91 & 41 \\ \hline
	\end{tabular}
\end{table}

\subsection{Impact of Various Stages}
\label{p4-Results}
Within this section, we provide a comprehensive presentation of the outcomes stemming from the sequence of steps employed in our approach. These steps are meticulously assessed, not only for their individual contributions but also for their collective impact on the overarching objective of enhancing the accuracy and efficiency of detection processes. It is paramount to emphasize that each step is characterized by a diverse array of potential options, which have been exhaustively surveyed in the preceding theoretical evaluation, as detailed in Section \ref{approach-realtime}.

\subsubsection{Reflection Removal}
The "Reflection Removal" process is a critical aspect of the methodology, and it involves the application of two widely recognized reflection removal approaches, namely "Image Crop" and "Unsupervised Detection." These techniques have been systematically employed across multiple datasets, encompassing Kvasir V1, Kvasir V2, Kvasir V3 \cite{pogorelov2017kvasir,pogorelov2018medico,borgli2020hyperkvasir}, and DowPK, all of which play a pivotal role in the realm of medical image analysis. A diverse array of features has been meticulously extracted from each dataset in three distinct versions: the original dataset, the dataset with reflections removed using Image Crop, and the dataset with reflections removed using the Telea method.

To gauge the effectiveness of these reflection removal strategies, the decision tree classifier was harnessed to calculate the F1-score across all three dataset versions. The objective was to identify the most promising variant for subsequent phases of the approach. The results were notably distinct for each dataset, as presented in Table~\ref{tbl:reflection}. Intriguingly, the application of Image Crop reflection removal consistently reduced the F1-score in all Kvasir datasets. Conversely, the Telea reflection removal method bolstered the F1-score when applied to Kvasir V1, Kvasir V2, and the DowPK dataset. The DowPK dataset exhibited no discernible improvements with Image Crop reflection removal but experienced enhancements when subjected to unsupervised reflection detection and removal using the Telea method.

\begin{table}
\centering
    \caption{Impact of reflection removal on the abnormalities detection F1-score.}
    \label{tbl:reflection}
    \begin{tabular}{|llll|}
        \hline
        Dataset & Original & Telea Ref. & Image Crop \\ \hline
        Kvasir V1 \cite{pogorelov2017kvasir} & 0.70 & 0.71 & 0.52 \\ \hline
        Kvasir V2 \cite{pogorelov2018medico} & 0.81 & 0.82 & 0.47 \\ \hline
        Kvasir V3 \cite{borgli2020hyperkvasir} & 0.82 & 0.82 & 0.61 \\ \hline
        DowPK & 0.62 & 0.68 & 0.63 \\ \hline
    \end{tabular}
\end{table}

The contrasting outcomes underscore the nuanced impact of reflection removal across different datasets, a phenomenon that can be attributed to two key characteristics of the data. Firstly, the presence of reflections within class-specific images played a pivotal role in determining the effectiveness of reflection removal. Classes with a larger number of images exhibiting reflections tended to show a comparatively lower impact from the removal process, resulting in fewer improvements. Secondly, the prevalence of reflections within the dataset itself was a crucial factor. Datasets with a higher incidence of reflections demonstrated a more pronounced response to reflection removal techniques.

For example, the DowPK and Kvasir V1 datasets \cite{pogorelov2017kvasir}, characterized by a smaller image count in comparison to Kvasir V2 and Hyper Kvasir datasets \cite{pogorelov2018medico,borgli2020hyperkvasir}, exhibited more substantial disparities in F1-scores when reflection removal was either applied or omitted, as evident in Table~\ref{tbl:reflection}. Additionally, the DowPK dataset, with a heightened density of reflections, experienced a more significant effect from the application of reflection removal techniques.

These insights furnish a deeper understanding of the heterogeneous impact of reflection removal across datasets, underlining the significance of factors such as the number of images per class and the abundance of reflections in the dataset. Researchers and practitioners are thus equipped with valuable knowledge to navigate the intricacies of this crucial preprocessing step in the pursuit of enhanced detection accuracy in the context of medical image analysis.

\subsubsection{Data Augmentation}
The utilization of data augmentation techniques has been instrumental in mitigating the class imbalance challenge, particularly in the context of versions 2 and 3 of the Kvasir dataset \cite{pogorelov2018medico,borgli2020hyperkvasir}, where class imbalance was notably pronounced. Data augmentation has been applied consistently across all datasets, and its impact has been particularly significant in rectifying the class imbalance issue, especially in versions 2 and 3 of the Kvasir dataset. While the Kvasir V1 dataset \cite{pogorelov2017kvasir}, devoid of severe class imbalance, exhibited minimal sensitivity to data augmentation, versions 2 and 3 of the Kvasir dataset benefited significantly from this strategy.

The results of applying data augmentation, when compared to un-augmented data, are presented in Table~\ref{tbl:augmentation}, showcasing the substantial improvements in detection accuracy achieved through this approach.
\begin{table}
\centering
    \caption{Impact of data augmentation on the detection F-1 Score.}
    \label{tbl:augmentation}
    \begin{tabular}{|lll|}
        \hline
        Dataset & Original & Augmentation \\ \hline
        Kvasir V1  \cite{pogorelov2017kvasir} & 0.71 & 0.71 \\ \hline
        Kvasir V2 \cite{pogorelov2018medico} & 0.82 & 0.84  \\ \hline
        Kvasir V3 \cite{borgli2020hyperkvasir} & 0.82 & 0.83  \\ \hline
        Dow & 0.68 & 0.79 \\ \hline
    \end{tabular}
\end{table}
As evident from the table, data augmentation had a marginal effect on the already well-balanced Kvasir V1 dataset, with no significant improvement in detection accuracy. Conversely, in versions 2 and 3 of the Kvasir dataset, where class imbalance posed a substantial challenge, data augmentation led to considerable enhancements in the F-1 score, elevating the overall detection performance. Similarly, in the DowPK dataset, data augmentation played a pivotal role in addressing class imbalance, resulting in a noteworthy boost in detection accuracy.

These findings underscore the potency of data augmentation as a powerful tool for ameliorating class imbalance issues and enhancing the performance of detection models, particularly in scenarios characterized by skewed class distributions. Researchers and practitioners can leverage this approach to bolster the robustness and reliability of medical image analysis systems across various datasets and clinical contexts.

\subsubsection{Feature Selection}
The feature selection process represents a pivotal component of the methodology, designed to optimize detection accuracy by identifying and prioritizing the most informative features. The approach employed here combines deep features with texture-based features, thereby harnessing the complementarity of these feature types to enhance detection performance. Notably, this combined approach yielded an accuracy of 0.99 and an F1-Score of 0.86 on the Kvasir V2 dataset \cite{pogorelov2018medico}. This outcome signifies a notable improvement when compared to using either deep features or texture-based features in isolation.

Feature Selection: A crucial aspect of the feature selection process revolved around identifying features with potentially misleading characteristics based on individual feature accuracy. To address this, a series of feature selection methodologies were employed. Multiple sets of features were randomly selected, and the optimal feature set, which demonstrated the best performance, was identified. This feature set comprises a combination of features encompassing Lire features, deep features, and Local Binary Patterns (LBP) with varying radii (ranging from 1 to 5). This combination yielded an F1-Score of 0.88 on the Kvasir V2 dataset \cite{pogorelov2018medico}.

The features that were considered for selection but ultimately exhibited misleading attributes are as follows:

 \begin{itemize}
		\item Color Histogram \cite{lux2008lire,huang1997image}
		\item Auto Color Correlation \cite{lux2008lire}
		\item Speeded up robust features (SURF) \cite{lux2008lire}
		\item Local Ternary Patterns (LTP) \cite{tan2010enhanced}
		\item Gray-Level Co-Occurrence Matrix (GLCM) \cite{haralick1973textural}
		\item Haralick Features (Statistics of GLCM) \cite{haralick1973textural}
	\end{itemize}
In contrast, the features retained in the selected feature set have proven to be instrumental in enhancing the detection accuracy. These features are drawn from various domains, including color, texture, and deep learning, resulting in a robust and highly effective feature set for medical image analysis.

This comprehensive feature selection process highlights the critical role of feature engineering in optimizing detection accuracy and underscores the value of combining diverse feature types to achieve superior results. Researchers and practitioners in the field can leverage these insights to enhance the robustness and reliability of their own medical image analysis systems.

\subsubsection{Classification Methods}

The classification phase represents the culmination of the methodology, where the final feature set is leveraged for robust and accurate detection using a variety of neural classifiers. Several classification techniques have been considered, each demonstrating commendable performance in the context of medical image analysis. Some of the noteworthy classifiers employed in this phase are as follows:

\textbf{Decision Tree Classifier}:
The decision tree classifier, known for its interpretability and efficiency, has yielded impressive results in terms of accuracy and F1-Score while requiring less computational time. This classifier was instrumental in assessing the impact of various preprocessing steps and feature combinations. The maximum F1-Score achieved using the Decision Tree classifier was 0.87 for the DowPK dataset and 0.85 for the Kvasir V2 dataset \cite{pogorelov2018medico}.

\begin{figure}
    \centering
    \includegraphics[width=15cm]{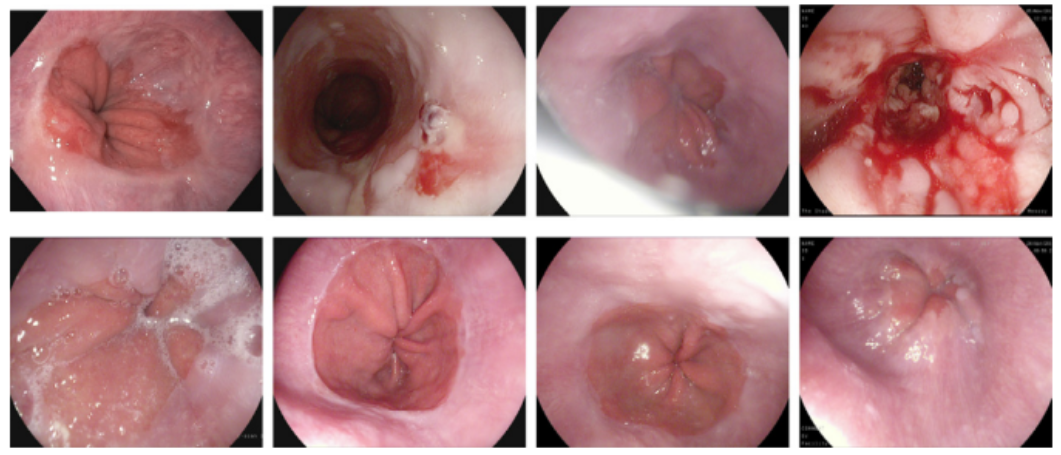}
    \caption{Images of Z-line and Esophagitis with high inter class similarities. (Row1: Z-line, Row2: Esophagitis)}
    \label{fig:eso-zline}
\end{figure}

\textbf{Random Forest Classifier}:
The Random Forest classifier, which leverages an ensemble of decision trees, further improved accuracy and F1-Score compared to a single decision tree while incurring a slightly longer detection time. The most notable F1-Score was achieved with the DowPK dataset, reaching 0.88, while the Kvasir V2 dataset \cite{pogorelov2018medico} exhibited an F1-Score of 0.85.
The analysis revealed high inter-class similarities between certain classes, particularly between the z-line and esophagitis shown in Figure~\ref{fig:eso-zline}. These classes appeared in some images with overlapping probability ranges. This overlap highlights the need for tailored class-specific thresholds to achieve accurate classification. Variations in the average probabilities of images belonging to different classes necessitate the use of distinct thresholds for each class, thereby accommodating the unique characteristics of each class.

\textbf{GA-Boost}:
To address the challenge of varied probability distributions within the dataset, a computational approach called GA-Boost was implemented. GA-Boost leverages a technique called a genetic algorithm to generate optimal thresholds specifically tailored for each class. These thresholds essentially act as decision points, separating data points belonging to one class from another.

Imagine a probability distribution as a bell curve.  For a two-class problem, the threshold would be a specific point on the curve that divides data points likely to belong to class 1 from those likely to belong to class 2.  However, in this case, the probability distributions for each class exhibited variations across the dataset.  A single, static threshold wouldn't work effectively.

GA-Boost tackles this challenge by mimicking the process of natural selection.  It starts with a random population of potential thresholds (each represented by a single floating-point value between 0.0 and 1.0).  These thresholds are then evaluated based on how well they classify the data (using a metric like F1-Score).  Individuals with higher F1-Scores are then selected for a simulated "breeding" process.

During this "breeding," the algorithm combines elements from two high-performing thresholds (using an addition operation with a modulo function) to create new offspring thresholds.  There's also a chance (20\% in this case) that a random mutation will be introduced into the offspring's threshold value.  This mutation helps to explore new regions of the solution space and prevent the algorithm from getting stuck in local optima.

The entire process (selection, breeding, mutation) is repeated across multiple generations (20 iterations in this case).  Over time, the population evolves, and the F1-Score metric is used to identify the best performing thresholds for each class.  These "winning" thresholds are then used for the final classification task, ensuring a more accurate fit to the specific characteristics of each class within the dataset.  The resulting optimal thresholds are presented in Table~\ref{tbl:thresholds}.

\footnotesize
\begin{longtable}{|p{4cm}|p{2cm}|}
    \caption{Thresholds computed for the Kvasir V2 \cite{pogorelov2018medico} Dataset Classes.}
    \label{tbl:thresholds} \\
    \hline
    Class Label & Threshold \\
    \hline
    \endfirsthead

    \hline
    Class Label & Threshold \\
    \hline
    \endhead

    % Add your content rows here
    retroflex-rectum & 0.6 \\ \hline
    out-of-patient & 0.2 \\ \hline
    ulcerative-colitis & 0.4 \\ \hline
    normal-cecum & 0.4 \\ \hline
    normal-z-line & 0.3 \\ \hline
    dyed-lifted-polyps & 0.6 \\ \hline
    blurry-nothing & 0.6 \\ \hline
    retroflex-stomach & 0.6  \\ \hline
    instruments & 0.4 \\ \hline
    dyed-resection-margins & 0.4 \\ \hline
    stool-plenty & 0.4 \\ \hline
    esophagitis & 0.2 \\ \hline
    normal-pylorus & 0.6 \\ \hline
    polyps & 0.4 \\ \hline
    stool-inclusions & 0.6 \\ \hline
    colon-clear & 0.6 \\ \hline
\end{longtable}
\normalsize

As reflected in Table~\ref{tbl:impacts}, the average results from ten runs of GA-Boost demonstrated improvements across multiple datasets.

\footnotesize
\begin{longtable}{|p{4cm}|p{1cm}|p{1cm}|p{1cm}|p{1cm}|}
    \caption{Impact on F1-score by applying various additions in the procedure on various datasets.}
    \label{tbl:impacts} \\
    \hline
    Dataset & Kvasir V1 \cite{pogorelov2017kvasir} & Kvasir V2 \cite{pogorelov2018medico} & Hyper Kvasir \cite{borgli2020hyperkvasir} & DowPK \cite{zeshan2023dowpk} \\
    \hline
    \endfirsthead

    \hline
    \multicolumn{5}{|c|}{\textbf{Impact on F1-Score Continued}} \\
    \hline
    Dataset & Kvasir V1 & Kvasir V2 & Hyper Kvasir & DowPK \\
    \hline
    \endhead

    % Add your content rows here
    No Preprocessing & 0.74 & 0.75 & 0.79 & 0.61 \\ \hline
    		No Augmentation (Reflection Removed) & 0.76 & 0.76 & 0.79 & 0.70 \\ \hline
    		Individual Feature AutoColorCorrelogram & 0.77 & 0.81 & 0.80 & 0.79 \\ \hline
    		Individual Feature ColorLayout & 0.67 & 0.75 & 0.74 & 0.75 \\ \hline
    		Individual Feature EdgeHistogram & 0.59 & 0.64 & 0.70 & 0.60 \\ \hline
    		Individual Feature Gabor & 0.38 & 0.30 & 0.32 & 0.31 \\ \hline
    		Individual Feature JCD & 0.75 & 0.76 & 0.78 & 0.72 \\ \hline
    		Individual Feature PHOG & 0.61 & 0.62 & 0.65 & 0.60 \\ \hline
    		Individual Feature Tamura & 0.49 & 0.48 & 0.52 & 0.42 \\ \hline
    		All Lire Features & 0.71 & 0.82 & 0.82 & 0.79 \\ \hline  
    		All Texture Features & 0.75 & 0.88 & 0.88 & 0.89 \\ \hline
    		Deep Features & 0.78 & 0.81 & 0.83 & 0.80 \\ \hline  
    		Selected Features (RF-Classifier) & 0.79 & 0.89 & 0.88 & 0.91 \\ \hline
    		3 Layer Neural Network & 0.89 & 0.90 & 0.90 & 0.91 \\ \hline
    		GA-Boost & 0.90 & 0.91 & 0.91 & 0.93 \\ \hline  

\end{longtable}
\normalsize

The application of GA-Boost led to noticeable enhancements in F1-Scores across the various datasets, further validating its utility in optimizing the classification process. These results underscore the adaptability of class thresholds to account for the distinct characteristics of each class, contributing to more accurate and robust medical image classification. Researchers and practitioners can employ this approach to fine-tune class thresholds and enhance classification performance in diverse medical image analysis tasks.

\subsection{Statistical Analysis}

The rigorous evaluation of the proposed approach involved conducting T-Tests on the results obtained from ten runs. The T-Tests were performed using the scipy.stats library, and the resulting p-values consistently fell within the range of 0.98 to 0.99 for all evaluation measures on all datasets. Additionally, the standard deviation of these evaluation measures ranged from 0.0001 to 0.001, as illustrated in Table~\ref{tbl:impacts}.

The T-Test results reinforce the statistical significance of the approach's consistently high performance in terms of detection accuracy, F1-Score, and Matthews Correlation Coefficient (MCC) across different datasets. This robust performance is complemented by a remarkable detection speed of 41 frames per second (FPS), rendering the approach highly efficient for real-time applications.

For a comprehensive comparison of the approach's results across diverse datasets, Table~\ref{tbl:impacts} offers an in-depth analysis. This comparative examination underscores the approach's consistent and superior performance in achieving high accuracy while maintaining efficient detection speeds.

Furthermore, a detailed analysis of the impact of various algorithmic steps is presented in Table~\ref{tbl:impacts}. Notably, the application of Telea reflection removal consistently improved results across most datasets, with the exception of the Hyper Kvasir dataset \cite{borgli2020hyperkvasir}. Similarly, the use of feature selection techniques led to varying improvements in the F1-Score across all datasets, emphasizing the efficacy of these techniques in enhancing the algorithm's performance.

The statistical rigor applied in evaluating the approach's performance, coupled with the consistently strong results across different datasets and algorithmic steps, solidifies its credibility and applicability in real-world medical image analysis tasks. Researchers and practitioners can rely on this approach to achieve accurate and efficient classification, with the flexibility to tailor class thresholds to accommodate variations in probability distributions.

\subsection{Results Analysis}
The results and their analysis provide valuable insights into the performance of the LiRE-CNN approach in detecting GI-tract abnormalities and landmarks. Let's summarize some of the key findings and observations:
\begin{itemize}
    \item \textbf{Reflection Removal Significantly Improves Accuracy:} The presence of reflections in the images had a notable impact on detection accuracy, leading to misclassifications. The application of reflection removal techniques, especially the Telea method, resulted in a significant enhancement of the F1-Score across multiple datasets, rectifying misclassifications and improving the overall performance.
    \item \textbf{Data Augmentation Mitigates Class Imbalance:} Class imbalance was observed, particularly in versions 2 and 3 of the Kvasir dataset \cite{pogorelov2018medico,borgli2020hyperkvasir}. Data augmentation effectively addressed this issue, leading to improved detection accuracy, especially for minor classes. This demonstrated the importance of data augmentation in handling class imbalances.
    \item \textbf{Feature Selection Enhances Accuracy:} A combination of deep features and texture-based features was found to be highly effective, achieving an accuracy of 0.99 with an F1-Score of 0.86 on the Kvasir V2 dataset \cite{pogorelov2018medico}. The application of feature selection techniques further improved accuracy by selecting the most relevant features while eliminating misleading ones.
    \item \textbf{Different Class Thresholds Improve Accuracy:} The application of the GA-Boost algorithm, which uses a genetic algorithm to determine tailored thresholds for different classes, led to significant improvements in the F1-Score. This demonstrated the importance of using class-specific thresholds to accommodate variations in probability distributions within the dataset.
    \item \textbf{Consistent and Efficient Real-Time Detection:} The LiRE-CNN approach consistently delivered high accuracy and F1-Scores across different datasets, including Kvasir V1 \cite{pogorelov2017kvasir}, Kvasir V2 \cite{pogorelov2018medico}, Hyper Kvasir \cite{borgli2020hyperkvasir}, and DowPK. The approach achieved an impressive F1-Score of 0.91 for the DowPK dataset, while maintaining a rapid detection speed of 41 frames per second (FPS). This efficiency is crucial for real-time applications in medical image analysis.
    \item \textbf{Challenges in Misclassifications:} The LiRE-CNN approach successfully addressed challenges related to reflections, class imbalances, and misleading features. However, some challenges remain, such as misclassifications between similar classes (e.g., dyed lifted polyps and dyed resection margins, esophagitis and normal z-line). These challenges highlight areas for further improvement.
    \item \textbf{Comparison with State-of-the-Art Approaches:} The LiRE-CNN approach outperformed several state-of-the-art approaches in terms of accuracy and F1-Score. It achieved a high F1-Score of 0.93 and an accuracy of 0.99 while maintaining a real-time detection speed of 41 FPS.
\end{itemize}
    
The results indicate that the LiRE-CNN approach is a promising solution for the accurate and efficient detection of GI-tract abnormalities and landmarks. Its ability to address common challenges and deliver consistent performance across diverse datasets makes it a valuable tool in the field of medical image analysis. Researchers and practitioners can benefit from this approach for real-time applications and improved accuracy in diagnosing GI-tract conditions.

\section{A Layered Decision Ensembles (TreeNet)}

\subsection{Evaluation Methodology}
\label{sec:eval}
The comprehensive evaluation methodology outlined above ensures a thorough assessment of the proposed model's performance. The consideration of detection speed, accuracy, efficiency with limited training data, and training time provides a holistic understanding of the model's capabilities in the context of medical image analysis. The following evaluation metrics were employed in the research:

\subsubsection{Training Data Requirement}
\label{sec:datareq}
The proposed model's ability to perform well with minimal training data is crucial for scenarios where obtaining large, labeled datasets is challenging. The evaluation includes metrics such as precision, recall, and F1-score, which are particularly important in assessing the model's performance on imbalanced datasets. The model's efficiency in learning intricate patterns from limited data is a key factor in its practical applicability in real-world medical scenarios.

\subsubsection{Training Time Requirements}
\label{sec:timereq}

The consideration of training time is essential for practical implementation, especially in time-sensitive medical settings. The advantages of resource efficiency and scalability associated with faster training times make the model more feasible for deployment in diverse healthcare environments. The evaluation of the model's performance with varying training times provides insights into its computational efficiency.

\subsubsection{Inference Time}
\label{sec:inferencetime}
Inference time is a critical aspect, especially in real-time applications. The ability of the model to provide rapid and accurate predictions is vital in medical image analysis scenarios where quick decisions are necessary. The evaluation of inference time, along with precision, ensures a balanced assessment of the model's efficiency and practical utility.

\subsubsection{Accuracy, Precision, Recall, MCC, and F1-Score}
\label{sec:p8-acc}
These metrics collectively provide a comprehensive evaluation of the model's classification performance. Accuracy, while intuitive, may have limitations in the presence of imbalanced datasets. Precision, recall, MCC, and F1-score offer a more nuanced understanding, considering aspects such as false positives, false negatives, and the model's sensitivity to imbalanced class distributions.

\subsection{Results of TreeNet}
\label{sec:p8-results}

The application of the proposed methodology to diverse benchmark datasets, employing texture features as descriptors, has resulted in noteworthy performance surpassing traditional approaches relying solely on Neural Networks or Decision Trees. Detailed results are presented in Table~\ref{tab:res_kvasir_v1} and Table~\ref{tab:res_kvasir_v2}, showcasing the methodology's superiority in various metrics.

The experimental results reveal several noteworthy aspects that shed light on the performance and behavior of the TreeNet model under varying conditions. In this section, we discuss three key observations that emerged from the evaluation of the model: the impact of data reduction on accuracy, the relationship between accuracy and training time, and the consistent precision and recall across different dataset sizes.

\subsubsection{Accuracy Reduction with Reduction of Data}
The observed robustness of the TreeNet model to variations in the size of the training dataset is a significant finding. The minimal decrease in accuracy, as indicated by the F1-score, even when only 50\% of the training data is utilized, suggests that TreeNet can maintain stable performance with limited training samples. This is particularly encouraging in scenarios where acquiring large, labeled datasets is challenging. The ability of TreeNet to perform well with reduced data could have practical implications in real-world applications, where obtaining extensive datasets may be resource-intensive.

\subsubsection{Impact of Accuracy on Training Time}
The correlation between accuracy and training time, particularly the substantial reduction in training time with a 50\% reduction in training data, highlights an important trade-off. This trade-off suggests that users can achieve a minor reduction in accuracy and F1-score in significantly shorter training durations by leveraging the TreeNet model. This finding is valuable in situations where computational resources or time constraints are critical factors. Users can make informed decisions about the acceptable level of accuracy trade-off based on the available resources and time considerations.

\subsubsection{Precision and Recall Relationship}
The observed stability in the relationship between precision and recall across different dataset sizes is a notable and consistent aspect of the TreeNet model. The ability of TreeNet to maintain a balanced trade-off between precision and recall indicates a high level of generalization. This generalization is a crucial attribute in classification tasks, especially in medical image analysis, where both false positives and false negatives can have significant implications. The TreeNet model's consistent performance in balancing precision and recall reinforces its reliability and effectiveness across diverse scenarios.

\subsection{Analysis of Results}
\label{sec:discussion}

The results of our experiments underscore the resilience and efficiency of the TreeNet model in handling variations in training data size. The observed trends in accuracy, training time, and the stability of precision and recall further enhance the credibility of TreeNet as a robust and adaptable tool for image classification tasks.

\begin{table}
    \centering
    \small
    \caption{Results on Benchmark Dataset of Kvasir V1 using various data proportions \cite{pogorelov2017kvasir,riegler2017multimedia}}
    \label{tab:res_kvasir_v1}
    \begin{tabular}{|p{0.1cm}|p{0.6cm}|p{0.6cm}|p{0.4cm}|p{0.4cm}|p{0.2cm}|p{0.2cm}|p{0.5cm}|p{0.2cm}|}
        \hline
        \setcounter{customcounter}{1}
        \textbf{Sr. No.} & \textbf{Chunk} & \textbf{Size} & \textbf{Time} & \textbf{Acc} & \textbf{P} & \textbf{R} & \textbf{MCC} & \textbf{F1} \\ \hline
        \thecustomcounter \stepcounter{customcounter} & 1 & 4000 & 2506 & 0.75 & 0.78 & 0.75 & 0.72 & 0.74 \\ \hline
        \thecustomcounter \stepcounter{customcounter} & 0.9 & 3600 & 1886 & 0.75 & 0.78 & 0.75 & 0.72 & 0.74 \\ \hline
        \thecustomcounter \stepcounter{customcounter} & 0.8 & 3200 & 1918 & 0.75 & 0.78 & 0.75 & 0.72 & 0.74 \\ \hline
        \thecustomcounter \stepcounter{customcounter} & 0.7 & 2800 & 1631 & 0.74 & 0.76 & 0.74 & 0.71 & 0.73 \\ \hline
        \thecustomcounter \stepcounter{customcounter} & 0.6 & 2400 & 1515 & 0.73 & 0.75 & 0.73 & 0.70 & 0.72 \\ \hline
        \thecustomcounter \stepcounter{customcounter} & 0.5 & 2000 & 1428 & 0.73 & 0.75 & 0.73 & 0.70 & 0.72 \\ \hline
        \thecustomcounter \stepcounter{customcounter} & 0.4 & 1600 & 1141 & 0.73 & 0.75 & 0.73 & 0.70 & 0.72 \\ \hline
        \thecustomcounter \stepcounter{customcounter} & 0.3 & 1200 & 968 & 0.71 & 0.72 & 0.71 & 0.68 & 0.71 \\ \hline
        \thecustomcounter \stepcounter{customcounter} & 0.2 & 800 & 786 & 0.68 & 0.70 & 0.68 & 0.64 & 0.67 \\ \hline
        \thecustomcounter \stepcounter{customcounter} & 0.1 & 400 & 582 & 0.65 & 0.67 & 0.65 & 0.61 & 0.65 \\ \hline
        \thecustomcounter \stepcounter{customcounter} & 0.05 & 200 & 408 & 0.63 & 0.64 & 0.63 & 0.58 & 0.62 \\ \hline
        \thecustomcounter \stepcounter{customcounter} & 0.01 & 40 & 201 & 0.50 & 0.53 & 0.5 & 0.45 & 0.45 \\ \hline
    \end{tabular}
\end{table}

\begin{table}
    \centering
    \small
    \caption{Results on Benchmark Dataset of Kvasir V2 using various data proportions \cite{pogorelov2018medico}}
    \label{tab:res_kvasir_v2}
    \begin{tabular}{|p{0.1cm}|p{0.6cm}|p{0.6cm}|p{0.4cm}|p{0.4cm}|p{0.2cm}|p{0.2cm}|p{0.5cm}|p{0.2cm}|}
        \hline
        \setcounter{customcounter}{1}
        \textbf{Sr. No.} & \textbf{Chunk} & \textbf{Size} & \textbf{Time} & \textbf{Acc} & \textbf{P} & \textbf{R} & \textbf{MCC} & \textbf{F1} \\ \hline
        \thecustomcounter \stepcounter{customcounter} & 1 & 5293 & 7428 & 0.71 & 0.87 & 0.71 & 0.70 & 0.77 \\ \hline
        \thecustomcounter \stepcounter{customcounter} & 0.5 & 2645 & 3866 & 0.70 & 0.87 & 0.69 & 0.68 & 0.76 \\ \hline
        \thecustomcounter \stepcounter{customcounter} & 0.4 & 2117 & 3482 & 0.68 & 0.87 & 0.68 & 0.67 & 0.75 \\ \hline
        \thecustomcounter \stepcounter{customcounter} & 0.3 & 1588 & 2931 & 0.66 & 0.87 & 0.66 & 0.66 & 0.73 \\ \hline
        \thecustomcounter \stepcounter{customcounter} & 0.2 & 1060 & 2172 & 0.58 & 0.87 & 0.58 & 0.59 & 0.69 \\ \hline
        \thecustomcounter \stepcounter{customcounter} & 0.1 & 536 & 1506 & 0.52 & 0.85 & 0.52 & 0.52 & 0.61 \\ \hline
        \thecustomcounter \stepcounter{customcounter} & 0.01 & 92 & 537 & 0.25 & 0.61 & 0.25 & 0.27 & 0.27 \\ \hline
    \end{tabular}
\end{table}

The hybrid approach outlined in the methodology demonstrates notable adaptability and effectiveness in addressing the challenges of medical image analysis, particularly in scenarios marked by limited data availability and class imbalances. The detailed results provided in Table~\ref{tab:res_kvasir_v1} and Table~\ref{tab:res_kvasir_v2} showcase superior performance metrics, indicating the potential of this hybrid system to contribute significantly to the field of healthcare diagnostics.

\subsubsection{Inference Time}

The system's achievement of 37 Frames Per Second (FPS) for inference time is a noteworthy result. In real-world applications, especially in medical contexts where quick decisions are crucial, achieving such rapid inference times is highly valuable. This speed positions the hybrid system as a practical and efficient solution for medical image analysis tasks, facilitating timely and accurate diagnostics.

\subsubsection{Model Performance Metrics}

The evaluation metrics employed, including accuracy (Acc), precision (P), recall (R), weighted F1-score (F1), and Matthews correlation coefficient (MCC), provide a comprehensive assessment of the model's classification capabilities. These metrics offer a nuanced understanding of the model's predictive prowess under varying data proportions. The emphasis on metrics such as precision and recall is particularly relevant in medical image analysis, where the consequences of false positives and false negatives can be significant.

\subsubsection{Data Proportion and Training Size}

The introduction of the 'Chunk' parameter, representing the proportion of training data used, and the 'size' parameter, quantifying the number of instances utilized for training, add granularity to the evaluation. The dynamic nature of these parameters allows for a detailed exploration of the model's performance across different data scenarios. The requirement for at least one instance from each class in the training process ensures a representative and inclusive learning experience, contributing to the model's robustness.

\subsubsection{Temporal Efficiency}

The inclusion of the 'Time' metric, measuring the duration of model training in seconds, adds a crucial dimension to the evaluation. The temporal efficiency of the training process is essential in practical applications, influencing computational resource requirements. The detailed assessment of training time contributes to a comprehensive characterization of the model's performance and feasibility for real-world deployment.

\begin{figure}
	\includegraphics[width=1.0\linewidth]{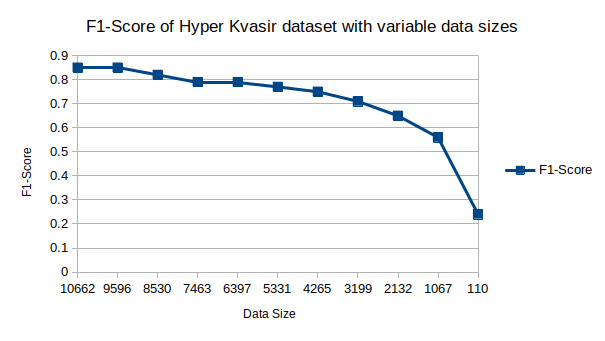}
	\caption{F1-score of the Hyper Kvasir dataset using various chunks of the training data \cite{borgli2020hyperkvasir}}
	\label{fig:res_hyperkvasir}
\end{figure}

\begin{figure}
	\includegraphics[width=1.0\linewidth]{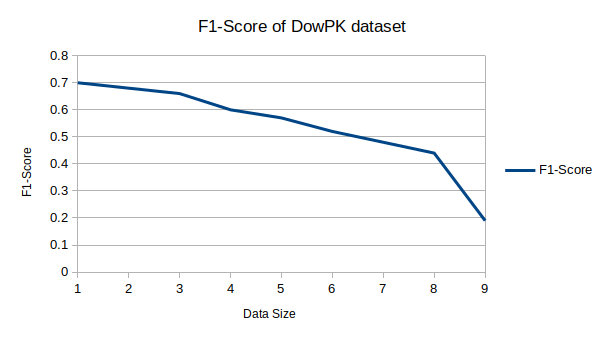}
	\caption{F1-score of the DowPK dataset using various chunks of the training data}
	\label{fig:res_dowpk}
\end{figure}

The evaluation of both the Hyper Kvasir \cite{borgli2020hyperkvasir} and DowPK \cite{zeshan2023dowpk} datasets reveals compelling trends, as depicted in Figure~\ref{fig:res_hyperkvasir} and Figure~\ref{fig:res_dowpk}. The noteworthy observation across both datasets is the consistent behavior indicating that a substantial reduction in training data has a comparatively minimal impact on the F1-Score of the results obtained on the test data.

\section{Segmentation using Depth-Wise Separable Convolution}

All experiments conducted as part of this study were carried out within the Google Colab platform, which offers a runtime session duration of up to 12 hours. It is worth noting that this 12-hour window is insufficient for effectively training deep learning models. To ensure a fair and consistent comparison across all experimental iterations, we maintained a uniform number of epochs for each. Furthermore, the dataset was partitioned into three distinct subsets, specifically: training, validation, and testing, with allocations of 80\%, 10\%, and 10\% of the data, respectively. In the context of this partition, a total of 800 images were designated for model training. However, this quantity falls short of the requisites for training deep learning models. To address this limitation, a data augmentation technique was thoughtfully applied exclusively to the training set, preserving the original configurations of the validation and test sets, which retained their sizes at 100 images each. The augmentation process entailed the application of 30 distinct transformations to the training set, resulting in an expanded dataset of 24,800 images. Importantly, these augmentations were also employed on the accompanying mask data, ensuring that the target variable underwent the same transformations as the input image.

In terms of the training regimen, the NAdam optimizer was selected, operating with a learning rate of 0.0001 and a batch size of 8. Throughout the training process, the loss values for both the training and validation sets were meticulously recorded for each epoch. These learning curves serve as invaluable tools for monitoring and comprehending the convergence behavior of the model, contributing essential insights to the overall experimental analysis.

\begin{figure}[!ht]
\centering
\includegraphics[scale=1]{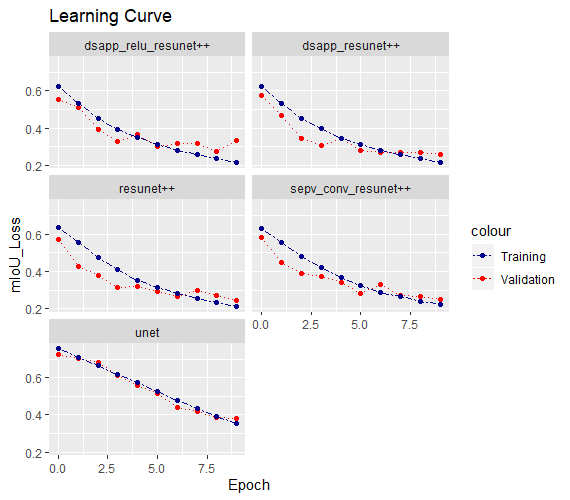}
\caption{Learning Curve}
\label{Learning Curve}
\end{figure}

The learning curves depicted in Figure ~\ref{Learning Curve}, underlining the performance trajectories of various architectural models, offer valuable insights into the convergence and training dynamics. Of particular significance is the architecture incorporating the Deep Atrous Spatial Pyramid Pooling (DASPP) bridge, wherein a discernible trend indicates possible convergence within a relatively short temporal span of 10 epochs, as discerned from the commencement of a subsequent increase in validation error. In contrast, the ResUNet++ and the separable convolved ResUNet, represented in the graph, exhibit intriguing nuances. In these cases, the learning curves reflect the potential for continued training beyond the 10-epoch threshold, as both the training and validation errors continue to exhibit a decreasing trend. The UNet architecture, while exhibiting a downward trajectory in its learning curve at the 10th epoch, distinguishes itself by its comparatively higher loss value, when juxtaposed with the loss observed in the ResUNet++ architecture.

This visual depiction of learning dynamics encapsulates a multitude of implications for the optimization and fine-tuning of these architectures. Understanding the different convergence patterns and loss values across the models contributes to an informed strategy for model selection and refinement. As such, these findings underscore the importance of monitoring learning curves as a fundamental tool for assessing the efficacy and potential improvements of neural network architectures during training, with a specific emphasis on the judicious utilization of architectural components, such as the DASPP bridge and the choice between the standard UNet model and its enhanced variants, like ResUNet++ and separable convolved ResUNet.

\begin{table}[!ht]
\begin{tabular}{|p{4cm}|p{1cm}|p{1cm}|p{1cm}|p{1cm}|}
\hline
\textbf{Model}                  & \textbf{Recall}  & \textbf{Precision} & \textbf{Dice}    & \textbf{mIoU}    \\ \hline
\textbf{Unet}                   & \textbf{75.23\%} & 84.52\%            & 71.91\%          & 59.53\%          \\ %\hline
\textbf{resunet++}              & 64.97\%          & 89.81\%            & \textbf{78.35\%} & \textbf{69.48\%} \\ %\hline
\textbf{sepv\_conv\_resunet++}  & 60.55\%          & \textbf{93.31\%}   & 77.25\%          & 67.56\%          \\ %\hline
\textbf{dsapp\_resunet++}       & 69.72\%          & 82.62\%            & 76.66\%          & 66.71\%          \\ %\hline
\textbf{dsapp\_relu\_resunet++} & 61.54\%          & 92.33\%            & 74.63\%          & 66.03\%          \\ \hline
\end{tabular}
\caption{Test Data Results}
\label{Test Data}
\end{table}

The table ~\ref{Test Data} provides a comprehensive overview of the performance exhibited by each model when subjected to testing data. Notably, the ResUNet++ model outshines its counterparts in terms of the dice coefficient and mean Intersection over Union (mIoU) metrics, manifesting superior performance. On the other hand, the model employing separable convolution exhibits comparable results in both Dice and mIoU metrics. Conversely, the model integrated with the DASPP bridge yields suboptimal results, indicating that further augmenting the model's depth fails to translate into improved performance.

Among the models under consideration, the model utilizing separable convolution emerges as the most parsimonious in terms of parameters, implying a more compact model size. This aspect carries profound implications for practical deployment, as a reduced number of parameters facilitates ease of integration into a production environment. These findings, including the pertinent size-related statistics, are thoughtfully aggregated in the table designated as "Model Size," offering insights into the resource efficiency of each model.

\begin{table}[!ht]
\centering
\begin{tabular}{|l|l|l|}
\hline
\textbf{Model}                  & \textbf{Params}    & \textbf{GFLOPs}  \\ \hline
\textbf{Unet}                   & 3,588,997          & 7,165,148          \\ \hline
\textbf{resunet++}              & 4,371,265          & 8,718,068          \\ \hline
\textbf{sepv\_conv\_resunet++}  & \textbf{3,047,265} & \textbf{6,070,057} \\ \hline
\textbf{dsapp\_resunet++}       & 5,024,705          & 10,024,898         \\ \hline
\textbf{dsapp\_relu\_resunet++} & 5,024,705          & 10,024,898         \\ \hline
\end{tabular}
\caption{Model Size}
\label{Model Size}
\end{table}

\section{Voting Neural Network (VNN) for Endoscopic Image Segmentation}
\subsection{Datasets}
The proposed methodology is rigorously evaluated using two real-world datasets, namely CVC-Clinic \cite{bernal2015wm} and MediaEval-2021 \cite{hicks2021medico}. These datasets serve as valuable resources for endoscopic image segmentation and are summarized below:
\subsubsection{CVC-Clinic Dataset}
    Image Characteristics: The CVC-Clinic dataset comprises images with uniform dimensions of 384 × 284 pixels, featuring three input channels. Ground truth images are provided with a single black channel, facilitating segmentation tasks.
    Annotation: The dataset consists of 612 images in the Tagged Image File Format (.tif), each meticulously annotated by human experts for segmentation purposes.

\subsubsection{MediaEval-2021 Dataset:}

    Image Characteristics: MedicoEval-2021 includes 1300 images, each accompanied by corresponding ground truth masks. Additionally, Hyper Kvasir \cite{borgli2020hyperkvasir} is encompassed within this dataset, encompassing 110,079 images and 374 videos. This diverse set encompasses various polyp classes along with their respective masks.
    Annotation and Examination: The datasets are meticulously examined by proficient endoscopists, ensuring high-quality annotations. However, effective utilization of these datasets in the proposed work necessitates robust preprocessing \cite{barik2010object} due to challenges such as specular reflections \cite{li2019specular}. These reflections can obscure or complicate the identification of polyp regions. Addressing this challenge with established techniques is crucial to enhance model accuracy and learning.
\subsection{Evaluation Measures}
various evaluation measures are employed to assess the effectiveness of segmentation in the context of the research. Some of the evaluation measures include:
\subsubsection{Accuracy}
Accuracy represents the ratio of correctly predicted pixels to the total number of pixels in the segmentation output.

$\text{Accuracy} = \frac{\text{True Positives} + \text{True Negatives}}{\text{Total Pixels}}$

\subsubsection{Precision}
Precision quantifies the accuracy of positive predictions, measuring the proportion of correctly identified positive pixels among all pixels predicted as positive.

$\text{Precision} = \frac{\text{True Positives}}{\text{True Positives} + \text{False Positives}}$

\subsubsection{Recall (Sensitivity)}
Recall assesses the ability of the model to capture all positive instances, indicating the proportion of correctly identified positive pixels among all actual positive pixels.

$\text{Recall} = \frac{\text{True Positives}}{\text{True Positives} + \text{False Negatives}}$

\subsubsection{F1 Score}
The F1 score is the harmonic mean of precision and recall, providing a balanced measure that considers both false positives and false negatives.

$F1 = \frac{2 \times \text{Precision} \times \text{Recall}}{\text{Precision} + \text{Recall}}$

\subsubsection{Intersection over Union (IoU) / Jaccard Index}
IoU measures the overlap between the predicted segmentation and the ground truth, representing the ratio of the intersection to the union of the two sets.

$IoU = \frac{\text{True Positives}}{\text{True Positives} + \text{False Positives} + \text{False Negatives}}$

\subsubsection{Dice Coefficient}
The Dice coefficient is another measure of the overlap between the predicted and true positive pixels, emphasizing a balance between precision and recall.

$Dice = \frac{2 \times \text{True Positives}}{\text{True Positives} + \text{False Positives} + \text{False Negatives}}$

\subsection{Results and Discussion}
In this research, several pre-trained neural network architectures designed for image segmentation were systematically evaluated based on two key performance parameters: accuracy and Dice Coefficient. The chosen models included Unet, UNet++, and ResuNet++ \cite{jha2019resunet++}, all of which utilized the cross-entropy loss function. These models were selected for further analysis and integration into a majority voting scheme due to their notable accuracy and overall positive evaluations in the context of the segmentation task.

The training process for these models spanned 200 epochs, during which the neural networks learned to extract features and optimize their parameters. The training was conducted using cross-entropy loss, a widely employed loss function in segmentation tasks. This process aimed to enhance the models' ability to accurately identify and delineate regions of interest within the images.

Upon completion of the training phase, the performance of the models was assessed on test data using two real-world datasets: CVC and MediaEval 2021 \cite{bernal2015wm}, \cite{hicks2021medico}. The results of these evaluations were compiled and are presented in Tables I and II, showcasing the effectiveness of the models in accurately segmenting endoscopic images. The utilization of the Dice Coefficient, which measures the overlap between predicted and ground truth regions, provided additional insights into the segmentation performance.

The choice of these specific neural network architectures and the majority voting approach suggests a strategic combination of models to harness their collective strengths. This ensemble approach aims to mitigate potential shortcomings of individual models, resulting in a more robust and accurate segmentation outcome. The detailed results presented in the tables offer a quantitative overview of the models' performance on the respective datasets, providing valuable insights for further analysis and comparison with state-of-the-art techniques.
In the assessment of the chosen models, a noteworthy aspect of the evaluation involved comparing their performance with and without the application of data preprocessing techniques. The incorporation of data preprocessing was found to have a positive impact on detection accuracy across all models.

Particularly, the preprocessing techniques were instrumental in addressing challenges posed by high reflections present in some images. These reflections were identified as a significant factor contributing to suboptimal segmentation when utilizing various deep learning segmentation algorithms on the original images.

For instance, the left-most image in Figure~\ref{architecture} initially exhibited an accuracy of 0.78 when processed without reflection removal. However, with the implementation of reflection removal techniques, the segmentation accuracy notably improved, reaching 0.96. This improvement underscores the efficacy of reflection removal in enhancing the models' ability to accurately identify and delineate regions of interest within the images.

\begin{figure}
\centering
\includegraphics[scale=0.30]{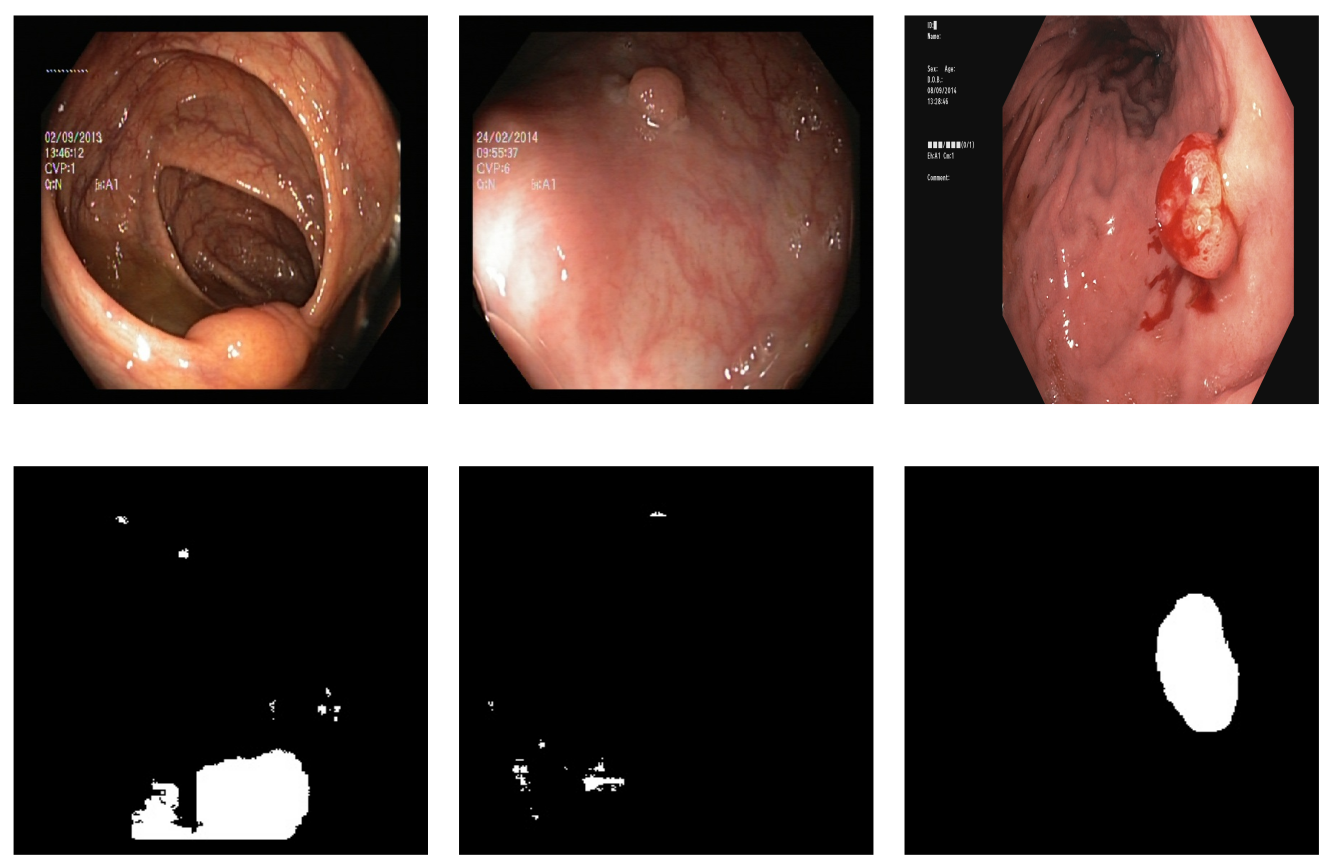}
\caption{Best segmentation results from MediaEval 2021 dataset \cite{hicks2021medico,khan2022voting}}
\label{architecture-res}
\end{figure}

Furthermore, Figure~\ref{architecture-res} illustrates selected images with high detection accuracy ranging from 0.95 to 0.98. These images showcase precision and recall values exceeding 0.80 and 0.90, respectively. Precision and recall are crucial metrics in segmentation tasks, providing insights into the model's ability to correctly identify positive instances and capture all actual positive instances, respectively.

The comparative analysis with and without data preprocessing highlights the importance of addressing image-specific challenges, such as reflections, to achieve optimal segmentation results. The presented results indicate that incorporating reflection removal techniques positively influenced the accuracy of the segmentation models, emphasizing the significance of thoughtful data preprocessing in enhancing the performance of deep learning-based segmentation algorithms.
In the dataset analysis, it was observed that certain images exhibited correct segmentation but with the inclusion of very minor non-polyp regions into the polyp and vice versa—incorporating minor regions into the non-polyp class. Conversely, some images were not accurately segmented, leading to a thorough examination of the factors contributing to incorrect segmentation.

Upon analyzing instances of incorrect segmentation, it was identified that the major contributor to this issue was a low recall value in the segmentation results. Several images with low detection accuracy and Jaccard Index were identified, and three representative examples are illustrated in Figure 5.

\begin{figure}
\centering
\includegraphics[scale=0.40]{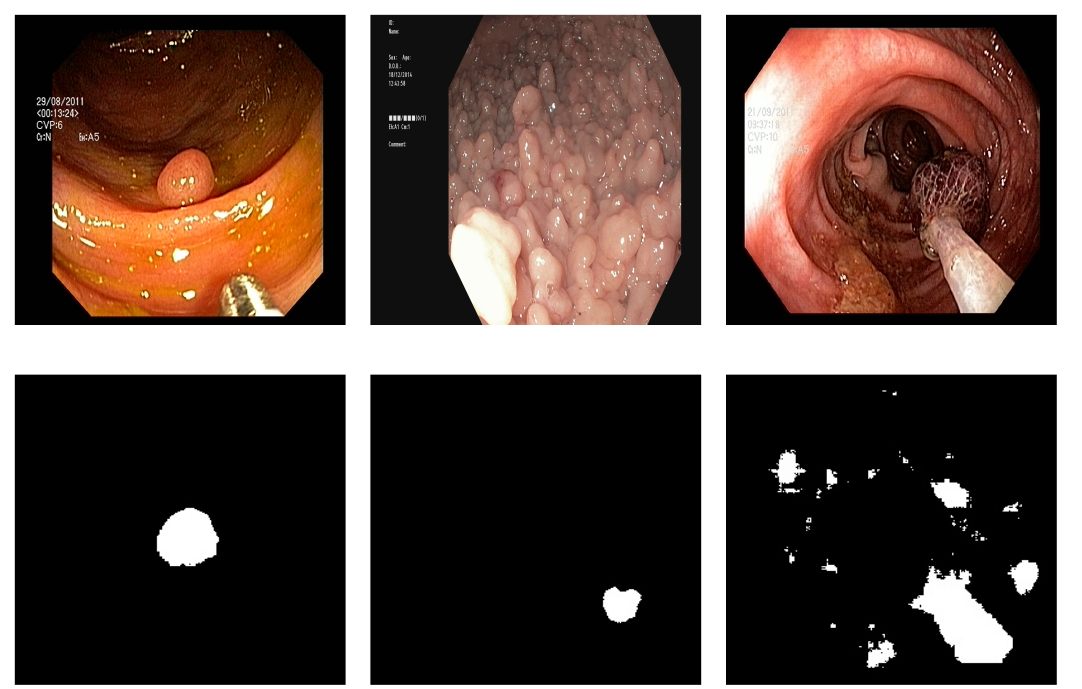}
\caption{Erroneous segmentation results from MediaEval 2021 dataset \cite{hicks2021medico,khan2022voting}}
\label{architecture}
\end{figure}

Image Analysis:

    Left-Most Image:
        This image depicted small polyps and other abnormalities, including swelled regions. The algorithm erroneously classified these swelled regions as polyps, resulting in low precision (0.43) with a high recall (0.99). The high recall indicates that the algorithm successfully identified the majority of actual positive instances (polyps), but the low precision suggests a significant number of false positives, leading to incorrect segmentations.

    Middle Image:
        This image contained a polyp covering more than 90\% of the image area—an exceptional case within the dataset. However, the algorithms struggled to detect the entire polyp, leading to low recall and accuracy. The precision was 1, indicating no false positives, but the recall was less than 0.1, signifying the failure to capture the entirety of the polyp region.

    Right-Most Image:
        This image exhibited low accuracy, precision, and recall. The algorithms detected an object inserted as polyps instead of the actual polyp, resulting in both precision and recall being less than 0.20. This suggests a significant misclassification of the object, contributing to an overall inaccurate segmentation.

\chapter{Conclusions and Future Directions}
\label{sec:conclusion}

\section{Conclusions}
This research embarked on a comprehensive exploration of various methodologies for medical image analysis, with a specific focus on enhancing the accuracy, precision, reliability, and speed of disease detection within the GI-Tract. By employing diverse approaches, the research aimed to contribute significantly to the development of more effective tools for diagnosing GI diseases.

The investigation progressed through a series of stages, each utilizing distinct techniques for GI image analysis. This iterative approach facilitated continuous improvement and a deeper understanding of the challenges and opportunities present in this field. The following sections will delve into the specific findings related to each explored methodology, highlighting their contributions to achieving the overall goals of improved accuracy, reliability, and speed in GI disease detection.

One key area of focus involved investigating techniques for image enhancement, specifically addressing the prevalent issue of light reflections in GI endoscopy images. These reflections can significantly hinder the visibility of abnormalities within the GI tract. The implemented reflection removal techniques yielded a noteworthy improvement in F1-score, ranging from 1\% to 6\% across the investigated datasets. This finding underscores the importance of clear and unobstructed images for accurate abnormality detection in GI disease diagnosis.

Furthermore, the research explored the potential of data augmentation techniques to address limitations associated with the size and diversity of medical image datasets, particularly in the GI domain. Data augmentation involves manipulating existing data to create artificial variations, essentially enriching the training data for machine learning models. By applying these techniques to various datasets, the research achieved a significant increase in F1-score, reaching up to 11\%. This suggests that data augmentation can be a valuable tool for improving the performance of abnormality detection models. The enriched training data fosters a more diverse range of image characteristics, enabling the models to learn more robust features and generalize better to unseen data.

It is important to acknowledge that the effectiveness of both image enhancement and data augmentation techniques might vary depending on the specific type and severity of the GI abnormalities being examined. As indicated in the class-wise F1-Score, certain abnormalities or their severity levels might benefit more from these techniques compared to others.

Ensemble learning techniques, which involve combining multiple models or decision-makers, have demonstrated the potential for improved performance compared to single models. The second objective of this research investigated this concept in the context of gastrointestinal (GI) disease detection. Specifically, it challenged the notion that deep learning models operate in isolation and explored the potential of combining traditional feature extraction techniques with deep learning for improved classification accuracy.

Chapter~\ref{sec:approaches} and Chapter~\ref{sec:resultsall} delved into various feature sets, encompassing both texture features and those generated by deep learning models themselves. The findings revealed a fascinating interplay between these methods. Interestingly, some texture features, when employed independently, proved effective for classification. However, the true power emerged from combining these features with deep learning models. This synergistic approach yielded a significant improvement in F1-score, achieving up to 11\% higher accuracy compared to using deep learning alone.

In Chapter~\ref{sec:resultsall}, these results highlight the potential of leveraging both traditional and deep learning techniques for GI disease detection. The research supports the idea that deep learning models can benefit from the complementary information provided by traditional features.

The limitations identified here point towards the inherent challenge of selecting the most effective feature sets and deep learning architectures. An exhaustive exploration of every possible combination is impractical.

The third objective of the research sought to enhance classification performance by identifying a subset of informative features. This objective addressed the well-documented challenge of the "curse of dimensionality." This phenomenon arises when a model is trained on a high number of features, leading to increased complexity and potentially hindering its ability to learn effectively. When faced with too many features, models can struggle to identify the most relevant information and may even overfit the training data, failing to generalize well to unseen examples.

To address this challenge, Section~\ref{approach-realtime} explored various feature selection techniques. Notably, a Random Forest-based approach yielded promising results, achieving an F1-score improvement of up to 4\% on the KvasirV1 dataset. This finding underscores the importance of feature selection as a method for optimizing model performance. By reducing dimensionality, feature selection allows the model to focus on the most informative characteristics within the data. This can not only improve classification accuracy but also potentially mitigate overfitting by reducing the model's complexity.

However, limitations exist that warrant further investigation. The effectiveness of specific feature selection techniques can vary depending on the dataset and the chosen features themselves. Future research could explore a wider range of these techniques and delve deeper into how they interact with different datasets and feature sets.

The fourth objective of this research investigated the potential of deep learning to achieve superior detection accuracy in GI image analysis. This objective specifically addressed the challenge of training models on finite datasets, a common hurdle in medical image analysis due to limitations in data availability.

Section~\ref{approach-homo} meticulously examined various deep learning methodologies, evaluating their individual performance. The research then investigated the impact of combining these methodologies with a data bagging approach. Data bagging involves training multiple models on subsets of data with replacement, a technique known to improve a model's generalizability, or its ability to perform well on unseen data. This combined strategy yielded impressive results, achieving an accuracy of up to 0.92 on the Kvasir V1 dataset.

These findings highlight the effectiveness of ensemble of deep learning for GI image analysis. Additionally, this showcase data bagging as a valuable tool to overcome the limitations associated with limited datasets. This paves the way for developing robust and generalizable deep learning models for GI disease detection, ultimately leading to more accurate diagnoses.

The fifth research objective centered on achieving real-time detection of abnormalities within gastrointestinal (GI) tract images. This objective acknowledged the critical role of real-time analysis in clinical settings, particularly for immediate decision-making. However, it also recognized the inherent trade-off between speed and accuracy inherent in real-time systems.

The research explored two distinct approaches to achieve real-time analysis. The first approach explored in Section~\ref{approach-realtime} prioritized speed by utilizing a combination of texture features and Gray Level Co-occurrence Matrix (GLCM) features. This method achieved a well-balanced performance, reaching an F1-score of 0.76 on the Kvasir V1 dataset. This score demonstrates its suitability for real-time applications where immediate analysis is essential.

The second approach explored in Section~\ref{approach-hetero} leveraged the power of deep learning by employing a lightweight neural network in conjunction with texture features. This method achieved significantly higher F1-scores, ranging from 0.88 to 0.90 on various Kvasir datasets. These findings highlight the potential of deep learning for achieving high accuracy even with real-time constraints.

It is important to acknowledge the limitations associated with both approaches. Real-time systems inherently face challenges in balancing speed and accuracy. Deep learning models, while powerful, often require significant computational resources, which can limit their real-time applicability.

The sixth objective of the research tackled the challenge posed by complex GI image datasets. These datasets often exhibit intricate characteristics, such as high similarity between certain classes and significant variations within a single class (intra-class dissimilarity). To address this challenge, research focused on identifying optimal threshold values using optimization algorithm. This approach aimed to enhance the model's ability to make accurate class-based decisions for each image by considering its similarity to different classes. This strategy acknowledges the inherent complexities of GI image data and strives to improve the accuracy of real-time detection systems in such scenarios.

The implemented method yielded promising results, achieving an F1-score of 0.90 on the KvasirV3 dataset. This finding underscores the critical role of considering class similarities and dissimilarities during the development of classification models for GI images. By establishing effective thresholds, the research paves the way for improved accuracy in real-time settings, particularly when dealing with datasets exhibiting such intricate characteristics.

It is important to acknowledge the inherent limitations associated with defining optimal thresholds. The research operated under the assumption that GI image datasets would exhibit this type of complexity.

The research effectively demonstrates the power of deep learning for enhanced detection accuracy. It also highlights the feasibility of real-time solutions that achieve a balanced trade-off between speed and accuracy. Furthermore, the research underscores the importance of addressing challenges arising from class similarities and dissimilarities in GI image analysis. These findings pave the way for further advancements in developing robust and efficient deep learning models for real-time GI disease detection, ultimately contributing to improved clinical decision-making and patient care.

In the filed of segmentation one key revelation discerned in our research pertains to the manifestation of depth-wise separable convolution within the model architecture. This specific convolutional technique demonstrates its mettle by effectively diminishing the model's footprint, thereby contributing to a streamlined and more efficient computational process. The imperative task that beckons forthwith is the meticulous tuning of hyperparameters and the conscientious extension of the number of training epochs. Such endeavors hold the promise of unraveling a deeper comprehension of the performance dynamics underpinning the utilization of depth-wise separable convolution, thereby offering invaluable insights for the development of future model architectures.

Another contribution in the domain of segmentation is a polyp segmentation system based on a deep convolutional neural network (CNN) autoencoder architecture. The system encompasses the capability to not only detect polyps but also classify them and generate comprehensive reports. For polyp detection in grayscale images, pre-trained CNN architectures are employed, and experiments are conducted on deep CNN architectures such as RESNET \cite{he2016deep}, RESNE++ \cite{jha2019resunet++}, U-NET \cite{ali2020depth}, and U-NET++ \cite{li2019specular} to achieve more accurate segmentation results.

These chosen architectures incorporate features from residual networks, including re-designed skip pathways and deep supervision. The re-designed skip pathways play a crucial role in minimizing the semantic gap between the feature maps of the encoder and decoder sub-networks, enhancing the overall segmentation performance.

The research encountered two primary challenges: limited dataset availability and the resulting impact on computer vision and deep learning outcomes. To address this limitation, data augmentation techniques are employed, including flipping, rotation, shift, cut mixing, brightness adjustment, zoom, and Gaussian blur. Through the application of these augmentation techniques, the dataset is expanded from 800 to 11,000 images. The dataset is then partitioned into 80\% for training, 10\% for validation, and 10\% for testing.

Following successful augmentation, multiple models, including UNet, ResUnet++ \cite{jha2019resunet++}, and Deeplabv3, are trained. The ensemble technique is employed to combine the results of these models, selecting the most accurate outcome. Boosting techniques are also leveraged to enhance the segmentation results.

The final results of the segmentation neural network are binary, with pixel values of either zero or 255. Prior to thresholding, these results can be utilized for combination as a soft majority voting approach. This innovative approach enhances the robustness and accuracy of the segmentation system, addressing challenges associated with limited data and leveraging the strengths of multiple deep learning architectures for improved polyp detection and segmentation.

\section{Future Directions}
\label{sec:future}
The promising outcomes and capabilities exhibited by the diverse approaches, in this study unveil opportunities for future research and development in the field of medical image analysis. Several directions for prospective investigations and enhancements are outlined below:

\subsection{Scope Extension to Diverse Medical Imaging Modalities}

The explored methodologies in this research hold promise beyond the realm of GI tract abnormalities. Future work could investigate how well techniques like TreeNet translate and adapt to various medical imaging modalities. Here are potential areas for exploration where data, objectives, and successful approaches from GI analysis might be applicable:
\begin{itemize}
    \item Analyzing X-rays, CT scans, or MRIs for disease detection in different organs like lungs, bones, or the brain.
    \item Analyzing digitized pathology slides to identify cancerous tissues or other abnormalities.
    \item Analyzing images generated by PET scans, which provide insights into biological processes at the molecular level.
\end{itemize}

Extending these techniques to diverse domains can offer valuable insights into their versatility and effectiveness in broader medical image analysis tasks. This could lead to the development of general-purpose frameworks capable of handling various medical imaging modalities, ultimately streamlining the diagnostic process across different medical specialties.

\subsection{Enhancement of Data Efficiency}
One of the persistent challenges in medical image analysis is the limited availability of labeled data.  While this research achieved promising results, future work should address this issue to make these techniques more practical in real-world scenarios. Here are some potential areas of exploration to enhance data efficiency:

\subsubsection{Semi-Supervised Learning}
Explore the use of semi-supervised learning techniques. These techniques can leverage a large amount of unlabeled data in conjunction with a smaller set of labeled data to improve model performance. This is particularly beneficial when acquiring labeled medical data can be expensive and time-consuming.

\subsubsection{Active Learning}
Investigate active learning approaches. Active learning allows the model to iteratively query for the most informative data points to be labeled. This can significantly reduce the amount of labeled data required for achieving good performance.  The model can focus its labeling efforts on the most uncertain or challenging examples, leading to more efficient utilization of limited labeling resources.

\subsubsection{Data Augmentation Techniques}
Further explore  data augmentation techniques specifically tailored for medical images. The augmentation techniques of the combination of medical images with general images like ImageNet can produce more generalized dataset for the training of the networks for better results.

\subsection{Interpretability and Explainability}
In medical applications, particularly those involving complex AI models, interpretability and explainability are paramount.  While TreeNet has proven effective, future work should focus on enhancing its interpretability to build trust and understanding among clinicians. Here are some potential areas of exploration:

\subsubsection{Feature Importance Analysis}
Develop techniques to  identify and explain the most important features that contribute to TreeNet's decisions. This can be achieved through methods like LIME (Local Interpretable Model-Agnostic Explanations) or SHAP (SHapley Additive exPlanations), which provide insights into  how each feature influences the model's predictions for a specific image.

\subsubsection{Visual Explanations}
Develop methods to generate visual explanations that highlight the image regions  most critical for TreeNet's decision-making process. This can involve techniques like attention visualization, which creates heatmaps highlighting the areas of the image that the model focuses on for classification.

\subsubsection{Counterfactual Analysis}
Explore counterfactual analysis techniques. This approach involves identifying the minimal changes to an image that would cause the model to change its prediction. This can provide clinicians with valuable insights into why a specific image was classified in a particular way and suggest potential alternative interpretations.

\subsubsection{Model-Agnostic Explanations}
Investigate model-agnostic explanations.  These techniques can be applied to any model, including TreeNet, and provide general explanations for its predictions without requiring access to the model's internal workings. Techniques like LIME or SHAP fall under this category.

By implementing these interpretability and explainability techniques, researchers can empower clinicians to understand how TreeNet arrives at its diagnoses. This fosters trust in the model's  findings and allows clinicians to make more informed decisions while leveraging the power of AI for improved patient care.

\subsection{Integration with Clinical Workflow}

For TreeNet and similar AI-powered medical image analysis tools to achieve widespread adoption, seamless integration into existing clinical workflows is crucial. Here are some potential avenues for future research in this direction:

\subsubsection{User-Friendly Interfaces}
Develop user-friendly interfaces that cater to clinicians' needs.  These interfaces should allow clinicians to  easily interact with TreeNet, upload images, and receive clear and concise results. The interface should  be intuitive and require minimal training for clinicians to become proficient in using it.

\subsubsection{Real-Time Integration with Endoscopy Systems}
Explore real-time integration of TreeNet with existing endoscopy machines. This would allow for real-time analysis of GI tract images during endoscopic procedures, providing immediate feedback and potentially aiding  clinicians in decision-making.  Security and privacy considerations  must be addressed to ensure patient data remains protected.

\subsubsection{Deployment on Low-Power Devices}
Investigate the deployment of TreeNet on low-power devices specifically designed for use in endoscopy or colonoscopy procedures. This would make TreeNet more portable and accessible in various clinical settings, even those with limited computational resources. Techniques like model compression or hardware acceleration can be explored to achieve this goal.

\subsubsection{Decision Support Systems}
Develop real-time decision support systems  that utilize TreeNet and other image analysis tools. These systems can  provide  clinicians with relevant information and  recommendations during diagnosis and treatment planning. This can include highlighting suspicious regions within the image, suggesting potential diagnoses, or offering treatment options based on the analysis results.

\subsection{Scalability and Deployment}
The ever-increasing volume of medical image data necessitates addressing scalability challenges for TreeNet and similar AI-powered tools. Future research should focus on optimizing TreeNet for deployment in large-scale healthcare systems while maintaining its performance and efficiency. 

\subsection{3D image Analysis}
While 2D medical images like X-rays and mammograms have long been a cornerstone of medical diagnosis, 3D imaging techniques are emerging as a powerful tool for healthcare professionals.  3D image analysis, the process of extracting meaningful information from these 3D datasets, holds immense potential for revolutionizing various aspects of medical care. Here's how 3D image analysis is shaping the future of medical diagnostics:

\subsubsection{Enhanced Visualization and Diagnosis}
3D image analysis allows for the creation of detailed, virtual representations of organs, tissues, and even entire anatomical structures. This provides clinicians with a superior view compared to traditional 2D images. Imagine a surgeon virtually dissecting a patient's heart in 3D to plan a complex surgery or a radiologist examining a tumor from all angles to assess its characteristics. This enhanced visualization can lead to more accurate diagnoses, improved surgical planning, and ultimately, better patient outcomes.

\subsubsection{Personalized Medicine}
3D image analysis can play a crucial role in personalized medicine. By analyzing detailed 3D models of a patient's anatomy, healthcare professionals can tailor treatment plans to the individual's specific needs.  For instance, 3D analysis of a patient's jawbone can be used to create custom-fit implants for dental procedures.  Similarly, 3D models of tumors can be used to plan targeted therapies that minimize damage to healthy tissues.

\subsubsection{Improved Treatment Monitoring and Evaluation}
3D image analysis allows for monitoring treatment progress over time by comparing serial 3D scans of the same patient. This can be particularly beneficial in monitoring the effectiveness of cancer treatment or tracking the healing process after surgery. By visualizing changes in the 3D structure, clinicians can adjust treatment plans as needed and evaluate the overall effectiveness of interventions.
\appendix
\chapter{Gastrointestinal (GI) tract Landmarks}
These landmarks are commonly referred to in medical contexts, particularly in the fields of gastroenterology and endoscopy. The normal appearance of these structures is essential for assessing the health of the gastrointestinal tract during diagnostic procedures. The gastrointestinal (GI) tract landmarks used in the research are as follows:
\section{Normal Z-Line}
    The Z-line is a junction between two types of epithelia in the esophagus. It marks the transition from the squamous epithelium of the esophagus to the columnar epithelium of the stomach. The "normal Z-line" refers to a typical appearance, without signs of pathology or abnormality.
    
\section{Normal Pylorus}
    The pylorus is the lower part of the stomach that connects to the small intestine. It contains the pyloric sphincter, a muscular valve that regulates the flow of partially digested food (chyme) from the stomach to the duodenum (the first part of the small intestine).

\section{Normal Cecum}
    The cecum is the first part of the large intestine (colon), located in the lower right abdomen. It connects to the ileum (the last part of the small intestine) and is involved in the absorption of fluids and salts.

\section{Retroflex Rectum}
    "Retroflex" refers to the act of bending backward. In endoscopy, the term "retroflex" is often used to describe the backward bending of the endoscope tip to view structures behind the instrument. In this case, "retroflex rectum" might refer to viewing the rectum by retroflexing the endoscope during a colonoscopy.

\section{Retroflex Stomach}
    Similar to retroflex rectum, "retroflex stomach" likely refers to the technique of retroflexing the endoscope to obtain a view of the stomach lining during an upper endoscopy (esophagogastroduodenoscopy or EGD).

\chapter{Pathological Findings}
    These pathological findings are often identified during procedures such as endoscopy or colonoscopy, which allow healthcare professionals to visually examine the gastrointestinal tract. The classification or grading of these findings helps in determining the severity of the condition and guiding appropriate treatment strategies. Some of the findings used in the research are as follows:

\section{Esophagitis}
    Esophagitis refers to inflammation of the esophagus. The classification into "a" and "b-d" may indicate different grades or stages of severity. Esophagitis can result from various causes, including reflux of stomach acid into the esophagus (gastroesophageal reflux disease or GERD), infections, or irritants.

\section{Polyps}
    Polyps are abnormal growths of tissue that can develop on the inner lining of various organs, including the gastrointestinal tract. In the context of the GI tract, polyps are often found in the colon (colorectal polyps) and may have the potential to become cancerous.

\section{Ulcerative Colitis (UC)}
    The disease typically involves the innermost lining of the colon, leading to the formation of ulcers and other inflammatory changes. The grading system, often using numbers 0 to 3, is commonly employed to describe the severity or extent of inflammation observed during medical assessments, such as endoscopic examinations.

\section{BBPS (Boston Bowel Preparation Scale)}
    The Boston Bowel Preparation Scale is a scoring system used to assess the quality of bowel preparation during a colonoscopy. The scores 0-1 and 2-3 represent different levels of bowel cleanliness.

\section{Hemorrhoids}
    Hemorrhoids, also known as piles, are swollen and inflamed veins in the rectum and anus that result in discomfort and bleeding. They can be internal or external.

\section{Barrett's}
    Barrett's esophagus is a condition in which the lining of the esophagus undergoes changes, becoming more like the lining of the small intestine. This condition is often associated with chronic gastroesophageal reflux disease (GERD), where stomach acid flows back into the esophagus.
    In Barrett's esophagus, the normal squamous epithelium lining of the lower esophagus is replaced by specialized columnar epithelium, which is more resistant to stomach acid but is considered a premalignant condition. Individuals with Barrett's esophagus have an increased risk of developing esophageal adenocarcinoma, a type of cancer.

\section{Barrett's Short Segment}
    Barrett's short segment refers to cases where the extent of Barrett's metaplasia is limited to a short segment of the esophagus, usually less than 3 cm. This is in contrast to long-segment Barrett's, where the affected segment is greater than 3 cm.

\chapter{Surgical Findings}
    The assessment of pathological and normal findings encompasses the pre-operative, intra-operative, and post-operative stages, particularly within the context of surgical interventions. In the course of these stages, attention is given to specific classifications, including dyed-lifted-polyps, dyed-resection-margins, and the ileum.
\section{Dyed-Lifted-Polyps}
    Refers to polyps that have undergone identification and elevation during diagnostic procedures, typically facilitated by the application of a dye or stain to enhance visualization.

\section{Dyed-Resection-Margins}
    Pertains to the margins of resected tissue during surgery, distinguished by the application of dye or staining for meticulous evaluation. This is particularly relevant in surgeries where precision is imperative, such as those addressing malignancies, where clear margins are critical.

\section{Ileum}
    Denotes the terminal portion of the small intestine, forming a connection with the cecum, the initial segment of the large intestine or colon. The ileum is integral to nutrient absorption.

\chapter{Endoscopic Procedures}
    Endoscopic procedures can be employed in various gastrointestinal scenarios, including situations where there is an impacted stool.

\section{Impacted-stool}
    An impacted stool refers to a hardened mass of feces that accumulates in the rectum or lower colon, causing difficulty in passing stool. Endoscopic interventions are utilized to visualize, assess, and manage such conditions. 

\section{Instruments}
    The category of "instruments" encompasses the various tools and devices used by surgeons during the course of a procedure. Instruments may include cutting tools, clamps, retractors, and other specialized devices tailored to the specific needs of the surgery.
    
\section{Colon-Clear}

    "Colon-clear" refers to a scenario where the endoscopic visualization reveals a clear and unobstructed view of the colon lining. This is typically ideal for a thorough examination and detection of abnormalities.

\section{Stool-Inclusions}

    This class indicates the presence of stool within the colon. The term "stool-inclusions" suggests that while there may be some stool present, it doesn't necessarily obstruct the view entirely.

\section{Stool-Plenty}

    "Stool-plenty" suggests a situation where the colon is heavily laden with stool, potentially hindering a clear view of the mucosal lining. This scenario may necessitate additional measures such as bowel preparation or, in some cases, interventions to clear the stool during the endoscopy.

\chapter{Non Diagnostic Classes}
These classes may be employed for documentation purposes, communication between healthcare professionals, or in research settings where specific observations need to be categorized. It's important to note that while these classes may not be intended for diagnostic purposes, they still provide valuable information about the conditions encountered during the endoscopic examination. Some of those classes are as follows:

\section{Blurry-Nothing}
    The term "blurry-nothing" suggests that, during the endoscopic procedure, the visual field appears unclear or indistinct, and there may not be identifiable structures or abnormalities. This class may be used to describe instances where the endoscopist encounters difficulty in obtaining a clear and focused view.

\section{Out-of-Patient}
    "Out-of-patient" could imply that the endoscopic examination did not take place within the typical outpatient setting. It might be used to classify cases where the procedure is not performed as part of a routine outpatient visit but, for example, in an emergency or inpatient setting.

%\appendix
\bibliography{k173501}
\bibliographystyle{unsrt}
\end{document}